\pdfoutput=1
\documentclass[a4paper,12pt,oneside]{amsart}
\usepackage[a4paper]{geometry}
\usepackage{amsaddr}
\usepackage[utf8]{inputenc} %direkte Eingabe von Sonderzeichen
\usepackage[T1]{fontenc} %teilt TeX mit, wie die Zeichenbelegung der verwendeten Schrift funktioniert.
\usepackage{lmodern} %Schrift, fast wie standart LaTeX aber in T1-Kodierung und Type-1-Format.
\usepackage[ngerman,english]{babel}

\usepackage{tikz}
\usetikzlibrary{calc}

\usepackage{caption} % linebreak in caption
\usepackage{amsmath}
\usepackage{amssymb}
\usepackage{amsthm}
\usepackage{nicefrac}    % \nicefrac
\usepackage{booktabs}    % \toprule, \midrule, \bottomrule
\usepackage{subfig}      % use multiple floats

\usepackage{tikz,pgfplots,pgfplotstable}
\usepgfplotslibrary{patchplots}
\usetikzlibrary{arrows,shapes,patterns}
\tikzset{dashdot/.style={dash pattern=on .4pt off 3pt on 4pt off 3pt}}
\usepackage{varwidth}

\usepackage{braket}
\usepackage[only,llbracket,rrbracket]{stmaryrd}

\usepackage{array}      % extends array and tabular environment
\usepackage{booktabs}   % improves tables, no vertical lines recommended
\usepackage[pdftex]{hyperref}

\theoremstyle{plain}
\newtheorem{theorem}{Theorem}[section]

\newtheorem{algo}[theorem]{Algorithm}

\theoremstyle{definition}

\newtheorem{remark}[theorem]{Remark}

\makeatletter

\newcommand{\Rmnum}[1]{\expandafter\@slowromancap\romannumeral #1@}
\makeatother

\newcommand*{\uptext}[1]{\text{\textup{#1}}}

\renewcommand*{\epsilon}{\varepsilon}
\renewcommand*{\rho}{\varrho}

\providecommand*{\vv}[1]{\ensuremath{\boldsymbol{#1}}}	%Vector
\providecommand*{\mv}[1]{\ensuremath{\boldsymbol{#1}}}	%Matrix

\providecommand*{\diverg}{\ensuremath{\operatorname{div}}} %div als Operator
\providecommand*{\dd}{\ensuremath{\mathrm{d}}}	%Integrations d als aufrechtes d
\providecommand*{\transp}{\ensuremath{t}}	% Operator to transpose a Matrix

\providecommand*{\abs}[1]{\ensuremath{\left \lvert #1 \right \rvert}}	%Betrag	|x|
\providecommand*{\norm}[1]{\ensuremath{\left \lVert #1 \right \rVert}}	%Norm		||x||
\providecommand*{\jump}[1]{\ensuremath{\left\llbracket #1 \right\rrbracket}}	%jump, needs stmaryrd package 
\newcommand{\setstyle}[1]{{\mathbb #1}}
\newcommand{\setR}{\setstyle{R}}

\newcommand{\setS}{\setstyle{S}}

\newcommand*{\admisr}{\ensuremath{ \mathcal{V}    }} % set of admissible volumes values
\newcommand*{\admis}{ \ensuremath{ \mathcal{A}    }} % set of admissible volumes values

\newcommand*{\sat}{  \ensuremath{ \uptext{sat}       }} % saturation
\newcommand*{\vap}{   \ensuremath{ \uptext{vap}        }} % vapour
\newcommand*{\liq}{   \ensuremath{ \uptext{liq}        }} % liquid
\newcommand*{\Vap}{   \ensuremath{ \uptext{Vap}        }} % vapour
\newcommand*{\Liq}{   \ensuremath{ \uptext{Liq}        }} % liquid
\newcommand*{\tmin}{  \ensuremath{ \uptext{min}        }} % text min 
\newcommand*{\tmax}{  \ensuremath{ \uptext{max}        }} % text min 
\newcommand*{\surf}{  \ensuremath{2\,\zeta\,\kappa}} % surface tension

\newcommand{\xGP}{-0.866, -0.339,  0.339, 0.866}

\newcommand{\U}{\ensuremath{\vv{U}}}

\begin{document}

\title[Sharp interface method for compressible liquid-vapor]%
{A sharp interface method for compressible liquid-vapor flow with phase transition and surface tension}

\author[S. Fechter, C.-D. Munz, C. Rohde]{Stefan Fechter and Claus-Dieter Munz}
\address{Institut f\"ur Aerodynamik und Gasdynamik, \\Universit\"at Stuttgart, Pfaffenwaldring 21, 70569 Stuttgart, Germany}

\author[C. Zeiler]{Christian Rohde and Christoph Zeiler}
\address{Institut f\"ur Angewandte Analysis und Numerische Simulation, \\Universit\"at Stuttgart, Pfaffenwaldring 57, 70569 Stuttgart, Germany}

\email{stefan.fechter@iag.uni-stuttgart.de}
\email{munz@iag.uni-stuttgart.de}
\email{Christian.Rohde@mathematik.uni-stuttgart.de}
% \email{Christoph.Zeiler@mathematik.uni-stuttgart.de}
\email{zeilerch@ians.uni-stuttgart.de}


\begin{abstract}
The numerical approximation of  non-isothermal  liquid-vapor flow within the compressible regime is a difficult task because  
complex physical effects at the phase interfaces can govern the global flow behavior.
We present  a sharp interface approach  which treats  the interface as a shock-wave like discontinuity.
Any mixing of fluid phases  is avoided by using the flow solver in the bulk regions 
only, and a  ghost-fluid approach close to the interface.
The coupling states for the numerical solution in the bulk regions are  determined by  the solution of local  multi-phase Riemann 
problems across the interface.  The Riemann solution accounts for the relevant physics by enforcing appropriate jump conditions 
at the phase boundary. A  wide variety of interface effects can be  handled in a thermodynamically consistent way. This   includes
surface  tension or mass/energy transfer by phase transition. Moreover, the local normal speed  of the interface, which is needed to calculate 
the time evolution of the interface, is given by the Riemann solution. The interface tracking itself is based on a level-set method.\\  
The focus in this paper is the description of the multi-phase Riemann solver and its usage within  
the  sharp interface approach.  One-dimensional problems  are selected to validate the approach. Finally, the three-dimensional 
simulation of a wobbling droplet and  a shock droplet interaction in two dimensions are shown. In both problems 
phase transition and surface tension determine the global bulk behavior.
\end{abstract}

\maketitle

%% main text
\section{Introduction} \label{sec:intro}

The   numerical modeling of multi-phase flow   is a very active field of research. In this paper we are interested
in  models for  fully compressible  
regimes with liquid and vapor  bulk phases  that  cope  correctly  with 
phase transition and surface tension effects. Our focus   is on 
the direct numerical simulation where single interfaces  separating the bulk dynamics  have to be resolved.
(See e.g.~\cite{saurel2008modelling,zein2010modeling} for alternative homogenized models.)

There are basically two  different approaches to model compressible multi-phase flows, the 
diffuse interface and the sharp interface approach. In the first, a smooth internal layer,  that 
has to be captured by the numerical  
method, stands for the interface. In particular,  artificial mixture states may occur.
Typically only one set of equations is solved in the whole computational domain, e.\,g.~the Navier-Stokes-Korteweg systems
(\cite{ANDMCFWHE1998,Rohde-05}). In the second approach, the sharp interface approach, 
the interface is represented as 
a discontinuity in the density field,  separating the computational domain in two bulk regions. 
The fluid flow in both of the bulk regions is  described by  the standard single-phase conservation equations.
The interface appears as an unknown interior boundary. Appropriate 
jump conditions couple the states of the bulk regions at the interface and have to ensure the well-posedness of the overall model.

From this description it becomes obvious where the problems occur in both approaches.
For problems in which the width of the physical interface is smaller than the typical grid cell size,
the diffuse profile in the diffuse interface approach has still to be thermodynamically consistent with the physics. 
But, within the diffused numerical interface non-physical mixing states may occur, for which an artificial equation of state
has to be defined. In the sharp interface approach the bulk phases are separated to avoid any mixing at the interface, which is in  
time-dependent problems a big challenge for the numerical framework.  
\bigskip

In this paper, we concentrate on  the sharp interface approach in the compressible flow regime including phase transition and surface tension.
For the  fluid flow in the bulk phases we restrict ourselves 
to a fluid that is described by the Euler equations with the conservative variables mass, momentum and total energy. The equation of state
(EOS) is used to calculate the primitive unknowns pressure and temperature.
In Section \ref{sec:model} of the paper we review this model and introduce appropriate coupling conditions at the phase interface. 
% In addition, we describe the local solution with given states from both bulk regions based on the Riemann problem at a phase interface. 
In the isothermal case it is well understood that the coupling conditions at a phase interface should consist of the 
mass conservation relation, a dynamical version of the Young-Laplace law for momentum balance,
and an additional Gibbs-Thomson relation to control the entropy production (see \cite{ABEKNO2006,rohde2015relaxation}). 
% \cite{Delhaye}
These conditions remain  valid in the  temperature dependent case that we consider here. 
However, since there is no mechanism for heat conduction, the release of latent heat has to be modeled in a different way. 
We suggest to use an algebraic jump condition.

In Section \ref{sec:riemannsolver}  we present a constructive algorithm to solve the generalized Riemann problem for the full Euler equations 
with initial data from different phases. The solution of the Riemann problem is supposed to include  in addition to the standard waves (shock, rarefaction,
contact wave) a  discontinuous wave 
that obeys  exactly the  relations from Section \ref{sec:model}. This wave  represents  the phase interface. 
The design of the Riemann solver is a complex issue not only due to the non-standard jump conditions across the phase boundary   but also since 
the hyperbolicity of the Euler equations breaks down and an 
elliptic spinodal region occurs for two-phase fluids, see e.\,g. \cite{menikoff1989riemann,MUE1985}. 
Let us note that the analysis of the  
Riemann problem for two-phase problems has been a very active field of research in the past decade 
(see  \cite{LEF2002} for a general theory  and  e.g.~\cite{HAT2004,JAEROHZER2012,MERROH2007,DEHAWA13,GODSEG2006,MUEVOS2006} for
specific examples) but is mostly restricted to either 
the isothermal case  or to  homogeneous coupling conditions that neglect surface tension, latent heat, and entropy production. The resulting coupling conditions are combined with a ghost-cell approach 
as the basic building block in our sharp resolution of a phase interface.
 
Our overall numerical approach relies on the idea to  use  local properties of the interface (local speeds, adjacent bulk states).  
The essential tool to compute these quantities    is the {generalized Riemann solver}  from Section \ref{sec:riemannsolver}
for input states from different bulk states. 
The other core building block of the numerical scheme     with sharp interface treatment is introduced in Section \ref{sec:bulksolver}.
It is 
a compressible bulk flow solver combined with    a tracking method for the interface. Precisely we use a  discontinuous Galerkin scheme 
with a finite-volume sub-cell resolution (see \cite{fechter2015multiphasedg}) for the flow and the level-set equation that governs 
the interface tracking. %The interface velocity is obtained from the multi-phase Riemann problem. 
To avoid any numerical smearing at the interface, the ghost-fluid approach from \cite{FEDASL1999}
is adapted to our situation. For much simpler isothermal  phase transition models  a similar approach can be found in 
\cite{CHACOQENGROH2012,DREROH2008,MERROH2007}. The focus and the novelty in this paper is the extension of the method to handle realistic flow regimes, that 
are governed by possibly non-isothermal  phase change and surface tension effects. \\  
The capability of our method to cope with such scenarios is  demonstrated in the final 
numerical Section \ref{sec:numerics} that provides also a 
validation of the numerical method. As an application to physically realistic droplet dynamics we consider evaporating droplets, a 
wobbling droplet, and a shock-droplet interaction.
Finally, we compare our model to experimental results from Simoes-Moreira\&Shepherd \cite{simoes1999evaporation} and Reinke\&Yadigaroglu \cite{reinke2001explosive}.

% {\tt TODO: gnerischer Einsatz anderer Riemannlöser und so, Link zu Multiscale modelling}
\section{The mathematical model and the basic numerical approach} \label{sec:model}
\subsection{Governing equations}
In a sharp interface method for multi-phase flow the basic assumption is that the 
width of the physical phase interface is much smaller than one grid cell. The implication is then to model 
the dynamical interface as a moving discontinuity within the flow field. Let us assume that this interface separates 
the computational domain $\Omega$ into the two domains $\Omega_\vap(t)$ and $\Omega_\liq(t)$
called bulk regions.
Here, ``vap'' stands for vapor and ``liq'' for liquid. We assume that the flow is inviscid and it 
is described in both bulk regions by the Euler equations
\begin{align}\label{eq:euler}
\begin{array}{rcccl}
  \rho_t &+& \diverg (\rho\, \vv v) &=&0,\\[1.5ex]
{(\rho\,\vv v)}_t
         & +
         & \diverg{(\rho\,\vv v \otimes \vv v + p\,\mv{I} )}
         & =
         & \vv{0},\\[1.5ex]
  {(\rho\, e )}_t &+& \diverg ( (\rho\,e+p) \vv v)  &=&0,       
    \end{array}
\end{align}
with suitable initial and boundary conditions.
Here, the variables are fluid density  $\rho=\rho(\vv{x},t)$,  velocity 
$\vv v=\vv v(\vv{x},t) = ( v_1(\vv{x},t),v_2(\vv{x},t), v_3(\vv{x},t))^\transp $ 
and the specific total energy $e = e(\vv x,t)$, which is related to 
the specific internal energy $\epsilon$ via $e=\epsilon+\frac{1}{2} \vv{v} \cdot \vv{v}$. 

The system \eqref{eq:euler} is closed by  an equation of state that  connects the pressure $p$
to the other variables and considers thermodynamic effects. In the bulk regions $\Omega_\vap(t)$ and $\Omega_\liq(t)$   
we assume that 
the fluid is in local thermodynamic equilibrium, which means that the  
state  of the fluid %at any time
%$t\in[0,\theta]$ and $\vv{x}\in\Omega_\vap(t) \cup \Omega_\liq(t)$ 
is determined by any two independent thermodynamic state variables. Note that we use in the sequel 
the same symbol  for  some thermodynamic variable, even if they are considered to depend on different
arguments. 
For instance we use 
\begin{align*}
 p=p(\rho,\epsilon)=p(\tau,S)=p(\tau,T), && \epsilon=\epsilon(\tau,S)=\epsilon(\tau,T),
\end{align*}
where $\tau=1/\rho$, $S$, and $T$, are specific volume, entropy, and temperature, respectively. 

%is the sum of the specific inner energy  $\epsilon=\epsilon(\vv{x},t)\in\setR$ and the specific kinetic energy $\frac{1}{2}\norm{v}^2$. 

\begin{figure}\centering
\begin{tikzpicture}

 \begin{axis}%
              [
               xmin=0.3, 
               xmax=5, 
               ymin=0.3, 
               ymax=1.1, 
               xlabel={specific volume $\tau$}, 
               ylabel={pressure $p$}, 
               xticklabels={,,},
               yticklabels={,,},
               x post scale = 1.5, 
               y post scale=0.7,
               ylabel style = {yshift=-25pt},
               xlabel style = {yshift=10pt} 
               ]
  % areas
  \addplot[pattern color=gray!50, pattern=north east lines, draw=none, area legend,  domain=0.7:1, smooth]  {  8* 9/4/(x^3) *(x-(1/3))^2 /(3*x-1) - 3/(x^2)   } -- ({axis cs:0,1}) -- ({axis cs:0,0})  --cycle ;
  \addplot[pattern color=gray!50, pattern=crosshatch dots, draw=none, area legend,  domain=1:3, smooth]  {  8* 9/4/(x^3) *(x-(1/3))^2 /(3*x-1) - 3/(x^2)   } -- ({axis cs:6,0}) -- ({axis cs:6,1})  --cycle ; 
 
  % isentropes
  \addplot[solid] plot[domain=0.51:5, smooth]  ({\x}, {(8*0.85)/(3*\x-1) - 3/(\x*\x)}  );
  \addplot[dashed] plot[domain=0.55:5, smooth]  ({\x}, {(8*0.9 )/(3*\x-1) - 3/(\x*\x)}  );
  \addplot[dashdotted] plot[domain=0.6:5, smooth]  ({\x}, {(8*0.95)/(3*\x-1) - 3/(\x*\x)}  );
  \addplot[ fill=white, domain=0.7:3, smooth]  {  8* 9/4/(x^3) *(x-(1/3))^2 /(3*x-1) - 3/(x^2)   } ;
 
  % saturation curve
  \addplot[thick,dashed,smooth] table[x expr={\thisrowno{2}}, y expr={\thisrowno{0}}] {Data/isotherms_sat.dat};
  \legend{$\admis_\liq$,$\admis_\vap$,$p\vert_{S_0}$, $p\vert_{S_1}$, $p\vert_{S_2}$, ,$p^\sat$}
 \end{axis}
\end{tikzpicture} 
   \caption{Phase diagram: The shaded areas mark the liquid and vapor phases, thin lines indentify the isentropes $\tau\to p(\tau,S)$ at constant specific entropy values $S_0<S_1<S_2$ and the thick dashed line identify the saturation curve.}
   \label{fig:VdW_p}
\end{figure}
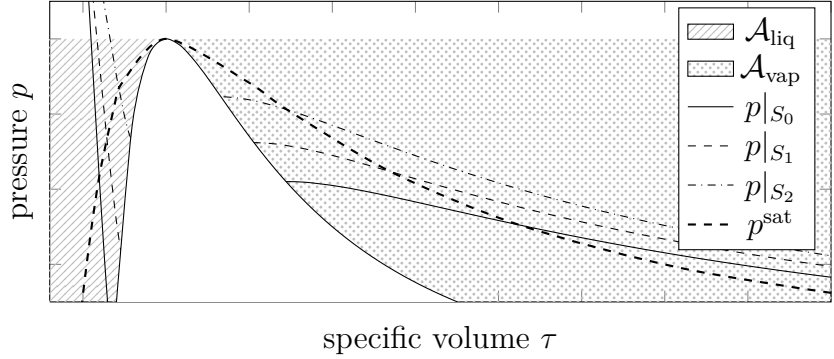

%\section{Two-Phase Thermodynamics}
%Let us consider a fluid in local thermodynamical equilibrium which means that the thermodynamical 
%state  of the fluid %at any time
%$t\in[0,\theta]$ and $\vv{x}\in\Omega_\vap(t) \cup \Omega_\liq(t)$ 
%is determined by any two thermodynamial variables. Note that  we will use in the sequel 
%the same symbol  for  some thermodynamical variable, even if it is considered to depend on different
%arguments. 
%For instance we use 
%\begin{align*}
% p=p(\rho,\epsilon)=p(\tau,S)=p(\tau,T), && \epsilon=\epsilon(\tau,S)=\epsilon(\tau,T),
%\end{align*}
%where $\rho$,  $\tau=1/\rho$, $p$, $\epsilon$, $S$, $T$, are the density, specific volume, 
%pressure, specific internal energy, entropy, temperature, respectively. 
Let us concentrate for a moment on the specific internal energy $\epsilon(\tau,S)$ and consider 
pressure and temperature to be given by  
\begin{align} \label{eq:U:derivation}
  p(\tau,S) = -\epsilon_\tau(\tau,S), &&  T(\tau,S) = \epsilon_S(\tau,S),
\end{align}
while the specific Helmholtz free energy $F$ and the specific Gibbs free energy $G$ are related through
\begin{align}\label{eq:U:potentials}
  F(\tau,S) = \epsilon(\tau,S) - S\,T(\tau,S), && G(\tau,S) = F(\tau,S) + \tau\,p(\tau,S).
\end{align} 
We consider fluids below the critical point such that 
%--~neglecting solid configurations~--
only liquid and vapor states exist.  The state space $\admis_{\liq}$ of the liquid region $\Omega_\liq(t)$ and 
the state space $\admis_{\vap}$ of the vapor region $\Omega_\vap(t)$ 
consist of states such that $\epsilon=\epsilon(\tau, S)$ is strictly convex.
% are characterized by a strictly convex function $\epsilon=\epsilon(\tau, S)$. 
%these bulk states we 
%assume that two disjunct open  sets $\admis_{\liq}$ and $\admis_{\vap}$ are given such that the function
%\begin{align} \label{eq:Uassumption}
% \epsilon: \left\{ \begin{array}{ccl}
%    \admis_{\liq}\cup\admis_{\vap} &\to& \setR, \\
%     (\tau, S) &\mapsto &\epsilon(\tau, S)   
%  \end{array} \right.
%  \qquad
%  \text{ is strictly convex}.
%\end{align}
%The sets $\admis_{\liq/\vap}$ define the liquid and the vapor phase. 
As a consequence, a positive real function for the speed  
of sound $c$ is obtained in both phases, because we have 
\begin{align}\label{eq:sound}
  c(\tau, S) := \tau \sqrt{-p_\tau(\tau,S)} =  \tau \sqrt{ \epsilon_{\tau\tau}(\tau,S)} >0 \qquad
 \left( (\tau,S) \in  \admis_{\liq}\cup\admis_{\vap} \right).
\end{align}
To illustrate the set-up, Figure~\ref{fig:VdW_p} shows the isentropes $\tau\mapsto p(\tau,S)$ for a constant value of  entropy $S$. 
Note in particular that the isentropes are monotone decreasing in the liquid and the vapor phase.
The area between $\admis_{\liq}$, $\admis_{\vap}$ is called the elliptic or spinodal region and refers to unstable thermodynamic states. 
% The connection of the states in the bulk phases is modeled by an undercompressive shock wave.
% and is excluded in this work. 
%{\tt sollte hier nichtdie elliptische Region erklaert werden.}

Furthermore we require that a saturation curves exist, that takes surface tension into account.
We assume that for any mean curvature value $\kappa\in\setR$, there is a liquid saturation curve  $\alpha\mapsto(\tau_{\liq}^\sat,S_{\liq}^\sat)\in\admis_{\liq}$ and a vapor saturation curve $\alpha\mapsto(\tau_{\vap}^\sat,S_{\vap}^\sat)\in\admis_{\vap}$, such that 
\begin{align}
  \begin{aligned} \label{eq:saturation}
      T\left(\tau_{\liq}^\sat(\alpha),\eta_{\liq}^\sat(\alpha)\right)&=T\left(\tau_{\vap}^\sat(\alpha),\eta_{\vap}^\sat(\alpha)\right), \\
      G\left(\tau_{\liq}^\sat(\alpha),\eta_{\liq}^\sat(\alpha)\right)&=G\left(\tau_{\vap}^\sat(\alpha),\eta_{\vap}^\sat(\alpha)\right),  \\
      p\left(\tau_{\liq}^\sat(\alpha),\eta_{\liq}^\sat(\alpha)\right)&-p\left(\tau_{\vap}^\sat(\alpha),\eta_{\vap}^\sat(\alpha)\right) = \surf
  \end{aligned} 
\end{align}
holds, where $\zeta\ge0$ is the constant surface tension coefficient.
Note that we can parameterize the curves with temperature, liquid/vapor pressure or specific Gibbs free energy.
Furthermore, we introduce the functions $T^\sat=T^\sat(\alpha)$, $p_{\liq/\vap}^\sat=p_{\liq/\vap}^\sat(\alpha)$ with $\alpha\in\{T$, $\mu$, $p_\liq$, $p_\vap\}$, that are the (saturated) temperature and pressure, respectively, evaluated on the saturation curve.

\medskip

For hyperbolic 
systems the notion of weak entropy solutions is widely believed to be 
the correct solution concept.
Hence, we look for integral solutions that 
%Thus  we search for functions  $\vv U=\vv U(\vv{x},t )$ which are weak solutions with $\vv U\in \admisr_{\liq/\vap}$
% for  almost all  $ (\vv{x},t) \in \Omega_{\liq/\vap}(t) \times [0,\theta]  $     and 
satisfy the entropy condition 
\begin{align}\label{eq:mathentropy}
(-\rho\,S)_t +\diverg \left(-\rho\,S\, \vv{v} \right)   \le 0 
\end{align}
weakly in the  bulk regions $\Omega_\vap(t)$ and $\Omega_\liq(t)$. 
%(not in the complete domain $\Omega$ where we have to take 
%into account the latent heat, see \eqref{eq:kinrel} below). 
Using the convexity of $\epsilon=\epsilon(\tau, S)$ in $\admis_\liq \cup \admis_\vap$ one readily checks that the
mathematical entropy $-\rho\,S$ is convex 
as a function  of $(\rho, \rho\,\vv v, \rho\,e)$ for all states in the bulk regions $\Omega_\vap(t)$ and $\Omega_\liq(t)$. 
%  and thus an mathematical entropy for \eqref{eq:euler} 
%(see also \cite[Chapter \Rmnum{2}]{GODSEG2006}).

% Furthermore we require that saturation states exist, i.e., there are smooth functions
% $T^\ast\mapsto(\tau_{\liq}^\sat,S_{\liq}^\sat)\in\admis_{\liq}$ and 
% $T^\ast\mapsto(\tau_{\vap}^\sat,S_{\vap}^\sat)\in\admis_{\vap}$ such that temperature, pressure and specific Gibbs free energy evaluated in that states satisfy
% \begin{align}\label{eq:saturation}
%  T^\ast=T_\liq&=T_\vap, &  p_\liq&=p_\vap, & G_\liq&=G_\vap.
% \end{align}

\subsection{Jump conditions for the liquid-vapour phase boundary}

We complete the mathematical model with the Euler model \eqref{eq:euler} for the bulk regions by coupling conditions for the phase boundary $\Gamma(t) := \bar{\Omega}_\vap(t)\cap\bar{\Omega}_\liq(t)$.
%It remains to provide coupling conditions for \eqref{eq:euler}   at the   free boundary   $\Gamma(t)$.  
Let $t\geq0$ be arbitrary but fixed and let some $\vv{\xi}\in \Gamma(t)$ be given. 
We denote the speed of $\Gamma(t)$ in the normal direction $\vv{n}=\vv{n}(\vv{\xi},t)\in \setS^{2}$ by $s=s(\vv{\xi},t)\in\setR$. 
The normal vector $\vv{n}$ is always chosen as pointing into the vapor domain $\Omega_\vap(t)$. 
Across the interface the following trace conditions, which represent 
the conservation of mass and 
the balance of momentum and energy in the presence of capillary surface forces and latent heat, are posed:
\begin{align}
  \jump{\rho\, (\vv v\cdot \vv{n} - s) } &= 0, \label{eq:rh:rho}\\
  \jump{\rho\, (\vv v\cdot \vv{n} - s)\,\vv{v}\cdot\vv{n} + p} &= \surf, \label{eq:rh:m}\\   
  \jump{ \vv{v} \cdot \vv{t}_1} = \jump{ \vv{v} \cdot \vv{t}_2}         &=0,      \label{eq:rh:mt}    \\ 
  %\jump{\rho\, (\vv v\cdot \vv{n} - s)\,e + p\,\vv v \cdot \vv{n} + q } &= s\,\surf, \label{eq:rh:e}  
 \textstyle \jump{G+T\,S + \frac{1}{2}(\vv v\cdot \vv{n} - s)^2 } & = L.   \label{eq:rh:e}
\end{align}
Thereby $\jump{a}:=a_\vap-a_\liq$ and $a_{\vap/\liq} :=\lim_{\epsilon\to 0, \epsilon>0} a(\vv{\xi} \pm \epsilon \vv{n})$ 
for some quantity $a$ defined in $\Omega_\vap(t) \cup \Omega_\liq(t)$. 
In \eqref{eq:rh:m} by $\kappa = \kappa(\vv{\xi},t)\in\setR$ we denote the mean curvature of $\Gamma(t)$ associated with orientation given through the choice of the normal $\vv{n}$. 
The constant surface tension coefficient is $\zeta\ge0$, and $\vv{t}_1,\vv{t}_2 \in \setS^{2}$ are a complete set of vectors tangential to $\vv{n}$. 
Note that in this study we assume constant surface tension such that tangential variation of $\zeta$ along the interface is ignored.
With \eqref{eq:rh:mt} we assume there is no slip tangential to the interface.

It remains to comment on \eqref{eq:rh:e}.   In the conditions  \eqref{eq:rh:rho}--\eqref{eq:rh:e} the effects of higher order heat fluxes 
are ignored. However, the amount of heat energy that
goes into evaporation  or that is liberated in the inverse  condensation process has to be taken into account. 
The interface source term $L$ in \eqref{eq:rh:e} accounts for that release or absorption of latent heat. 
We assume that it can be expressed  by
\begin{align}\label{eq:L}
 L=L(T^\ast)=T^\ast \left(S_\vap^\sat(T^\ast)- S_\liq^\sat(T^\ast)\right).
\end{align}
For isothermal phase transitions, expression \eqref{eq:L} is just the definition of latent heat. Here, it is assumed that \eqref{eq:L} holds also for non-isothermal transitions, with respect to some reference temperature $T^\ast$. This temperature can  
for instance be chosen equal to some ambient system temperature $T^\ast=T_\infty$ or to the average temperature of the bulk phases at the interface.
If it is reasonable to consider an ambient system pressure $p_\infty$ one may use $L=T^\sat(p^\infty) \jump{S^\sat(p_\infty)}$ 
instead of \eqref{eq:L}.

A discontinuous wave $\vv{U} = (\rho,\rho\,\vv v, \rho\,e)^\transp$ with 
\begin{equation}\label{def:wave}
 \U(\vv{x},t) =
\begin{cases}
 \U_{\liq} &: \vv{x}\cdot\vv{n} -s t\leq0 \\
 \U_{\vap} &: \vv{x}\cdot\vv{n} -s t>0
\end{cases} \qquad (\vv{n}\in\setS^2,\,   s\in\setR),
\end{equation}
is called a    planar phase interface if conditions  \eqref{eq:rh:rho}--\eqref{eq:rh:e} hold.
 In this work, we are interested only  in interfaces that are  
 non-characteristic and subsonic, i.\,e.\  the adjacent states satisfy 
\begin{equation}\label{eq:subsonic}
   \abs{\rho_\liq (\vv v_\liq\cdot \vv{n} - s) } =  
   \abs{\rho_\vap (\vv v_\vap\cdot \vv{n} - s) }
   <  \min \{ c_\liq, c_\vap \}.
\end{equation}
Otherwise it will not be possible to solve the Riemann problem close to equilibrium states (see Remark~\ref{rem1} below).\\
For the isothermal case it is  known (see e.g.\ \cite{ABEKNO2006,TU93}) that well-posedness of the free boundary value problem requires an additional condition. 
The same holds for \eqref{eq:euler} with van der Waals fluids (see \cite{LEFTHA2003}). We assume that this is also necessary for more general
equations of state fulfilling \eqref{eq:U:derivation}.\\  
One possible choice is yet another algebraic coupling condition.
In the literature, kinetic relations have been suggested (see  \cite{ABEKNO2006,TU93}), which control the entropy change explicitly.
Under the presence of latent heat, the local entropy balance corresponding to \eqref{eq:mathentropy} reads
\begin{align*}
  \jump{\rho\, (\vv v\cdot \vv{n} - s)\, (S - S^\sat(T^\ast)) } &= \eta, 
\end{align*}
where $\eta\geq0$ is the interface entropy production. This construction lets the  entropy production vanish, if the mass flux 
\begin{align} \label{eq:massflux}
  j = \rho_\liq (\vv{v}_\liq \cdot \vv n- s) =    \rho_\vap (\vv{v}_\vap\cdot  \vv n-s)
\end{align} 
across the interface vanishes. On the other, the second law of Thermodynamics implies that $\eta$ is not negative. 
Thus a simple constitutive law is $\eta\,T^\ast=k^\ast\,j^2$
that uses the previously defined reference temperature $T^\ast$ and an entropy production constant $k^\ast \geq0$.
With \eqref{eq:rh:e} the kinetic relation
\begin{align}\label{eq:kinrel}
  \textstyle\jump{G + \frac{1}{2}(\vv v\cdot \vv{n} - s)^2} + \jump{T\,S}-T^\ast\jump{S}         = -k^\ast\,j 
\end{align}
follows.

We conclude the description of the mathematical model under consideration with the following remarks:
\begin{remark}[Kinetic relation and supersonic phase interfaces]\label{rem1}
\hfill
\begin{enumerate}
  \item Note that in the isothermal case equation \eqref{eq:kinrel} reduces to \\
        $\jump{G + \frac{1}{2}(\vv v\cdot \vv{n} - s)^2} = k^\ast\,j$ if $T^\ast$ 
        is chosen as the ambient temperature.
        That is the suggested kinetic relation in \cite{ABEKNO2006,TU93}.
        The local energy balance \eqref{eq:rh:e} is usually not considered, so that release and absorption of latent heat appears implicitly.

\item 
There are also solutions of \eqref{eq:rh:rho}--\eqref{eq:rh:mt} connecting densities in 
different phases where  the  corresponding  shock 
wave is supersonic (i.e.~\eqref{eq:subsonic} is in particular  violated). Then an additional kinetic relation like \eqref{eq:massflux}
is superfluous to solve e.g.~the Riemann problem. However it appears not to be physical to have supersonic phase interfaces.
\end{enumerate}
  
\end{remark}

\begin{remark}[Static phase boundaries and latent heat]\label{rmk2}
\hfill
\begin{enumerate}
 \item A static phase boundary ($s=0$) satisfies
 \begin{align*}
   \jump{\vv v} &= \vv 0, & \jump{p} &= \surf, &\jump{G+T S} &= L(T^\ast), & T^\ast\,\jump{S} &= L(T^\ast). 
 \end{align*}
 \item One particular static phase boundary is given for $T^\ast=T_\liq=T_\vap$. With \eqref{eq:L} one finds
 \begin{align*}
   \jump{\vv v} &= \vv 0, & \jump{p} &= \surf, &\jump{G} &= 0, & \jump{T} &= 0.
 \end{align*}
 This corresponds to the saturating conditions \eqref{eq:saturation}.
 
 \item Ngan \& Truskinovsky show in \cite{NGANTRU1998}, that static phase boundaries satisfy $\jump{p} = \surf$, $\jump{G+T S} = 0$, $\jump{S}=0$ in the limit of zero heat conductivity. We find the same conditions neglecting the latent heat ($L\equiv 0$).
 Note that, assuming $\jump{S}=0$, usually leads to high temperature jumps.
 
 Riemann problems for the two-phase model with $L\equiv0$ and $\zeta=0$ are considered in \cite{MUEVOS2006,HAT2004,CHEHAT2015}.

\end{enumerate}
\end{remark}

\subsection{Numerical modeling}

In the following we briefly sketch the numerical modeling to motivate the next sections with a detailed description. 
The structure of the numerical approximation follows the mathematical model in the previous subsections. 
For the solution of the Euler equations \eqref{eq:euler} in the liquid or vapor region, a 
flow solver is needed that allows the treatment of a real equation of state. Let's assume this is a finite volume scheme 
with any numerical flux. In the bulk regions no interface phenomena have to be resolved and the numerical flux may be based on 
any finite volume flux calculation. We use a Godunov-type scheme with an approximate Riemann solver. Away from the position of the phase interface 
the solution of the Riemann problem is the usual (single-phase) one with four constant states.   
In addition to the flow solver for every bulk region we need a coupling procedure that takes care of the jump conditions \eqref{eq:rh:rho}--\eqref{eq:rh:e}
at the interface, and we need information about the position of the phase interface. Hence, the numerical 
modeling consists of the following four building blocks for the sharp-interface resolution of two-phase flows:  
\begin{enumerate}
	\item Interface tracking method,
	\item Compressible flow solver for the bulk phases,
	\item Two-phase Riemann solver that takes surface tension and phase transition into account,
	\item Coupling method of the bulk regions at the interface.
\end{enumerate}
The interface tracking provides us with the actual position of the interface at each time step. From the flow field we can 
interpolate values to this position from both sides, from the liquid and from the vapor region. The solution of a two-phase 
Riemann problem then estimates the velocity of the phase interface. This information is needed to calculate the time 
evolution of the interface in the tracking step. The two-phase Riemann problem solver also provides the time evolution of the states at the interface. These states from the left and 
right hand side at the interface are used for the coupling to the flow solver in the bulk phases. 
To avoid any mixing of liquid and vapor, a ghost fluid technique \cite{FEDASL1999} is applied, by which the grid cells that 
contain the interface have two states -- a liquid and a vapor state as given by the solution of the two-phase Riemann problem. 
With this information the bulk solver calculates the flow field in the liquid and the vapor 
region separately.
The ghost cell approach avoids unphysical mixed states, for which an thermodynamic consistent mixing equation of state has to be designed. 

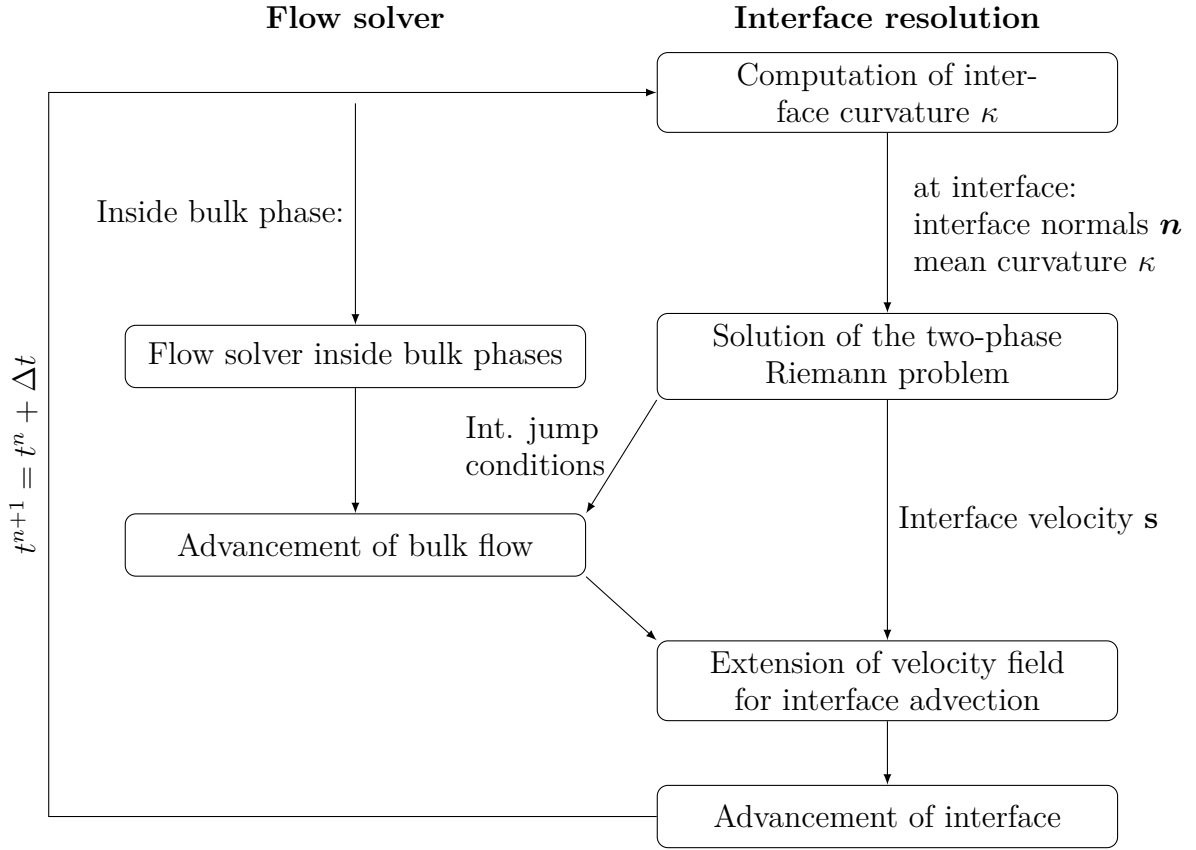
\begin{figure}
 \centering
 \begin{tikzpicture}[scale=1,node distance=1.8cm]
 \tikzstyle{box} = [rectangle,draw, text width=14.0em, text centered, rounded corners, minimum height=2em]
 \tikzstyle{myarrow} = [->,>=latex]
 \node (1a) {};
 \node[box,right of=1a,node distance=7cm] (1) {Computation of inter\-face curvature $\kappa$};
 \node[box,below of=1a,node distance=3.5cm] (2) {Flow solver inside bulk phases};
 \node[box,right of=2,node distance=7cm] (3) {Solution of the \mbox{two-phase} Riemann problem};
 \node[box,below of=2,node distance=2.5cm] (4) {Advancement of bulk flow};
 \node[below of=4] (5a) {};
 \node[box,right of=5a,node distance=7cm] (5) {Extension of velocity field for interface advection};
 \node[box,below of=5] (6) {Advancement of interface};
 \draw[myarrow] (1a.south)--node[midway,left] {Inside bulk phase:} (2.north);
 \draw[myarrow] (1.south)-- node[pos=0.5,right,xshift=0.2cm] {\parbox{4cm}{at interface: \newline interface normals $ \vv n$ \newline mean curvature $\kappa$}} (3.north);
 \draw[myarrow] (3.south west)-- node[pos=0.4,left] {\parbox{2.0cm}{Int. jump \newline conditions}} (4.north east);
 \draw[myarrow] (3.south)-- node[midway,right] {Interface velocity $\mathbf{s}$} (5.north);
 \draw[myarrow] (2.south)--(4.north);
 \draw[myarrow] (4.south east)--(5.north west);
 \draw[myarrow] (5.south)--(6.north);
 \draw[myarrow] (6.west)--++(-8.0cm,0) -- node[left,midway]{\rotatebox{90}{$t^{n+1} = t^n+\Delta t$}} ([xshift=-8.0cm]1.west)-- (1.west) ;
 \node[above of=1a,node distance=1cm] (10) {\textbf{Flow solver}};
 \node[right of=10,node distance=7cm] {\textbf{Interface resolution}};
\end{tikzpicture}
 \caption{Program structure for the simulation of compressible two-phase flows with surface tension and phase transition; Left: Flow solver in the bulk phases, Right: Interface treatment.
 % Interface related effects are resolved by a local Riemann solver and couple the bulk flow solver by numerical fluxes.
 }
 \label{fig:algo}
\end{figure} 

The solution strategy is visualized in Figure~\ref{fig:algo}.
On the left-hand side, the part of the flow solver within the bulk phases are listed and the right side shows all interface related tasks. 
Assuming that the location and the geometry of the interface is known, the interface curvature is evaluated. The state 
in the liquid and vapor bulk phase at the interface as well as the curvature are used to determine the local solution 
based on a two-phase Riemann solver. This local solution is used to couple the bulk regions within a ghost 
fluid approach as well as for the definition the interface velocity that is needed to transport the interface, 
as already mentioned before.

In the following, Section \ref{sec:riemannsolver}, we first describe the most important building block in our list: The two-phase Riemann solver
for the coupling of the bulk regions at the interface position. The treatment of the bulk phases is described in \ref{sec:dgsem} followed by a short description of the interface tracking method in \ref{sec:levelset}. A detailed description of the coupling method at the interface is given in section \ref{sec:coupling}.

\section{The two-phase Riemann problem}\label{sec:riemannsolver}

The coupling procedure that takes care of the jump conditions \eqref{eq:rh:rho}-\eqref{eq:rh:e}
at the interface uses the solution of a two-phase Riemann problem whose solution includes phase change and surface tension. This solution also provides 
the information about the interface velocity for the interface tracking method.  
The initial data of the Riemann problem is evaluated at the approximated position of the phase interface.
This is described in detail in the following Section \ref{sec:coupling}. Here we consider first the solution 
of the two-phase Riemann problem in the interface-normal direction. 
% the overall algorithm provides for any point of the discrete
%phase boundary states as input data for the Riemann solver. 

In a local coordinate system, the Riemann problem states are 
$\vv V_\Liq \in \admisr_\liq$, 
$\vv V_\Vap \in \admisr_\vap$ 
and an associated curvature value $\kappa\in (\kappa_\tmin,\kappa_\tmax)$. 
The one-dimensional liquid and vapor state spaces are defined as 
\begin{align*}
   \admisr_{\liq/\vap}:=
   \Set{(\tau,  v, \epsilon) \in \setR^3| (1/\tau, S(\tau,\epsilon))  \in\admis_{\liq/\vap} },
\end{align*}
where $v=\vv v \cdot \vv n$ is the normal component of the fluid velocity with respect to the phase boundary.
With the  Riemann solver we compute the interface bulk states $\vv V_\liq\in \admisr_\liq$, $\vv V_\vap\in \admisr_\vap$ which result from the local interaction of the input data based on a chosen kinetic relation and on the reference temperature for the latent heat.
From the technical point of view, the output of this section will be mappings of the type

\begin{align}\label{micromapp}
M:\left\{   
\begin{array}{rcl}
 \admisr_\liq \times  \admisr_\vap \times \setR &\to&
 \admisr_\liq  \times  \admisr_\vap \times\setR\\
  (\vv V_\Liq, \vv V_\Vap, \kappa)   &\mapsto& (\vv V_\liq,\vv V_\vap, s).
\end{array}
\right.
\end{align}
The Riemann problem under consideration is now 
\begin{align}\label{eq:euler:n}
 \begin{pmatrix}
  \rho \\ \rho\,v  \\ \rho\,(\epsilon+\frac{1}{2} v^2)
 \end{pmatrix}_t
 +
 \begin{pmatrix}
  \rho\,v \\ \rho\,v^2 + p  \\ (\rho\,(\epsilon+\frac{1}{2} v^2) +p) v   
 \end{pmatrix}_x
 =
 \begin{pmatrix}
  0\\0\\0
 \end{pmatrix},
\end{align}
% It is subject to the initial condition 
\begin{align*}
 \begin{pmatrix}
  \tau \\ v \\ \epsilon
 \end{pmatrix}(x,0) &=
 \begin{cases}
   \vv V_\Liq &\text{for } x\leq0,\\
   \vv V_\Vap &\text{for } x>0.
 \end{cases}
\end{align*}

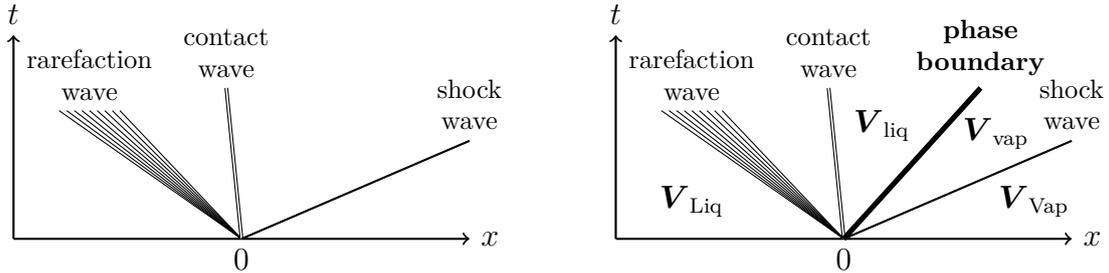
\begin{figure}
    \begin{tikzpicture}[scale=1]    % Zeichenbereich
      % draw axis
      \draw[->,thick] (-3,0) -- (3,0) node[right] {$x$};
      \draw[->,thick] ( -3,0) -- (-3,2.7) node[above] {$t$};
      \node[below] (O) at ( 0, 0)  { $0$ };      
      % rarefaction
      \coordinate (r) at (-2,1.7);
      \foreach \dx in {-0.4, -0.3,...,0.4} 
      \draw[-,thin] (0,0) -- ($ (r) + (\dx,0)$);
      \node[above,align=center,font=\footnotesize] at (r) {rarefaction\\wave};
      % contact wave
      \draw[double,align=center,font=\footnotesize] ( 0,0) -- (-0.2,2) node[above] {contact\\wave};   
      % phase boundary
      %\draw[-,line width=2pt,align=center,font=\footnotesize\bf] ( 0,0) -- (1.8,2) node[above] {phase\\boundary};      
      % shock
      \draw[-,thick,align=center,font=\footnotesize] ( 0,0) -- (3,1.3) node[above] {shock\\wave};      
      % states
      %\node (a) at (  -2,0.5)  { $\vv{U}_\Liq$ };
      %\node (b) at (-0.8,1.5)  { $\vv{U}_\sh$ };
      %\node (c) at ( 0.5,1.5)  { $\vv{U}_\liq$ };
      %\node (d) at (   2,1.4)  { $\vv{U}_\vap$ };
      %\node (e) at ( 2.5,0.5)  { $\vv{U}_\Vap$ };      
    \end{tikzpicture}  
\hfill
    \begin{tikzpicture}[scale=1]    % Zeichenbereich
      % draw axis
      \draw[->,thick] (-3,0) -- (3,0) node[right] {$x$};
      \draw[->,thick] ( -3,0) -- (-3,2.7) node[above] {$t$};
      \node[below] (O) at ( 0, 0)  { $0$ };      
      % rarefaction
      \coordinate (r) at (-2,1.7);
      \foreach \dx in {-0.4, -0.3,...,0.4} 
      \draw[-,thin] (0,0) -- ($ (r) + (\dx,0)$);
      \node[above,align=center,font=\footnotesize] at (r) {rarefaction\\wave};
      % contact wave
      \draw[double,align=center,font=\footnotesize] ( 0,0) -- (-0.2,2) node[above] {contact\\wave};   
      % phase boundary
      \draw[-,line width=2pt,align=center,font=\footnotesize\bf] ( 0,0) -- (1.8,2) node[above] {phase\\boundary};      
      % shock
      \draw[-,thick,align=center,font=\footnotesize] ( 0,0) -- (3,1.3) node[above] {shock\\wave};      
      % states
      \node (a) at (  -2,0.5)  { $\vv{V}_\Liq$ };
      %\node (b) at (-0.8,1.5)  { $\vv{V}_\sh$ };
      \node (c) at ( 0.5,1.5)  { $\vv{V}_\liq$ };
      \node (d) at (   2,1.4)  { $\vv{V}_\vap$ };
      \node (e) at ( 2.5,0.5)  { $\vv{V}_\Vap$ };      
    \end{tikzpicture} 
    
    \caption{Left: Typical wave structure for the exact single phase Riemann problem with $3$ waves. It consists of a rarefaction wave followed by a contact wave and a shock wave.    \\
    Right: Typical wave structure for the exact two-phase Riemann problem with $4$ waves. The additional wave is the subsonic phase boundary. }
    \label{fig:fan}
\end{figure}

We expect that the solution of this  two-phase Riemann problem consists of four waves, one wave 
being an subsonic phase boundary with adjacent states $\vv V_\liq,\vv V_\vap$, see
Figure~\ref{fig:fan} right for some illustration. Exact Riemann solvers of this type can be found in \cite{CHEHAT2015,MUEVOS2006}. Note however that they do not cover the general kinetic relation \eqref{eq:kinrel} nor surface tension or latent heat.

\medskip

We now provide the exact solution of the Riemann problem using an iterative scheme.
The purely hyperbolic solution consists of three waves: two shock or rarefaction waves and a contact wave, as it is shown in Figure \ref{fig:fan}, left.
Because we want to solve a two-phase problem for the kinetic relation \eqref{eq:kinrel}, we will rely on a different wave fan.
We propose to solve the problem by adding an additional phase boundary (see Figure \ref{fig:fan}, right). 
This artificial phase boundary is a discontinuous wave that is supposed to satisfy the jump conditions  \eqref{eq:rh:rho} to \eqref{eq:rh:e} and \eqref{eq:subsonic}.
In this way we preserve the jump conditions at least for fixed surface tension $\kappa$. All other waves are satisfy the standard Rankine-Hugoniot conditions and Riemann invariants. 

The phase boundary is subsonic but might be faster or slower than the contact wave. Thus the phase of the state between the phase boundary and the contact wave is not known in advance.
Any iterative scheme, that relies on one fan configuration will fail if the phase of the middle state changes, since ratios of specific volume and entropies are very  different in the liquid and the vapor. Furthermore approximate states may belong to the spinodal region where the equation of state can not be evaluated. Note that we do not allow mixture states.

\begin{figure}
\centering
    \begin{tikzpicture}[scale=1.3]    % Zeichenbereich
      % draw axis
      \draw[->,thick] (-3,0) -- (2.7,0) node[right] {$x$};
      \draw[->,thick] ( -3,0) -- (-3,2.7) node[above] {$t$};
      \node[below] (O) at ( 0, 0)  { $0$ };  
      
      \coordinate (RB) at ( 23:3.0);
      \coordinate (RC) at ( 53:2.5);
      \coordinate (PB) at ( 93:2.0);
      \coordinate (LC) at (133:2.0);
      \coordinate (LB) at (163:2.3);

      \draw[thick] (0,0) -- (LB)              node [above, align=center,font=\footnotesize] {elementary\\wave};
      \draw[thick] (0,0) -- (RB)              node [above, align=center,font=\footnotesize] {elementary\\wave};
      \draw[line width=2pt] (0,0) -- (PB)     node [above, align=center,font=\footnotesize\bf] {phase\\boundary};
      \draw[double] (0,0) -- (LC)             node [above, align=center,font=\footnotesize] {contact\\wave};
      \draw[double] (0,0) -- (RC)             node [above, align=center,font=\footnotesize] {contact\\wave};

      % states
      \node at ( 10:2.0)  {$\vv{V}_\Vap$ };
      \node at ( 38:1.5)  {$\vv{V}_4$ };
      \node at ( 73:1.5)  {$\vv{V}_3$ };
      \node at (113:1.5)  {$\vv{V}_2$ };
      \node at (148:1.5)  {$\vv{V}_1$ };
      \node at (170:2.0)  {$\vv{V}_\Liq$ };
    \end{tikzpicture}
    \caption{Wave fan as it is assumed in Algorithm~\ref{algo:micromfull}. Depending on the wave configuration, either the left contact wave vanishes ($\vv V_1=\vv V_2$), or the right contact wave vanishes ($\vv V_3=\vv V_4$).}
    \label{fig:fan4}
\end{figure}
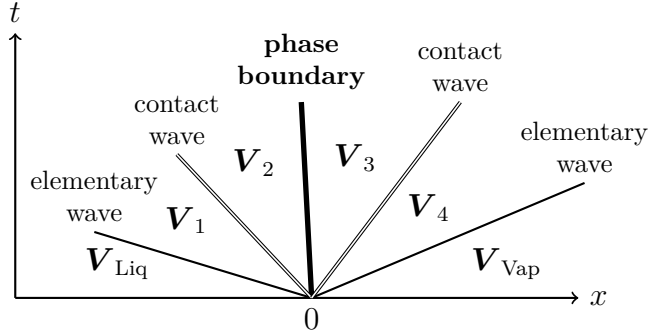

We propose to introduce an additional contact wave to overcome that problem. 
We start with four intermediate states $\vv V_1, \vv V_2 \in \admisr_\liq$ and $\vv V_3, \vv V_4 \in \admisr_\vap$ as it is shown in Figure~\ref{fig:fan4}, which will be determined by the root of a target function. 
The states $\vv V_\Liq$ and $\vv V_1$ are connected by a left elementary wave (rarefaction or Lax shock wave) and  
$\vv V_4$ and $\vv V_\Vap$ are connected by a right elementary wave.
The states $\vv V_2$ and $\vv V_3$ are connected by the phase boundary. 
When the phase boundary propagates faster than the characteristic speed, then $\vv V_1$ and $\vv V_2$ are connected by the contact wave $\vv V_3 = \vv V_4$. In the other case $\vv V_3$ and $\vv V_4$ are the adjacent state of the contact wave and $\vv V_1 = \vv V_2$.

Let us now determine the target function whose root is the solution of the two-phase Riemann problem. Shock, rarefaction and contact waves are computed as in \cite[Chapter \Rmnum{2} Gas dynamics and reacting flows]{GODSEG2006}. The phase boundary satisfies \eqref{eq:rh:rho}--\eqref{eq:rh:e}, \eqref{eq:subsonic}, \eqref{eq:kinrel}. The thermodynamic unknowns are chosen to be the specific volume and temperature, because we will later on use thermodynamic libraries that compute thermodynamic quantities in terms of $(\tau,T)$.

\begin{algo}[Evaluation of the target function]\label{algo:destination}
Let the constant mean curvature $\kappa\in \setR$, surface tension $\zeta\geq0$, entropy production constant $k^\ast\geq0$, reference temperature $T^\ast>0$ and initial Riemann states $\vv V_\Liq$, $\vv V_\Vap$ be given as constant input parameters.

Furthermore, assume that arguments of the target function
\begin{align}\label{eq:target}
 F: \setR_+^8 \to \setR^8, \qquad (\tau_1, T_1,\, \tau_2, T_2,\, \tau_3, T_3,\, \tau_4, T_4) \mapsto (r_1, r_2, \ldots, r_8)
\end{align}
are given as actual guess. 

The following steps determine the residuals $ r_1,r_2, \ldots, r_8$ for given arguments of $F$. 
The algorithm returns the residuals and, in addition, an error flag, 
$\vv V_i=(\tau_i, v_i, \epsilon_i)^\transp$ for $i=1,\ldots,4$, the mass flux $j$ and the propagation speed $s$ of the phase boundary.

\begin{description}%\raggedright
 \item[Step~1] Evaluate pressures, specific entropies, inner and Gibbs free energies
 \begin{align*}
  p_i&:=p(\tau_i,T_i), & 
  S_i&:=S(\tau_i,T_i), &
  \epsilon_i&:=\epsilon(\tau_i,T_i), & 
  G_i&:=G(\tau_i,T_i)  
 \end{align*}
 for $i = \Liq,\Vap,1,\ldots, 4$.
 
 Abort if $(\tau_1,S_1), (\tau_2,S_2)\notin\admis_\liq$ or $(\tau_3,S_3), (\tau_4,S_4)\notin\admis_\vap$ and return the error flag. 
 
 \item[Step 2:] The left elementary wave determines $v_1$ and $r_1$.\\
     If $p_\Liq > p_1$, the left wave is a rarefaction wave and 
     \begin{align*}
      v_1 &= v_\Liq + \int_{\tau_\Liq}^{\tau_1} - \sqrt{- p_\tau(\tau,S_\Liq)}\, \dd \tau, \\
      r_1 &= S_1 - S_\Liq.
     \end{align*}
     If $p_\Liq \leq p_1$, the left wave is a shock wave and 
     \begin{align*}
      v_1 &= v_\Liq - \sqrt{\abs{ (p_\Liq-p_1)\,(\tau_1-\tau_\Liq)}}, \\
      r_1 &= \epsilon_1 - \epsilon_\Liq + \frac{1}{2}(p_\Liq+p_1)\,(\tau_1-\tau_\Liq).
     \end{align*}   

 \item[Step 3:] The right elementary wave determines $v_4$ and $r_2$.\\
     If $p_4 < p_\Vap$, the right wave is a rarefaction wave and 
     \begin{align*}
      v_4 &= v_\Vap - \int_{\tau_\Vap}^{\tau_4} - \sqrt{- p_\tau(\tau,S_\Vap)}\, \dd \tau, \\
      r_2 &= S_\Vap - S_4.
     \end{align*}
     If $p_4 \leq p_\Vap$, the right wave is a shock wave and 
     \begin{align*}
      v_4 &= v_\Vap + \sqrt{\abs{ (p_4-p_\Vap)\,(\tau_\Vap-\tau_4)}}, \\
      r_2 &= \epsilon_\Vap - \epsilon_4 + \frac{1}{2}(p_4+p_\Vap)\,(\tau_\Vap-\tau_4).
     \end{align*}
     
  \item[Step~4] The contact waves determine (normal) fluid velocity and pressure in the adjacent states. Assign 
     \begin{align*}
       r_3 &:= p_2-p_1, & r_4 &:= p_4-p_3, & v_2 &:= v_1, & v_3 &:= v_4.
     \end{align*}

  \item[Step~5] The mass flux through the phase boundary and its speed are
     \begin{align*}
      j := \frac{v_3-v_2}{\tau_3-\tau_2}, && s:=v_2-j\,\tau_2=v_3-j\,\tau_3.
     \end{align*}
     
  \item[Step~6] The left contact wave propagates with speed $v_3$, the phase transition with $s$ and the right contact wave with speed $v_4$. The wave fan configuration is therefore known and the additional contact wave can be rejected, set
     \begin{align*}
      r_5 := \begin{cases}
             T_4-T_3 &: j<0,\\
             (T_2-T_1)\,(T_4-T_3) &: j=0,\\
             T_2-T_1 &: j>0.
            \end{cases}
     \end{align*}

   \item[Step~7] The phase transition connects  $\U_2$ and $\U_3$, thus
    \begin{align*}
    \begin{array}{lccccc} 
     r_6:=& j\,(v_3-v_2)                              &+& p_3-p_2           &-& \surf,\\ 
     r_7:=& (h_3-h_2)+\frac{1}{2} j^2\,(\tau_3-\tau_2)&-& L(T^\ast),         & & \\
     r_8:=& (h_3-h_2)+\frac{1}{2} j^2\,(\tau_3-\tau_2)&-& T^\ast\,(S_3-S_2) &+& k^\ast\,j,
    \end{array}
    \end{align*}
    with $h_i = G_i+T_i\,S_i$, $i=2,3$. 

\end{description}
\end{algo}

The target function can be solved with a standard multidimensional root-finding algorithm. Note however that the function $F$ is not globally differentiable.
Furthermore it may happen that the actual guess leads to a state in the spinodal region or the wrong phase. That is recognized by Step 1 in Algorithm \ref{algo:destination}  and an error flag is returned. We apply a damped Quasi-Newton method, that reduces the time step until all states are in the correct phases. The residual computed in Step 6 demands that the additional nonphysical contact wave vanishes, if the root-finding algorithm converges. 

\begin{algo}[Two-phase Riemann solver]\label{algo:micromfull}
  Let the arguments $(\vv V_\Liq$, $\vv V_\vap$, $\kappa) \in \admisr_\liq \times  \admisr_\vap \times \setR$
  of function $M$ in \eqref{micromapp} and the constants  $\zeta\geq0$, $k^\ast\geq0$, $T^\ast>0$ be given.

  \begin{description}
   \item[Step~1] 
     Compute temperature $T_{\Liq/\Vap}$ from the initial states $\vv V_{\Liq/\Vap}$ and assign the constant input parameters of Algorithm~\ref{algo:destination}.

    \item[Step~2] Assign an initial guess $\tau_i$, $T_i$, $i=1,\ldots,4$, for the root-finding algorithm such that 
     \begin{align*}
       \left(\tau_1,\eta(\tau_1,T_1)\right) &\in \admis_\liq, & \left(\tau_2,\eta(\tau_2,T_2)\right) &\in \admis_\liq, \\ 
       \left(\tau_3,\eta(\tau_3,T_3)\right) &\in \admis_\vap, & \left(\tau_4,\eta(\tau_4,T_4)\right) &\in \admis_\vap
     \end{align*}
     holds.
   
   \item[Step~3] Solve $F(\tau_1, T_1,\, \tau_2, T_2,\, \tau_3, T_3,\, \tau_4, T_4)= \vv 0$
      and return $\vv V_\liq = \vv V_2$, $\vv V_\vap = \vv V_3$, $s$.
  \end{description}
\end{algo}

% section about the bulk solver based on the DG scheme
\section{Numerical solution strategy for compressible two-phase flow}
\label{sec:bulksolver}
In this section we describe how to use the solution of the two-phase Riemann problem for a sharp resolution of the interface and how the different 
building blocks interact.  
First, we give an overview about the flow solver for the liquid and vapor bulk phases. This is kept short because our flow solver is already described 
in other papers, \cite{Kopriva2010,hindenlang2012dgsem,fechter2015multiphasedg}, and, furthermore, other flow solvers may be used for this task. 
A description of the interface tracking method based on a level-set approach follows. The main topic of this section is then the consistent coupling 
of the bulk phase solutions at the interface in a way such that it remains sharp. This is established by the use of the 
interface Riemann solver in combination with a ghost fluid method.

\subsection{Flow solver for the bulk flow}
\label{sec:dgsem}
The flow solver in our simulations is a discontinuous Galerkin spectral element method (DGSEM) with a local finite-volume sub-cell refinement at the interface \cite{fechter2015multiphasedg}. Within the sub-cells a second order finite volume scheme is applied to increase the interface resolution and stability of the numerical scheme. Note that in both numerical methods the degrees of freedom (DOF) in one cell are equal. Thus, for the description of the numerical resolution in Section \ref{sec:numerics} we use the general term DOF that is independent of the approximation order used in the DG scheme. One DOF represents one grid sub-cell in the context of finite-volume schemes and one polynomial coefficient in the DG context.

%Directly at the phase interface we use a finite-volume method within the DG framework that efficiently handles the discontinuity of the phase interface within the computational domain. 
We restrict ourselves to a standard finite volume method for the description of the numerical strategy, which solves numerically the Euler equations \eqref{eq:euler} in the bulk phases with a general equation of state.  
For fluid flow without phase transition, this numerical strategy is described in detail in \cite{fechter2015multiphasedg}. 
% We add some remarks at the end of this section.  
%solver and use an EOS tabulation method, as introduced in \cite{dumbser2013cavitation,bolemann2015eostabulation}, to speed-up the EOS evaluation during the simulation runtime.

%In a first step we discretize the computation domain $\Omega$ into non-overlapping grid cells $Q$. 
%
%The system   \eqref{eq:euler} is usually written in vector form for the vector of the conserved variables
The finite volume scheme approximates 
%as $\vv U_t + \operatorname{div}{\vv{F}(\vv U)}=\vv{0}$,
%with an appropriately defined flux $\vv{F}$.
%
the integral formulation of the conservation equation over every grid cell $Q$:
%For the derivation of the finite volume method we use the weak formulation of \eqref{eq:euler} that is obtained by an integration over the grid cell $Q$ and the application of the divergence theorem
\begin{align}
\frac{d}{dt}\int\limits_Q \vv U\ dV + \int\limits_{\partial Q}  \left(\vv F  \cdot \vv N\right)\ dS = 0.
\label{eq:weakformulation}
\end{align}
where $\vv U=(\rho,\rho\,v_1,\rho\,v_2, \rho\,v_3, \rho\,e)^\transp$ denotes the vector of the conserved variables, $\vv F$ denotes the flux tensor, and $\mathbf{N}$ denotes the outwards pointing normal vector on the grid cell surface $\partial Q$. At the grid cell boundaries within the bulk phases we use numerical flux calculations composed of an approximate solution of the single-phase Riemann problem. In the simulations presented in this paper, the HLLC scheme is applied (see e.g., \cite{Toro}). 
% For more details see the book of Toro \cite{Toro}. 
At the grid sub-cell boundaries that form the phase interface we apply the solution of the exact interface Riemann solver. This coupling is described in section \ref{sec:coupling}.
% In a next step we introduce the (approximate) Riemann solution into equation \eqref{eq:weakformulation}
%\begin{align}
%\int\limits_Q \vv U_t\ dV + \int\limits_{\partial Q}  \left(\vv F\cdot\vv N\right)^\ast\ dS = 0 
%\label{eq:withNumericalFlux}
%\end{align}
%and mark the resulting numerical Riemann flux with $^\ast$. 
The integrals in \eqref{eq:weakformulation} are approximated by the midpoint rule. For time integration, we use a standard explicit Runge-Kutta method. 
%In section \ref{sec:coupling} the application of the Riemann solvers within $\Omega$ is introduced. Within the bulk phases standard single-phase Riemann solvers are used while at the interface position the two-phase Riemann solver is used.

\subsection{Interface tracking}
\label{sec:levelset}
For the sharp interface approximation we need an interface tracking method that evolves the interface and provides the actual location. Besides the position, we also need the interface curvature to evaluate surface tension effects. 
%, are evaluated. Thus, this approach mimics the influence of the coarse numerical approximation of the phase interface.
Here, we rely on a level-set method \cite{Sussman1994}. The level-set function $\Phi$ is initialized as an approximated signed distance function. The time evolution of $\Phi$ is calculated by 
\begin{equation}
  \frac{D\Phi}{Dt}=\frac{\partial \Phi}{\partial t} +  \vv s \nabla(\Phi) = 0 \;.
  \label{eq:levelset}
\end{equation}
Note that we reformulated the level-set equation into the divergence form and source term that is more convenient for our numerical approach. It is approximated by the same numerical method as the solver for the bulk flow.
For the level-set equation we need a velocity field $\vv s$ that coincides at the interface with the local interface velocity that is determined by the interface Riemann solver.

The approximation of the level-set equation \eqref{eq:levelset}, as well as the estimation of the interface normals $\vv n=\nicefrac{\nabla\Phi}{|\nabla\Phi|}$ and the curvature $\kappa = \nabla\cdot(\nicefrac{\nabla\Phi}{|\nabla\Phi|})$, is described in detail in \cite{fechter2015multiphasedg} and is based on a-posteriori reconstruction of the level-set function. These quantities are estimated using the level-set gradients in the vicinity of the interface.

\subsection{Coupling at the interface}
\label{sec:coupling}
%In the scope of the numerical two-phase method, two different Riemann solvers are applied within the computational domain to provide a sharp interface resolution. Within the bulk phases standard single-phase Riemann solvers are used to estimate the numerical flux $\vv F^\ast$. At the computational interface, that coincides with the nearest grid cell boundary to the tracked interface position (as determined by the level-set method), a special two-phase Riemann problem is solved (see the description in section \ref{algo:destination}). Due to the non-linearity of the Riemann solver, it is applied normal to the interface. This method guarantees a sharp interface for all times.

% We start with the description of the interface coupling procedure in a one-dimensional context and extend it in a second step to multi dimensions.

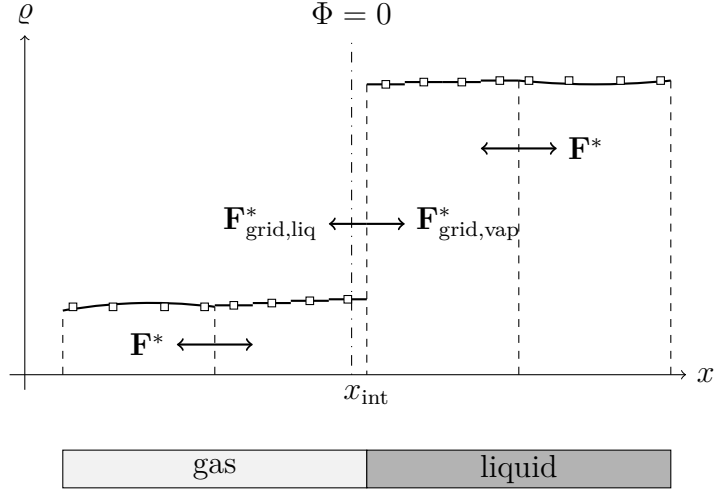
\begin{figure}
 \centering
 \begin{tikzpicture}[scale=1.]
  % x axis
  \draw[->] (-0.7,0) -- (8.2,0) node[right] {$x$};
  \draw[->] (-0.5,-.2) -- (-0.5,4.5) node[above] {$\rho$};
  % rho values
  \draw[thick] plot [smooth, tension=1] coordinates{(0.,0.85) (1,0.95) (2.0,0.9)};
  \draw[thick] (2.0,0.92)--(2.5,0.92);
  \draw[thick] (2.5,0.95)--(3.0,0.95);
  \draw[thick] (3.0,0.98)--(3.5,0.98);
  \draw[thick] (3.5,1.0) --(4.0,1.0);
  \draw[thick] (4.0,3.85)--(4.5,3.85);
  \draw[thick] (4.5,3.88)--(5.0,3.88);
  \draw[thick] (5.0,3.88)--(5.5,3.88);
  \draw[thick] (5.5,3.9) --(6.0,3.9);
  \draw[thick] plot [smooth, tension=1] coordinates{(6.,3.9) (7,3.85) (8.0,3.9)};
  % integration points in DG cell
  \foreach \x in \xGP {
        \draw[fill=white] (\x+1,0.9) +(-0.05,-0.05) rectangle +(0.05,0.05);
        \draw[fill=white] (\x+7,3.9) +(-0.05,-0.05) rectangle +(0.05,0.05);
  }
  \draw[fill=white] (2.25,0.92) +(-0.05,-0.05) rectangle +(0.05,0.05);
  \draw[fill=white] (2.75,0.95) +(-0.05,-0.05) rectangle +(0.05,0.05);
  \draw[fill=white] (3.25,0.98) +(-0.05,-0.05) rectangle +(0.05,0.05);
  \draw[fill=white] (3.75,1.00) +(-0.05,-0.05) rectangle +(0.05,0.05);
  \draw[fill=white] (4.25,3.85) +(-0.05,-0.05) rectangle +(0.05,0.05);
  \draw[fill=white] (4.75,3.88) +(-0.05,-0.05) rectangle +(0.05,0.05);
  \draw[fill=white] (5.25,3.88) +(-0.05,-0.05) rectangle +(0.05,0.05);
  \draw[fill=white] (5.75,3.90) +(-0.05,-0.05) rectangle +(0.05,0.05);
  % ghost states
%   \draw[fill=white] (4.25,1.00) circle (0.05);
%   \draw [->] (3.75,1.0) arc[x radius=0.25cm, y radius =.25cm, start angle=-180, end angle=0];
%   \draw[fill=white] (3.75,3.85) circle (0.05);
%   \draw [->] (4.25,3.85) arc[x radius=0.25cm, y radius =.25cm, start angle=0, end angle=180];
  %
  \draw[dashdot] (3.8,0)--(3.8,4.5) node[above] {$\Phi=0$};
  \draw[dashed] (0,0) -- (0,0.8);
  \draw[dashed] (2,0) -- (2,0.9);
  \draw[dashed] (4,3.85) -- (4,0) node[below] {$x_\text{int}$};
  \draw[dashed] (6,0) -- (6,3.9);
  \draw[dashed] (8,0) -- (8,3.9);
  \draw[thick,<->] (2.5,0.4) -- (1.5,0.4) node[left] {$\mathbf{F}^\ast$};
  \draw[thick,->] (4,2.0) -- (4.5,2.0) node[right] {$\mathbf{F}^\ast_\text{grid,vap}$};
  \draw[thick,->] (4,2.0) -- (3.5,2.0) node[left] {$\mathbf{F}^\ast_\text{grid,liq}$};
  \draw[thick,<->] (5.5,3.0) -- (6.5,3.0) node[right] {$\mathbf{F}^\ast$};
%   \node at (1,-0.5) {DG};
%   \node[] at (3,-0.5) {\parbox{2cm}{\centering\footnotesize FV \\ sub-cells}};
%   \node[] at (5,-0.5) {\parbox{2cm}{\centering\footnotesize FV \\ sub-cells}};
%   \node at (7,-0.5) {DG};
  \draw[fill=gray!10] (0,-1.5) rectangle (4.0,-1.) node[pos=0.5,black] {gas};
  \draw[fill=gray!60] (4.0,-1.5) rectangle (8,-1.) node[pos=0.5,black] {liquid};
\end{tikzpicture}
 \caption{Sketch of the numerical coupling procedure at the interface using the information of the two-phase Riemann solver at the interface and the application of a ghost-fluid approach. 
 %Within the bulk phases the numerical flux $\vv F^\ast$ is used, at the interface position the fluxes $\vv F_\text{grid,liq}$ and $\vv F_\text{grid,vap}$ are used.
 }
 \label{fig:onedcoupling}
\end{figure}

\begin{figure}
 \centering
 \begin{tikzpicture}[scale=2]
   \draw[fill=gray!10] (3,3) rectangle (7,7);
   \draw[fill=gray!60] (3,3)--(3,5)--(3.75,5)--(3.75,4.75)--(4.25,4.75)--(4.25,4.5)--(4.5,4.5)--(4.5,4.25)--(4.75,4.25)--(4.75,3.75)--(5,3.75)--(5,3)--cycle;
   %FV cells
   \def\xcell{4};
   \def\ycell{3};
     \foreach \x in {0.25,0.5,0.75}{
      \draw (\xcell,\ycell+\x)--(\xcell+1,\ycell+\x);
      \draw (\xcell+\x,\ycell)--(\xcell+\x,\ycell+1);
     }
   \def\xcell{5};
   \def\ycell{3};
     \foreach \x in {0.25,0.5,0.75}{
      \draw (\xcell,\ycell+\x)--(\xcell+1,\ycell+\x);
      \draw (\xcell+\x,\ycell)--(\xcell+\x,\ycell+1);
     }
   \def\xcell{3};
   \def\ycell{5};
     \foreach \x in {0.25,0.5,0.75}{
      \draw (\xcell,\ycell+\x)--(\xcell+1,\ycell+\x);
      \draw (\xcell+\x,\ycell)--(\xcell+\x,\ycell+1);
     }
   \def\xcell{3};
   \def\ycell{4};
     \foreach \x in {0.25,0.5,0.75}{
      \draw (\xcell,\ycell+\x)--(\xcell+1,\ycell+\x);
      \draw (\xcell+\x,\ycell)--(\xcell+\x,\ycell+1);
     }     
   \def\xcell{4};
   \def\ycell{4};
     \foreach \x in {0.25,0.5,0.75}{
      \draw (\xcell,\ycell+\x)--(\xcell+1,\ycell+\x);
      \draw (\xcell+\x,\ycell)--(\xcell+\x,\ycell+1);
     }
 % general stuff 
 \draw[dashdot,thick,black] ([shift=(0:2)]3,3) arc (0:90:2);
 \foreach \x in {3,4,5,6} {
    \draw (3,\x)--(7,\x);
    \draw (\x,3)--(\x,7);
 }
   \draw[thick,black] (3,5)--(3.75,5)--(3.75,4.75)--(4.25,4.75)--(4.25,4.5)--(4.5,4.5)--(4.5,4.25)--(4.75,4.25)--(4.75,3.75)--(5,3.75)--(5,3);

  \draw[fill=white] (5,3.375) +(-0.03,-0.03) rectangle +(0.03,0.03);
  \draw[fill=white] (5,3.125) +(-0.03,-0.03) rectangle +(0.03,0.03);
  \draw[fill=white] (5,3.625) +(-0.03,-0.03) rectangle +(0.03,0.03);
  \draw[fill=white] (4.875,3.75) +(-0.03,-0.03) rectangle +(0.03,0.03);
  \draw[fill=white] (4.75,3.875) +(-0.03,-0.03) rectangle +(0.03,0.03);
  \draw[fill=white] (4.75,4.125) +(-0.03,-0.03) rectangle +(0.03,0.03);
  \draw[fill=white] (4.625,4.25) +(-0.03,-0.03) rectangle +(0.03,0.03);
  \draw[fill=white] (4.5,4.375) +(-0.03,-0.03) rectangle +(0.03,0.03);
  \draw[fill=white] (4.375,4.5) +(-0.03,-0.03) rectangle +(0.03,0.03);
  \draw[fill=white] (4.25,4.625) +(-0.03,-0.03) rectangle +(0.03,0.03);
  \draw[fill=white] (4.25,4.625) +(-0.03,-0.03) rectangle +(0.03,0.03);
  \draw[fill=white] (4.125,4.75) +(-0.03,-0.03) rectangle +(0.03,0.03);
  \draw[fill=white] (3.875,4.75) +(-0.03,-0.03) rectangle +(0.03,0.03);
  \draw[fill=white] (3.75,4.875) +(-0.03,-0.03) rectangle +(0.03,0.03);
  \draw[fill=white] (3.625,5) +(-0.03,-0.03) rectangle +(0.03,0.03);
  \draw[fill=white] (3.375,5) +(-0.03,-0.03) rectangle +(0.03,0.03);
  \draw[fill=white] (3.125,5) +(-0.03,-0.03) rectangle +(0.03,0.03);

  \draw[thick,->] (4.25,4.625) -- (4.95,5.425) node[right] {$\vv n$};

\end{tikzpicture}
 \caption{Generalized coupling procedure for the multi-dimensional case. The tracked interface position is represented by the dash-dot line and the shifted interface position by the thick solid line. At each integration point on the shifted phase interface (marked with a white rectangle) the two-phase Riemann solver is applied into the normal direction $\vv n$ of the interface.}
 \label{fig:twodcoupling}
\end{figure}
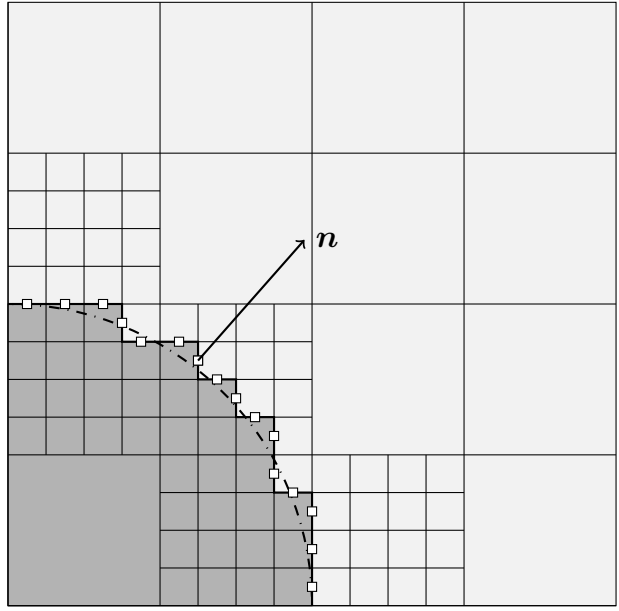

\paragraph{Coupling procedure at the interface}
%Let's assume that we know the position of the interface $\Gamma(t)$ from the level-set function.
The methodology is sketched in Figure~\ref{fig:onedcoupling} within a one-dimensional domain. 
The interface tracking algorithm provides the position of the phase interface as the zero point $\Phi=0$. In Figure~\ref{fig:onedcoupling}, this position is marked by $x_\text{int}$. To avoid numerical diffusion we shift  the phase interface to the grid cell boundary of the nearest neighboring grid cell.  
The extension to multiple dimensions is sketched for the two dimensional case in Figure~\ref{fig:twodcoupling}.
At the shifted position we solve the two-phase Riemann problem as described in the previous section. The Riemann solver is applied along the approximated interface normal $\vv n$, evaluated at the integration point of the grid boundary. The solution of the two-phase Riemann problem provides three quantities: The state of the vapor phase, the state of the liquid phase at the interface and the interface velocity $s$ in the interface normal direction $\vv n$. The states are then used in the flux calculation $\vv F_\liq^\ast$ and $\vv F_\vap^\ast$ in the liquid and vapor bulk phases. Hence, in Figure~\ref{fig:onedcoupling}, the interface liquid state is the left state for the flux calculation in the liquid bulk region, while the gas state at the interface is the right state for the flux calculation in vapor bulk region.   
The grid-normal part of the fluxes $\vv F_\text{grid,liq}^\ast$ and $\vv F_\text{grid,liq}^\ast$ is then used for the flux summation within the bulk phases. This procedure accomplishes the sharp resolution of the interface and imposes the jump conditions described in \eqref{eq:rh:rho}-\eqref{eq:rh:e}. But, of course, the guarantee of the exact conservation of mass, momentum and energy at any time is lost.    

Within the bulk phases standard approximate Riemann solvers are applied, whose fluxes are named by $\vv F^\ast$ in Figure~\ref{fig:onedcoupling}.

\paragraph{Computation of the two-phase solution}
The application of the two-phase Riemann solver at a quadrature point of the approximated interface position $\Gamma(t)$ may be split into six steps:
\begin{enumerate}
 \item In each bulk phase we extract the states.
 For simplicity, we name these states at the interface $\vv U_\Liq$ in the liquid phase and $\vv U_\Vap$ in the vapor phase. 
 \item For the states $\vv U_\Liq$ and $\vv U_\Vap$ the mapping \eqref{micromapp} to the two-phase Riemann solver framework is applied to compute 
	\begin{align*}
	 v_\Liq &= \vv v_\Liq \cdot \vv n, \\
	 v_\Vap &= \vv v_\Vap \cdot \vv n, \\
	 \epsilon_\Liq &= e_\Liq - \frac{1}{2}\norm{\vv v_\Liq}^2, \\
	 \epsilon_\Vap &= e_\Vap - \frac{1}{2}\norm{\vv v_\Vap}^2.
	\end{align*}
       The states $\vv V_\Liq = (\rho_\Liq^{-1}, v_\Liq, \epsilon_\Liq)^\transp$ and $\vv V_\Vap = (\rho_\Vap^{-1}, v_\Vap, \epsilon_\Vap)^\transp$ are then the input data for the two-phase Riemann solver of Algorithm~\ref{algo:micromfull}. The interface normal vector $\vv n$ is estimated using the level-set field (see Section \ref{sec:levelset}) evaluated at the position of the surface integration point on the approximated phase interface.
 \item With the data $\vv V_\Liq$ and $\vv V_\Vap$ the solution of the two-phase Riemann problem (as described in section \ref{algo:destination}) is solved.
 \item If Algorithm~\ref{algo:micromfull} converges, the solution states according to the mapping \eqref{micromapp} are then 
	\begin{align*} 
	  \vv v_\liq &= v_\liq \,\vv n + \sum_{k=1}^{2} (\vv v_\Liq\cdot\vv t^k) \,\vv t^k, &
	  \vv v_\vap &= v_\vap \,\vv n + \sum_{k=1}^{2} (\vv v_\Vap\cdot\vv t^k) \,\vv t^k,\\
	  \vv U_\liq &= \frac{1}{\tau_{\liq}}\left( 1, \vv v_\liq, \epsilon_\liq + \frac{\norm{\vv v_\liq}^2}{2} \right), &
	  \vv U_\vap &= \frac{1}{\tau_{\vap}}\left( 1, \vv v_\vap, \epsilon_\vap + \frac{\norm{\vv v_\vap}^2}{2} \right) 
	\end{align*}
       using the normal $\vv n$ and tangential vectors $\vv t^k$ of the interface approximation.
       %FIXME: if not?
 \item We use the solution states $\vv U_\liq$ and $\vv U_\vap$ for the flux calculation and calculate the numerical fluxes $\vv F^\ast_\liq = \vv F(\vv U_\liq)$ and $\vv F^\ast_\vap = \vv F(\vv U_\vap)$ for the liquid and vapor phase. 
 \item In a last step, we project the numerical flux onto the grid normal direction $\vv N$
	\begin{align*}
	\vv F_\text{grid,liq}^\ast &= \vv F_\liq^\ast \cdot \vv N, \\
	\vv F_\text{grid,vap}^\ast &= \vv F_\vap^\ast \cdot \vv N.
	\end{align*}
       For the advancement of the bulk flow we consider only the grid normal component of the interface flux. This is consistent with the evaluation of the Riemann solution normal to the grid cell boundary in the single-phase case. Note that we use the flux $\vv F_\text{grid,liq}^\ast$ solely for the propagation of the liquid phase and the flux $\vv F_\text{grid,vap}^\ast$ solely for the vapor phase. 
%       This flux choice ensures the resolution of a sharp interface for all times.
\end{enumerate}

\paragraph{Interface update}
With the solution of the fluxes at the surface integration points of the numerical interface, we have all information needed to advance the bulk flow in time. Using the information about the interface advection velocity we update the position of the interface and compute the corresponding interface normal vectors $\vv n$ as well as the interface curvature $\kappa$. These approximations are the input for the calculation of the two phase Riemann solution in the next time step.

\paragraph{Interface movement}
If the tracked interface position, as defined by the zero level of the level-set function, has moved across one grid cell, the numerical interface is updated accordingly. This implies that the state in this cell has to be re-defined according to its new fluid affiliation. As no information is available, the state is defined by using information from the surrounding grid cells
\begin{align}
  \vv U_{\textnormal{new}} = \frac{1}{\sum_{i=1}^\text{nCell}\delta_{i,\text{new}}}\sum_{i=1}^\text{nCell}\delta_{i,\text{new}} \vv U_i 
\end{align}
with
\begin{align}
\delta_{i,\text{new}}= \begin{cases}
                             1 \text{ if fluid(i) = fluid(new)}\,, \\
                             0 \text{ else.}
                           \end{cases}
\end{align}
This state averaging is done for all surrounding nCell cells of the same fluid phase (in 3D: nCell $\in \{1,\ldots,6\}$).  As the extrapolation occurs in small cells at the interface, it has only a minor impact on the solution.

\section{Validation of the Riemann solver, the numerical method, and numerical examples}
\label{sec:numerics}
 In the first part of this section  the influence of parameters on  the solution of the Riemann problem  as obtained by 
Algorithm \ref{algo:micromfull}
is  studied. In particular, entropy production coefficient $k^\ast$,  reference temperature $T^\ast$  and surface tension $\zeta$ are
varied. This is followed by a convergence study of the  numerical method     and a verification of the 
long-time behavior  within a closed, domain. In a second step the mass transfer modeling is validated
based on experiments for rapid evaporation processes. In all these cases, we restrict ourselves to one-dimensional simulations. 
To validate the multi-dimensional implementation the results for a two-dimensional simulation of an evaporating droplet are compared to 
a one-dimensional simulation in cylindrical coordinates. The approximation of surface tension effects is
validated by an oscillating droplet test case and the features of the numerical two-phase method are shown by a shock-
droplet interaction.
\medskip

The numerical validation is performed for the fluid n-dodecane. An
EOS was introduced by Lemmon\&Huber \cite{lemmon2004thermodynamic} for this fluid that predicts the thermodynamic fluid behavior within  
wide pressure and temperature ranges. Due to n-dodecanes retrograde behavior, adiabatic evaporation waves may appear, which have been investigated
experimentally, e.\,g.~by Simoes-Moreira\&Shepherd \cite{simoes1999evaporation}. 

% FIXME: further reference on DFT??
We estimate the entropy production constant $k^\ast$ in the kinetic relation \eqref{eq:kinrel} at the interface using  density functional theory 
estimate of Waibel \cite{waibel2013mscthesis}. 
The values are obtained numerically for the fluid octane. %  and are validated by molecular dynamics simulations.
In the following we assume that the computed value   $k^\ast=47.39\,\nicefrac{\text{m}^4}{\text{kg\,s}}$ 
is a proper estimate of $k^\ast$ for the whole family of alkanes. 
%The used standard value of the entropy production coefficient is set to for the fluid n-dodecane at $T=500$\,K and a pressure of $p=1$\,bar. 

\subsection{Parameter study for  the  Riemann solver}\label{sec:1driemann}

We apply the Riemann solver as introduced with Algorithm \ref{algo:micromfull} in Section \ref{sec:riemannsolver}. All results refer to the EOS for n-dodecane.

\subsubsection{Influence of entropy production coefficient $k^\ast$}
\label{exp:dodecane1:RP}
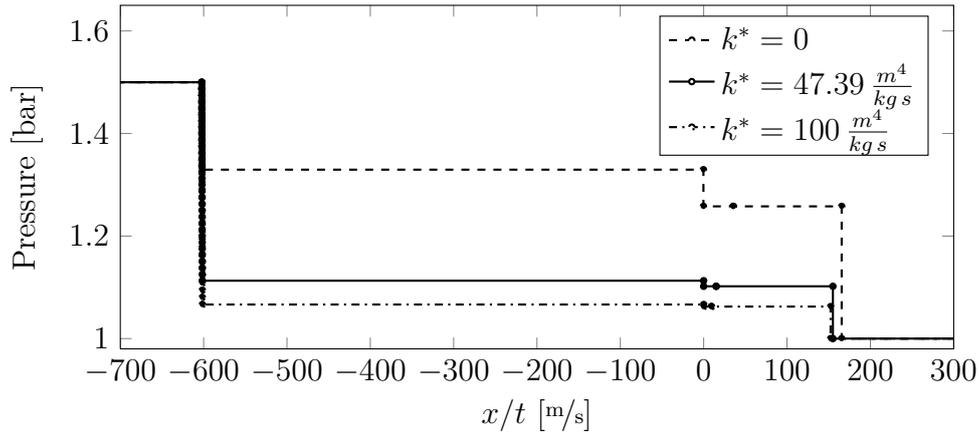
\begin{figure}
  \begin{tikzpicture}\centering
    \begin{axis}[ xmin=-700, xmax=300, 
                  ymin= 0.98, ymax=1.65,
                  xlabel={$x/t$ [$\nicefrac{\text{m}}{\text{s}}$]},
                  ylabel={Pressure [bar]}, 
                  x post scale = 1.6, y post scale = 0.8,
                  legend pos= north east,legend cell align=left
                  ]
    \addplot [thick, mark=*, mark size=1pt, dashed]	table[x expr={\thisrowno{0}}, y expr={1e-5*\thisrowno{4}}, col sep=comma] {Data/dodecane_fan/fan_k0.csv};
    \addlegendentry{$k^\ast=0$};
    \addplot [thick, mark=*, mark size=1pt, solid ] 	table[x expr={\thisrowno{0}}, y expr={1e-5*\thisrowno{4}}, col sep=comma] {Data/dodecane_fan/fan_k47.39.csv};
    \addlegendentry{$k^\ast=47.39\,\frac{m^4}{kg\,s}$};    
    \addplot [thick, mark=*, mark size=1pt, dashdotted]	table[x expr={\thisrowno{0}}, y expr={1e-5*\thisrowno{4}}, col sep=comma] {Data/dodecane_fan/fan_k100.csv};
    \addlegendentry{$k^\ast=100\,\frac{m^4}{kg\,s}$};    
    \end{axis}
  \end{tikzpicture}
  \caption{Pressure solution for  the Riemann problem in  \ref{exp:dodecane1:RP} with different values of the entropy production. 
  To visualize the positions of  the rarefaction, evaporation and  shock wave marking with dots is used.}
%   The position of the contact wave (not visible in the pressure 
%   field) is marked by a circle.} 
  \label{fig:dodecane:RP}
\end{figure}

The first example addresses the influence of the entropy production coefficient $k^\ast$ on the solution of the Riemann problem. 
The initial conditions   under consideration are 
\begin{align}
  (\rho,v,p,T)(x,0) &= \begin{cases}
                     (584.08\,\nicefrac{\text{kg}}{\text{m}^3},0\,\nicefrac{\text{m}}{\text{s}}, 1.5\,\text{bar}, 500\,\text{K}) &\text{ if }x\le 0,\\
                     (  4.38\,\nicefrac{\text{kg}}{\text{m}^3},0\,\nicefrac{\text{m}}{\text{s}}, 1.0\,\text{bar}, 500\,\text{K}) &\text{ otherwise, }
                 \label{eq:case1_ini}
                 \end{cases}
\end{align}
with $p_\text{sat}(T=500\,\text{K}) = 1.29\,\text{bar}$. We have give  here  and below 
the initial data in  (overdetermined)  primitive form to plain the 
thermodynamical conditions.
The surface tension coefficient and the reference temperature 
are chosen to be
\[
\zeta = 0 ,\quad  T^\ast=500\,K, \quad L(T^\ast)= 249410\nicefrac{J}{kg} .  
\]
The pressure of the Riemann problem solution as described in Section \ref{sec:riemannsolver} 
is shown in Figure \ref{fig:dodecane:RP} for different values of entropy production $k^\ast$ (one of them being the generic value from
the introductory remark). 
The solutions start with a  rarefaction wave followed by  an evaporation wave (phase boundary), a contact discontinuity,
and finally a shock wave.
%Note that the contact discontinuity is not visible in the pressure 
%distribution. Hence, we marked the position of the contact wave by a ??? \corr{Was anderes nehmen. Das erkennt niemand}.
Note that the slope of the rarefaction wave is quite strong in the chosen spatial scaling.
The results indicate  how the variation in $k^\ast$ influences   the predicted mass flux rates 
and the interface speed. In fact, the  speeds of the evaporation wave are
\begin{align*}
 s = \begin{cases}
      -0.3  \,\nicefrac{\text{m}}{\text{s}} &\text{for } k^\ast=0\,\nicefrac{\text{m}^4}{\text{kg\,s}} \text{ (dashed line)}, \\
      -0.015\,\nicefrac{\text{m}}{\text{s}} &\text{for } k^\ast=47.39\,\nicefrac{\text{m}^4}{\text{kg\,s}} \text{ (solid line)}, \\  
      +0.048\,\nicefrac{\text{m}}{\text{s}} &\text{for } k^\ast=100\,\nicefrac{\text{m}^4}{\text{kg\,s}} \text{ (dash dotted line)}.
     \end{cases}
\end{align*}

%Note that the liquid pressure is below the saturation pressure of $1.29$\,bar and the state $(\tau_\liq,S_\liq)$ is metastable.
%\noindent Note that increasing $k^\ast$ increases also the speed of the evaporation wave.

\subsubsection{Influence of reference temperature $T^\ast$ and surface tension $\zeta$}
\label{exp:dodecane2:RP}
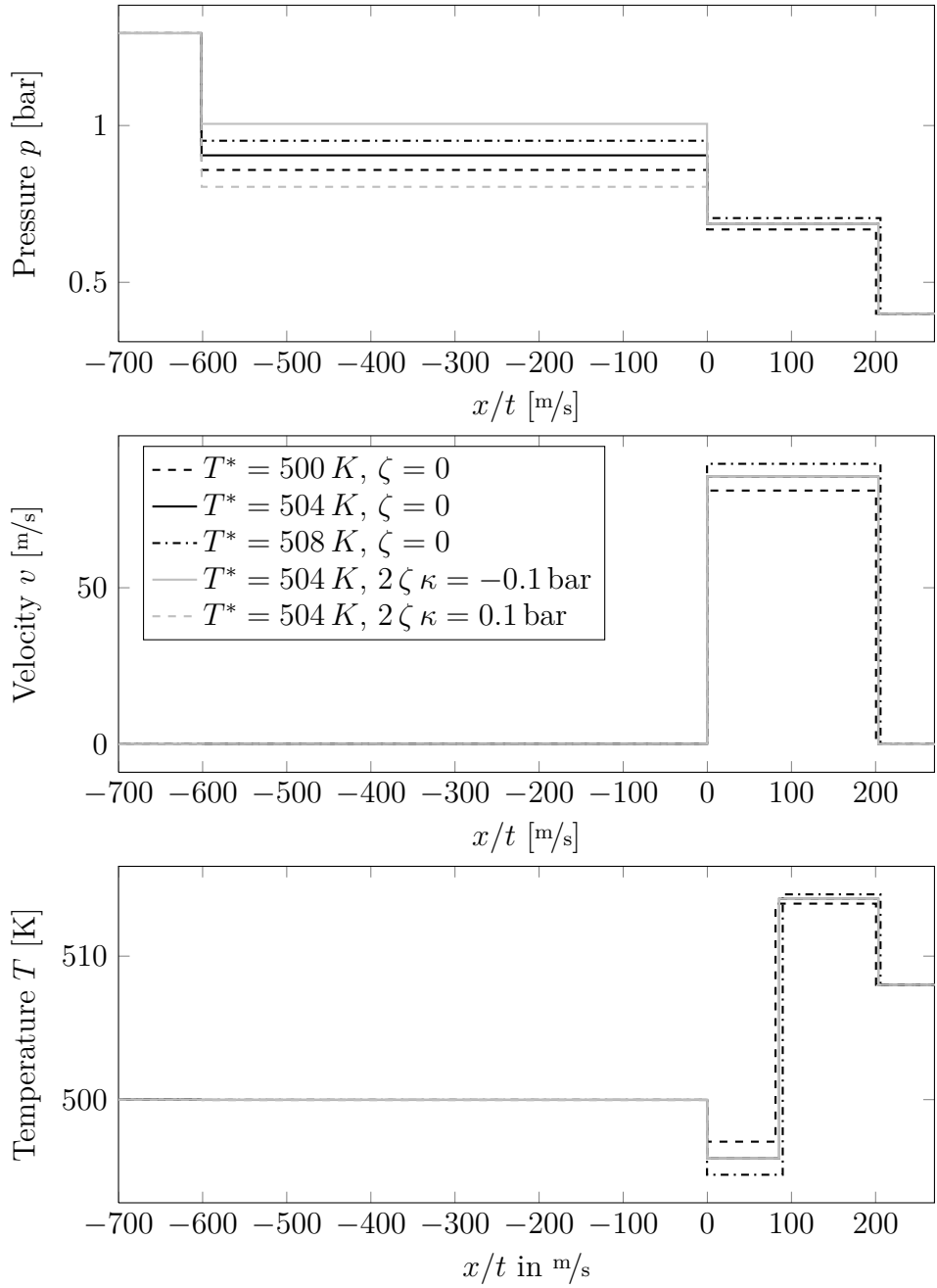
\begin{figure}
  \begin{tikzpicture}\centering
    \begin{axis}[ xmin=-700, xmax=270, 
                  xlabel={$x/t$ [$\nicefrac{\text{m}}{\text{s}}$]},
                  ylabel={Pressure $p$ [bar]}, 
                  x post scale = 1.6, y post scale = 0.8,
                  legend pos= north east,legend cell align=left
                  ]
    \addplot [thick, dashed]	table[x expr={\thisrowno{0}}, y expr={1e-5*\thisrowno{4}}, col sep=comma] {Data/dodecane_fan/fan_Tl.csv};
    \addplot [thick, solid ] 	table[x expr={\thisrowno{0}}, y expr={1e-5*\thisrowno{4}}, col sep=comma] {Data/dodecane_fan/fan_Tmean.csv}; 
    \addplot [thick, dashdotted]	table[x expr={\thisrowno{0}}, y expr={1e-5*\thisrowno{4}}, col sep=comma] {Data/dodecane_fan/fan_Tr.csv};
    \addplot [thick, lightgray]	table[x expr={\thisrowno{0}}, y expr={1e-5*\thisrowno{4}}, col sep=comma] {Data/dodecane_fan/fan_surf-.1bar.csv};
    \addplot [thick, dashed, lightgray]	table[x expr={\thisrowno{0}}, y expr={1e-5*\thisrowno{4}}, col sep=comma] {Data/dodecane_fan/fan_surf+.1bar.csv};
   
    \end{axis}
    \end{tikzpicture}
    
    \begin{tikzpicture}\centering
    \begin{axis}[ xmin=-700, xmax=270, 
                  xlabel={$x/t$ [$\nicefrac{\text{m}}{\text{s}}$]},
                  ylabel={Velocity $v$ [$\nicefrac{\text{m}}{\text{s}}$]}, 
                  x post scale = 1.6, y post scale = 0.8,
                  legend pos= north west,legend cell align=left
                  ]
    \addplot [thick, dashed]	table[x expr={\thisrowno{0}}, y expr={\thisrowno{2}}, col sep=comma] {Data/dodecane_fan/fan_Tl.csv};
    \addplot [thick, solid ] 	table[x expr={\thisrowno{0}}, y expr={\thisrowno{2}}, col sep=comma] {Data/dodecane_fan/fan_Tmean.csv}; 
    \addplot [thick, dashdotted]	table[x expr={\thisrowno{0}}, y expr={\thisrowno{2}}, col sep=comma] {Data/dodecane_fan/fan_Tr.csv};
    \addplot [thick, lightgray]	table[x expr={\thisrowno{0}}, y expr={\thisrowno{2}}, col sep=comma] {Data/dodecane_fan/fan_surf-.1bar.csv};
    \addplot [thick, dashed, lightgray]	table[x expr={\thisrowno{0}}, y expr={\thisrowno{2}}, col sep=comma] {Data/dodecane_fan/fan_surf+.1bar.csv};
    \legend{ {$T^\ast=500\,K$, $\zeta=0$},
             {$T^\ast=504\,K$, $\zeta=0$},
             {$T^\ast=508\,K$, $\zeta=0$},
             {$T^\ast=504\,K$, $\surf=-0.1$\,bar},
             {$T^\ast=504\,K$, $\surf=0.1$\,bar} };
    \end{axis}    
    \end{tikzpicture}
  
    \begin{tikzpicture}\centering
    \begin{axis}[ xmin=-700, xmax=270, 
                  xlabel={$x/t$ in $\nicefrac{\text{m}}{\text{s}}$},
                  ylabel={Temperature $T$ [K]}, 
                  x post scale = 1.6, y post scale = 0.8,
                  legend pos= north east,legend cell align=left
                  ]
    \addplot [thick, dashed]	table[x expr={\thisrowno{0}}, y expr={\thisrowno{5}}, col sep=comma] {Data/dodecane_fan/fan_Tl.csv};
    \addplot [thick, solid ] 	table[x expr={\thisrowno{0}}, y expr={\thisrowno{5}}, col sep=comma] {Data/dodecane_fan/fan_Tmean.csv}; 
    \addplot [thick, dashdotted]	table[x expr={\thisrowno{0}}, y expr={\thisrowno{5}}, col sep=comma] {Data/dodecane_fan/fan_Tr.csv};
    \addplot [thick, lightgray]	table[x expr={\thisrowno{0}}, y expr={\thisrowno{5}}, col sep=comma] {Data/dodecane_fan/fan_surf-.1bar.csv};
    \addplot [thick, dashed, lightgray]	table[x expr={\thisrowno{0}}, y expr={\thisrowno{5}}, col sep=comma] {Data/dodecane_fan/fan_surf+.1bar.csv};
   
    \end{axis}    
  \end{tikzpicture}  
  \caption{Numerical results for the Riemann problem for Example \ref{exp:dodecane2:RP} with different reference 
  temperatures and surface tensions. All figures correspond to the same legend. } \label{fig:dodecane2:RP}  

\end{figure}

For the second test problem we vary the reference temperature $T^\ast$ and the constant curvature $\kappa$. 
We consider the  initial states  
\begin{align}
%   \label{eq:case2_ini} 
  (\rho,v,p,T)(x,0) &= \begin{cases}
                   (584.01\,\nicefrac{\text{kg}}{\text{m}^3},0\,\nicefrac{\text{m}}{\text{s}}, 1.39\,\text{bar}, 500\,\text{K}) &\text{ if }x\le 0,\\
                   (1.65\,\nicefrac{\text{kg}}{\text{m}^3},0\,\nicefrac{\text{m}}{\text{s}}, 0.4\,\text{bar}, 508\,\text{K}) &\text{ else }
                 \end{cases}
%                 \nonumber\\
\end{align}
with $p_\text{sat}(T=500\,\text{K}) = 1.29\,\text{bar}$. 
%At the temperature $T=503.15$\,K the pressure $1.39$\,bar corresponds to the saturation pressure. 
The entropy production rate is \eqref{eq:kinrel} with $k^\ast=50\,\nicefrac{\text{m}^4}{\text{kg\ s}}$ and the surface tension coefficient is $\zeta=0.0089\,\nicefrac{\text{N}}{\text{m}}$.\\
The solution has the same wave pattern as in the previous example.
% \ref{exp:dodecane:RP}. 
Figure \ref{fig:dodecane2:RP} shows the solution for different values of the reference temperature and surface tension.
Differences in the reference temperature result only  in slightly different solutions. The corresponding results are plotted 
in black as straight or dashed lines. 
The structure of the solution is quite the same, but the values of the constant states are different. The speed of the evaporation wave 
varies and is 
\begin{align*}
 s = \begin{cases}
      -0.28  \,\nicefrac{m}{s} &\text{for } T^\ast=500\,\text{K}, L(T^\ast)=  249410\,\nicefrac{J}{kg} \text{ (dashed line)}, \\
      -0.33  \,\nicefrac{m}{s} &\text{for } T^\ast=504\,\text{K}, L(T^\ast)=  246784\,\nicefrac{J}{kg} \text{ (solid line)}, \\  
      -0.37  \,\nicefrac{m}{s} &\text{for } T^\ast=508\,\text{K}, L(T^\ast)=  244115\,\nicefrac{J}{kg} \text{ (dash dotted line)}.
     \end{cases}
\end{align*}
The gray lines in Figure \ref{fig:dodecane2:RP} correspond to solutions with different $\zeta$.

As expected, the capillarity forces affect the pressure of the solution in particular. Temperature and fluid velocity nearly coincide with the 
solutions obtained for $\zeta =0$. 
The amount of surface tension in the first case (gray solid line) corresponds to a droplet of $3.5\,\mu$m diameter. 
The gray dashed line corresponds to a bubble with the same diameter as only the sign is altered.

\subsection{Validation of the numerical method on exact solutions of the two-phase Riemann problem}
Up to now we considered only the exact multi-phase Riemann solver \ref{sec:riemannsolver} for varying parameters. 
Next we  validate the numerical method from Section \ref{sec:bulksolver} in the one-dimensional case. 
The test problem is the  same shock-tube problem with
resolved evaporation effects as used in Section \ref{exp:dodecane1:RP} for 
entropy production coefficient $k^\ast=47.39\,\frac{\text{m}^4}{\text{kg\,s}}$, and reference temperature   $T^\ast=500$\,K.  
The  benchmark solution is now computed from Algorithm \ref{algo:micromfull}.
%The initial conditions are given in \eqref{eq:case1_ini} 

% Hence, the initial conditions for the fluid n-dodecane are chosen according to
% \begin{align}
%   (\rho,v,p,T) &= \begin{cases}
%                    (584.08\,\nicefrac{\text{kg}}{\text{m}^3},0\,\nicefrac{\text{m}}{\text{s}}, 1.5\,\text{bar}, 500\,\text{K}) &\text{ if }x\le 0.5,\\
%                    (  4.38\,\nicefrac{\text{kg}}{\text{m}^3},0\,\nicefrac{\text{m}}{\text{s}}, 1.0\,\text{bar}, 500\,\text{K}) &\text{ else. }
%                  \end{cases}
%                  \label{eq:case3_ini} 
% %                 \nonumber\\
% \end{align}
% with $ p_\text{sat}(T=500\,\text{K}) = 1.29\,\text{bar}$, 
The initial states induce an  overheating of approximately
$7$\,K on the liquid side and an overheating of $11$\,K on the vapor side, 
which results in a strong evaporation rate at the interface.  

\begin{figure}
  \centering
  \subfloat[Density]{\includegraphics[width=.45\textwidth]{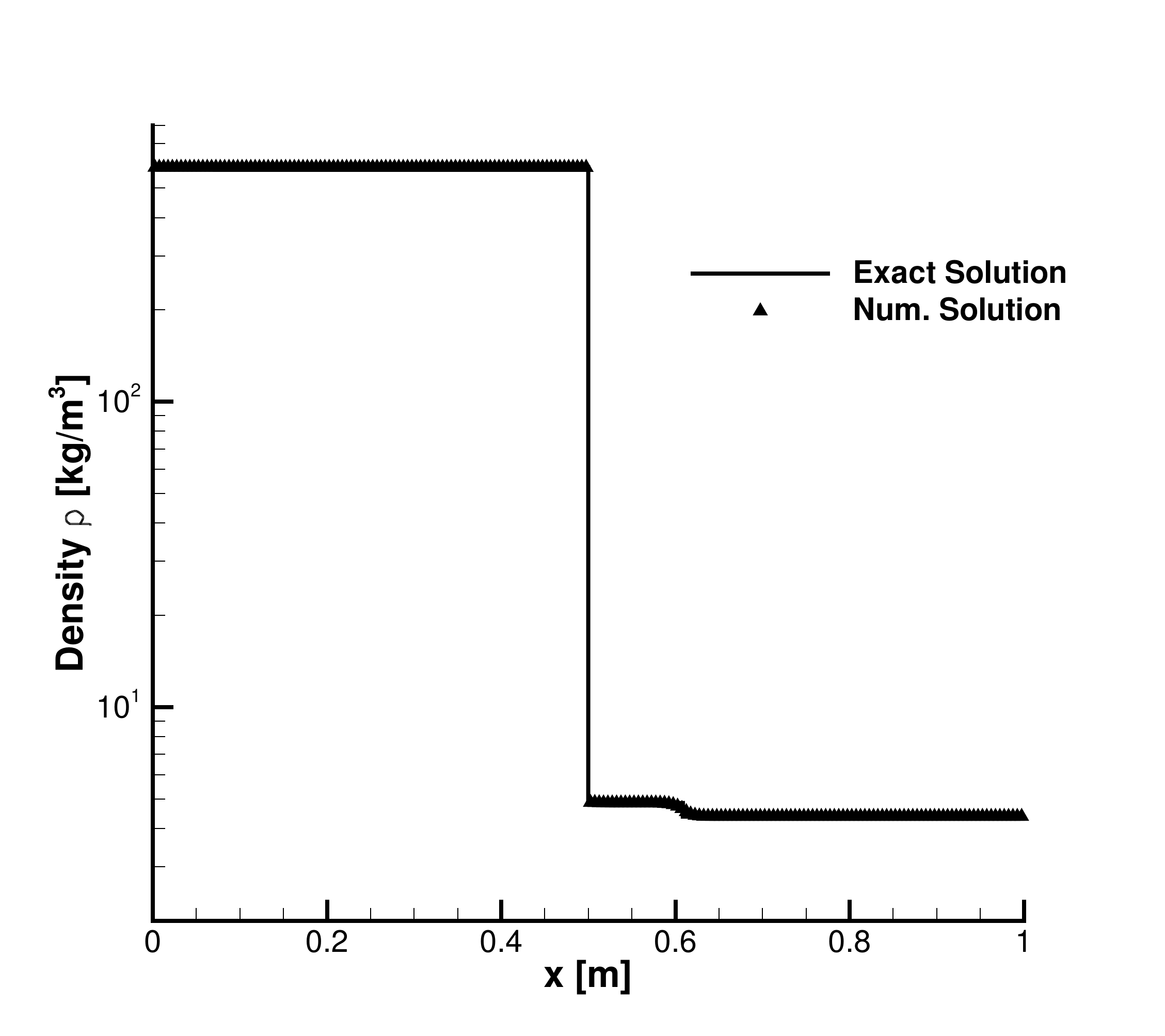}} \hfil
  \subfloat[Pressure]{\includegraphics[width=.45\textwidth]{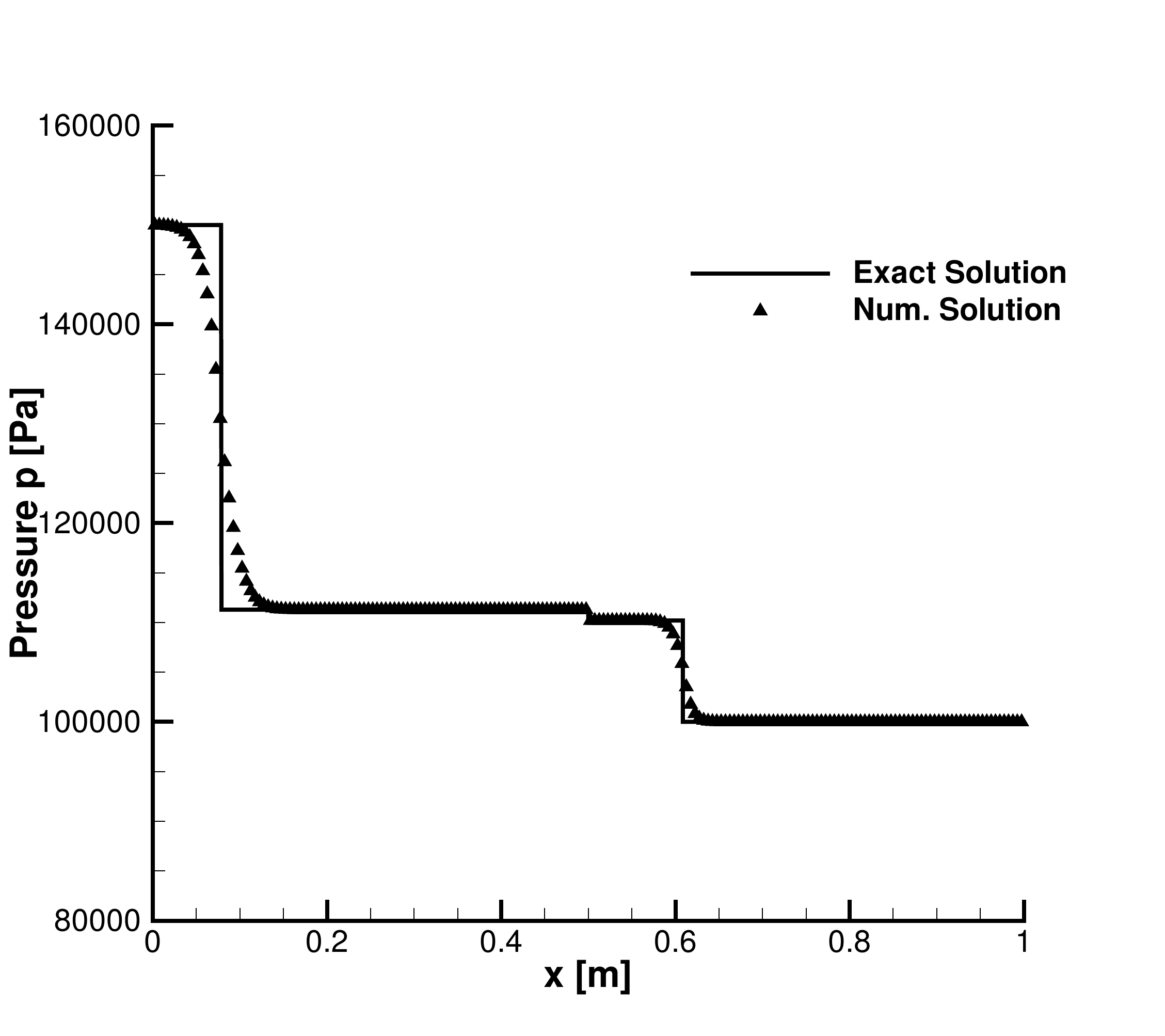}} \\
  \subfloat[Velocity]{\includegraphics[width=.45\textwidth]{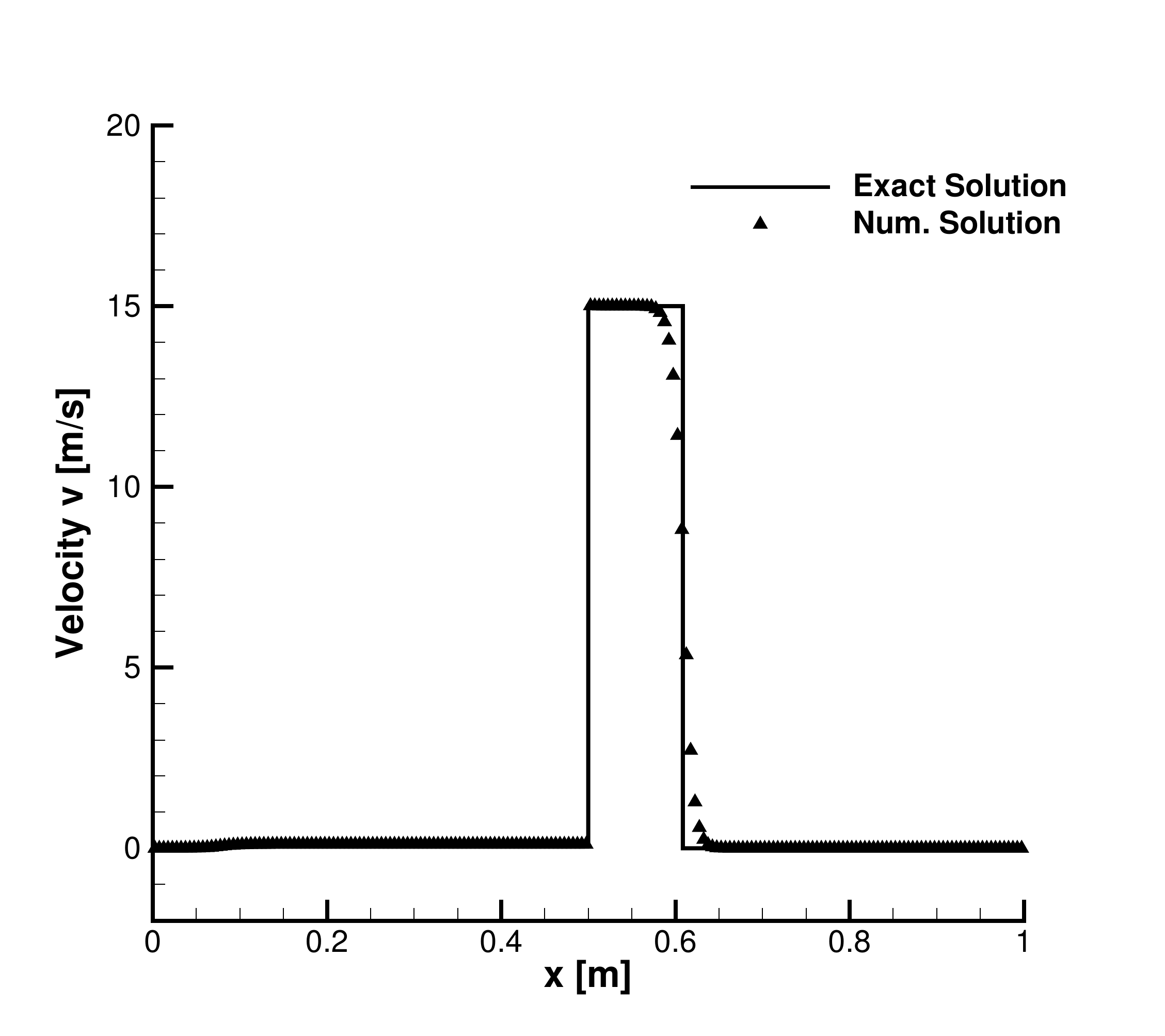}} \hfil
  \subfloat[Temperature]{\includegraphics[width=.45\textwidth]{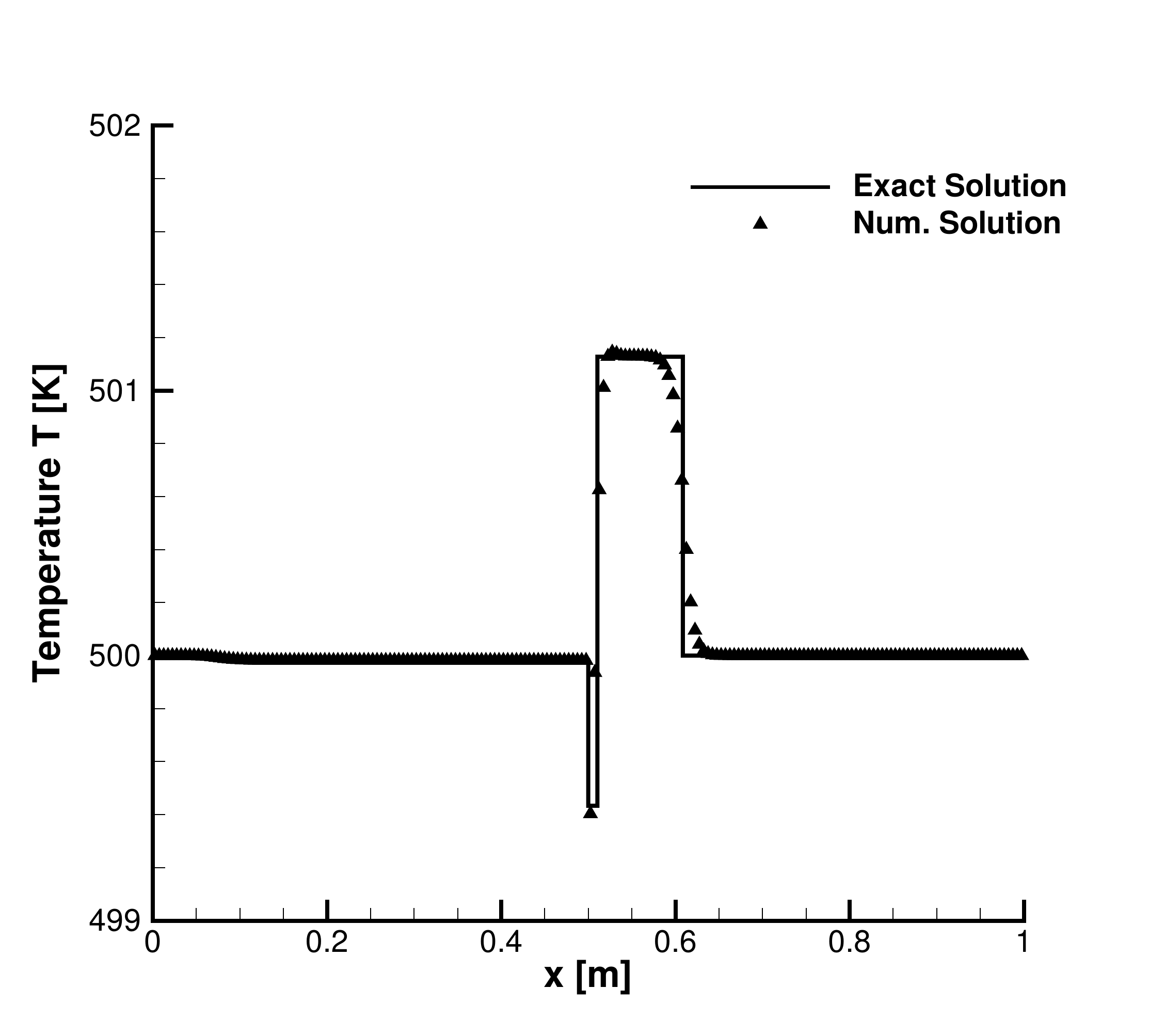}} \\
  \caption{Results for the one-dimensional n-dodecane shock tube problem at 
  time $t=0.7\,\mu$s.  Approximate numerical solution  and  exact solution of the Riemann solver.}
  \label{fig:1driemann}
\end{figure}

In Figure~\ref{fig:1driemann} the comparison between the numerical (small circles) and the benchmark (continuous line) solution  is 
plotted at $t=700\,\mu$s. The numerical solution was obtained using 200 degrees of freedom (DOF) in the unit interval. %\corr{dann passen die Anfangsdaten nicht}
% by use of a second order finite volume TVD scheme \corr{Was für Verfahren werden denn jetzt im  Bulk benutzt?}. 
%The initial data are directly imposed as initial discontinuity. The numerical values are plotted as  
%small circles, while the exact one is plotted as a continuous line. 
The exact solution and the approximate two-phase Riemann 
problem solution coincide very well. The wave pattern is also correctly recognized by the numerical scheme. A shock wave runs 
to the right hand side and is smeared out by the numerical dissipation over a couple of grid cells. The phase boundary is sharply resolved 
and can be clearly seen either in the density,  or  in the velocity plot. 
%The contact discontinuity can not be seen in
%the density plot due to the small jump.
Both the contact and phase boundary are close together -- the right jump is the contact while the left one is the sharp phase boundary.
The left running  (steep) rarefaction wave is clearly visible in the pressure plot.    
%The application of the interface Riemann solver at the position of the phase interface does reproduce the exact solution of the Riemann problem including all waves present in the Riemann wave fan.

The error norms and the order of convergence are shown in Table~\ref{tab:1driemann}. Although the phase boundary is sharply resolved 
across one cell the experimental order of convergence is   only  $0.5$ with respect to the $L^2(0,1)$-norm.
This is due to the  occurrence of the contact wave in the Riemann pattern. 

\begin{table}
 \centering
 \begin{tabular}{lcc}
  \toprule
  \# DOF & $L^2(0,1)$-error & EOC \\
  \midrule
  40  & 5.4111e-03 &        \\
  80  & 3.7964e-03 & 0.5113 \\
  160 & 2.7627e-03 & 0.4585 \\
  320 & 1.9751e-03 & 0.4842 \\
  640 & 1.4203e-03 & 0.4757 \\
  1280& 1.0240e-03 & 0.4720 \\
  \bottomrule
 \end{tabular}
 \caption{Convergence order for the one-dimensional Riemann 
 problem using the numerical method. Shown is the ${L}^2$-error for  the specific volume 
 $\tau=\nicefrac{1}{\rho}$. The DOF characterizes the resolution of the numerical scheme.}
 %\corr{Vorher Zellen, jetzt bitte auch. Auflösung traurig}.}
 \label{tab:1driemann}
\end{table}

\subsection{Validation of the numerical method by long-time behavior}
% Calculation of the long-time behavior with wall BC =>Maxwell states should be the solution
The purpose of the this test case is to study the long-time behavior of the numerical method for a one-dimensional domain.
If appropriate boundary conditions (see below) are applied 
it is expected that  the  exact solution  converges  for  $t \to \infty$  to a   
static two-phase equilibrium. %A two-phase equilibrium solution for $d=1$ consists of two constant states 
%$\vv U^+_\text{eq}$ and $\vv U^-_\text{eq}$ that  satisfy the %(see \ref{Bedingungen})
%thermodynamic ($\llbracket G\rrbracket = 0$) and the hydrodynamic ($\llbracket p \rrbracket =  0$) equilibrium conditions for a stationary interface.

The long-time behavior is investigated using a closed one-dimensional computational domain. In the numerical method we apply wall boundary conditions that imply that the velocity is zero at the boundaries.
% \corr{Was heisst das?}.
Thus, the total mass and energy inside the computational domain is conserved.\\
We performed a simulation for the initial conditions \eqref{eq:case1_ini} over a
physical time of $5\,s$. %and compared the states at the phase interface to the thermodynamic 
%equilibrium states $p^\sat(T_\text{int})$ and $\rho^\sat(T_\text{int})$ evaluated using \eqref{eq:saturation} for the temperature at the interface position $T_\text{int}$.
% \corr{Wie bekomme ich die denn raus? Eigentlich nur Residuum?}.
For the final time, the numerical solution remains static and the states are summarized in Table~\ref{tab:longtimetest}. This equilibrium solution is compared with the saturation states $\rho_{\liq/\vap}^\sat(T)$, $p^\sat(T)$, evaluated for the liquid temperature $T=499.99$\,K.
%the solution to the state on the saturation curve. 

It can be observed, that for long simulation times the states at phase equilibrium are reproduced. 
Due to evaporation effects, the density increased in the vapor phase and decreased in the liquid phase 
to their respective saturation values in accordance with the theory.
The remaining temperature gap is due to the non-unique static interface, see Remark~\ref{rmk2}.

\begin{table}
 \centering
 \begin{tabular}{lllll} \toprule
  & \multicolumn{2}{c}{numerical method} & \multicolumn{2}{c}{Phase equilibrium} \\
  & vapor & liquid & vapor & liquid \\ \midrule
Density [$\nicefrac{\text{kg}}{\text{m}^3}$] & 5.802  & 584.027 &  5.801  & 584.024  \\
Pressure [bar]                               & 1.294  & 1.294   &  1.294  & 1.294    \\
Temperature [K]                              & 499.99 & 500.03  &  499.99 & 499.99   \\
 \bottomrule
 \end{tabular}
 \caption{Results for the long-time behavior test case compared to the values at phase equilibrium.}
%  The values for the phase equilibrium are obtained using CoolProp \cite{coolprop}.}
 \label{tab:longtimetest}
\end{table}

\subsection{Comparison to experiments with explosive evaporation/boiling}
% FIXME: strange behavior for high temperatures and low reservoir pressures 
%         => maybe related to attached waves? Approach here does not resolve these waves.
%         => exact RS not as robust as the HLLP 
The numerical model is compared to published literature data 
for explosive boiling\slash evaporation for the fluids n-dodecane, propane and butane.
Note that in these experiments fast phase change processes take place. 
We compare quantitatively the results of the present approach to the experiments of Simoes-Moreira\&Shepherd 
\cite{simoes1999evaporation} for the fluid n-dodecane and to the experiments of Reinke\&Yadigaroglu \cite{reinke2001explosive} for the fluids propane and butane. 
% The choice of these fluids is related to the availability of an accurate estimate for the entropy production constant for the alkanes. The accurate approximation of the mass transfer rate is the crucial part for the comparison to experiments.

\subsubsection{Shock tube experiments with n-dodecane}
% (1) Experiments of Simoes-Moreira with liquid dodecane and expansion to vacuum
We compare the phase velocity of the evaporation front to the measurements of Simoes-Moreira\&Shepherd \cite{simoes1999evaporation}. 
They investigated stable evaporation waves in superheated liquid n-dodecane 
(critical temperature $T_c=658.1$\,K, molar mass $M=0.1703\,\nicefrac{\text{kg}}{\text{mol}}$) 
% \corr{Was sind das für Werte und wo gehen sie in das Modell/Numerik ein?}
at different initial temperatures. 
These experiments have been chosen also by Saurel et al. \cite{saurel2008modelling} 
and Zein et al. \cite{zein2010modeling} for the numerical validation of their phase transition method. Their approach is based on 
the  homogenized Baer-Nunziato  model,  together with a thermodynamic relaxation procedure  to account for  phase transition effects.

For initial conditions it is assumed that the liquid is in the saturation state at the given initial temperature
of the experiment, as described in \cite{simoes1999evaporation}. On the vapor side, a constant (low) pressure of 
$p=0.05$\,bar is assumed. These initial conditions correspond to the experimental measurements.
% The temperature is set constant in the computational domain according to the initial temperature in the experiments.
% This is an approximation of the low reservoir pressure used in the experiments. 
%The reservoir pressure has only  an influence on the evaporation front velocity (see figure \ref{fig:simoesresult} bottom figure). 
%We assume an initially constant temperature within the whole domain. 
% These initial conditions correspond to the experimental conditions. We applied no pressure above the saturation conditions of the liquid contrary to the initial conditions of Zein et al. \cite{zein2010modeling}.

\begin{figure}
 \centering
  \pgfplotstableread{Data/datasimoes_chokedseries.dat}{\datasimoes}
  \begin{tikzpicture}[scale=1,]
   \begin{axis}[xlabel={Temperature $T$\,[K]},xmin=450,xmax=580,
               ylabel={Evaporation front velocity $s$\,[$\nicefrac{\text{m}}{\text{s}}$]},ymin=0,ymax=2.0,
               height=8cm,width=12cm,grid=major,
               title={Choked flow series},
               legend pos= north west,legend cell align=left,]
   \addplot [black,only marks,mark=*,mark size=3pt] table [x={temp}, y={ufr_exp}] {\datasimoes};
   \addlegendentry{Experiment \cite{simoes1999evaporation}};
   \addplot [black,dashed,mark=diamond*,mark size=3pt] table [x={temp}, y={ufr_saurel}] {\datasimoes};
   \addlegendentry{Saurel \cite{saurel2008modelling}};
   \addplot [black,dashdot,mark=square*,mark size=3pt] table [x={temp}, y={ufr_zein}] {\datasimoes};
   \addlegendentry{Zein \cite{zein2010modeling}};
   \addplot [black,mark=triangle*,mark size=3pt] table [x={temp}, y={ufr_num}] {\datasimoes};
   \addlegendentry{Present model};
   \end{axis}
   \end{tikzpicture}
  \vspace{1cm}
  \pgfplotstableread{Data/datasimoes_tempseries.dat}{\datasimoes}
  \begin{tikzpicture}[scale=1,]
   \begin{axis}[xlabel={Reservoire pressure $p_\text{res}$\,[bar]},xmin=0,xmax=1.4,
               ylabel={Evaporation front velocity $s$\,[$\nicefrac{\text{m}}{\text{s}}$]},ymin=0,ymax=0.6,
               height=6cm,width=12cm,grid=major,
               title={Isothermal test series ($T=503$\,K)},
               legend pos= north east,legend cell align=left,]
   \addplot [black,only marks,mark=*,mark size=3pt,unbounded coords=jump] table [x expr={\thisrow{P}/1e5}, y={ufr_exp}] {\datasimoes};
   \addlegendentry{Experiment \cite{simoes1999evaporation}};
   \addplot [black,mark=triangle*,mark size=3pt,unbounded coords=jump] table [x expr={\thisrow{P}/1e5}, y={ufr_num}] {\datasimoes};
   \addlegendentry{Present model};
   \end{axis}
  \end{tikzpicture}
  \caption{Comparison of evaporation front speeds $s$ to
  experimental measurements of Simoes-Moreira\&Shepherd \cite{simoes1999evaporation}. 
  Shown is the comparison for the choked series (top) and the isothermal series (bottom) at $T=503$\,K. 
  Included is the comparison to numerical studies of Saurel et al. \cite{saurel2008modelling} and Zein et al. \cite{zein2010modeling} 
  computed with a numerical relaxation method for the Baer-Nunziato model.}
   \label{fig:simoesresult}
\end{figure}
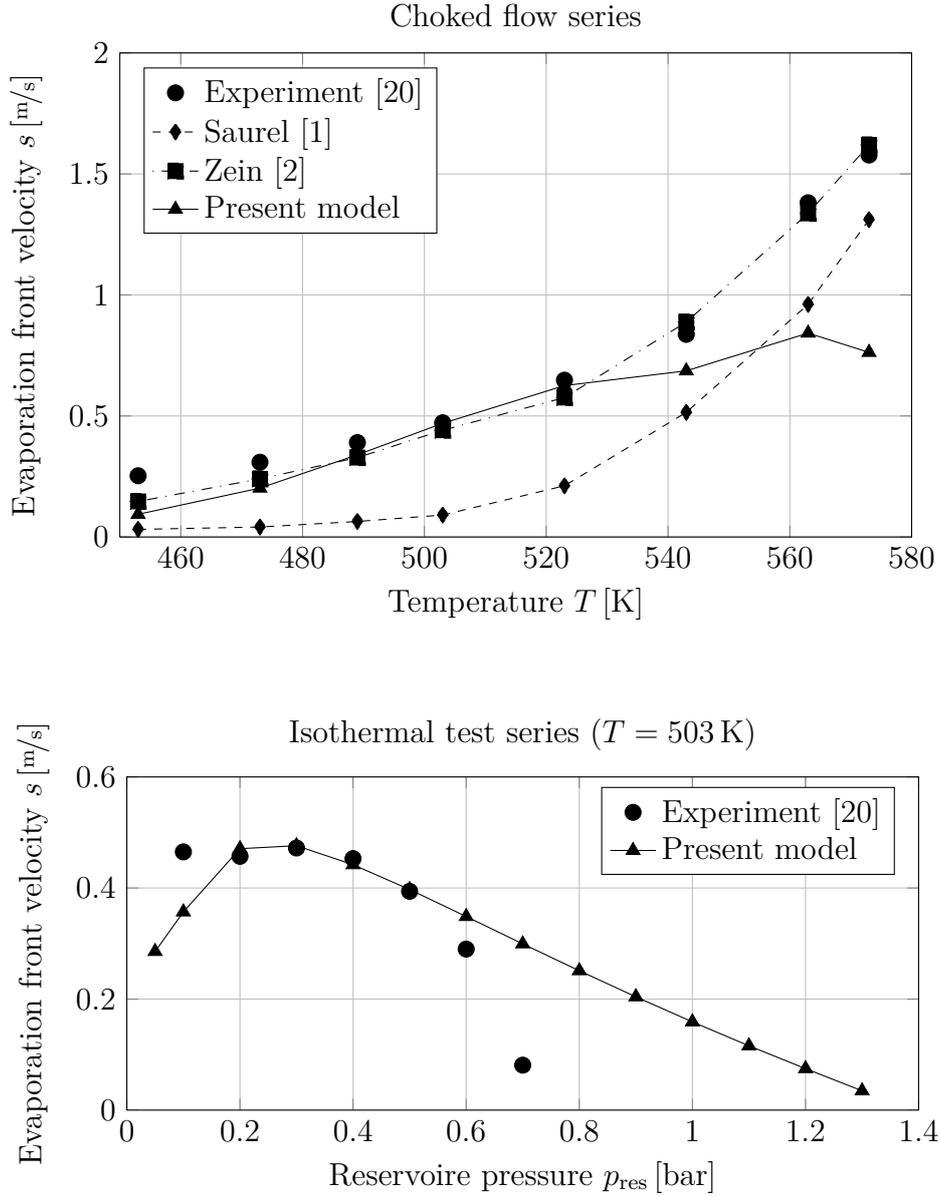

In the experiments of Simoes-Moreira\&Shepherd \cite{simoes1999evaporation}, the liquid n-dodecane is relaxed
into a reservoir with low pressure resulting in an evaporation wave with choked flow on the vapor side (``choked series'') for 
various initial liquid temperatures. In the experiments, the pressures on both sides of the evaporation wave as well as the speed
of the evaporation wave are measured. However, it was not possible to measure directly the densities in the bulk phases. 
In our numerical simulations the evaporation front speed  $s$ is estimated as the speed  of the phase boundary based 
on the jump conditions at the interface.
The estimated wave speeds of the evaporation wave are visualized in Figure~\ref{fig:simoesresult} (top figure) 
for different initial liquid temperatures including a comparison to the simulations
of Saurel et al. \cite{saurel2008modelling} and Zein et al. \cite{zein2010modeling}. 
%Note that in the evaporation models of Saurel and Zein a different numerical method was used.

For low investigated temperatures the experimental results are well reproduced underestimating 
slightly the measured front speed of the evaporation wave. For higher temperatures there is a slighter 
increase of the speed compared to the experimental as well as the other numerical results.
It seems that the assumed micro-scale model is not accurate for large interface pressure jumps. 
This might be related to the modeling of the entropy production coefficient $k^\ast$ for these conditions or
the discontinuous temperature at the interface.

The comparison for the isothermal series at $T=503$\,K in \cite{simoes1999evaporation} is shown in 
the lower diagram of Figure~\ref{fig:simoesresult} using different reservoir pressures $p_\text{res}$. 
Note that above a reservoir pressure of $p_\text{res}>0.7$\,bar no stable 
evaporation waves could be observed experimentally. 
For reservoir pressures $p_\text{res}< 0.6$\,bar our approach reproduces 
the experimentally observed wave speeds well up to pressures of $p_\text{res}<0.6$\,bar. 
However, for very low reservoir pressures 
($p_\text{res}\leq 0.1$\,bar) the front speed decreases again, as opposed to the measurements. 
This might be related to the already discussed inaccuracy for large pressure jumps. 
For higher pressure we observe a linear decrease of the evaporation front velocity. 
This should be due to the assumed micro-scale model because 
static phase boundaries under non-saturated conditions would require a different kinetic relation, see \cite{ABEKNO2006}.
%The experimental trend can be reproduced by introducing a threshold to the kinetic relation \cite{ABEKNO2006}.
% FIXME: strange behavior for p->0 with decreasing wave velocity due to very high mass transfer rates

\subsubsection{Butane and propane evaporation experiments}
% (2) experiments of Reinke and Yadigaroglu (various fluids: butane, propane, R-134a and water)
%  => exapansion to conditions at 1 bar
Reinke\&Yadigaroglu \cite{reinke2001explosive} investigated the explosive boiling of various 
liquids and developed a correlation for the evaporation front speed $s$ from their experiments. 
They expanded experimentally a saturated liquid at a superheated temperature to a state at ambient conditions 
($p_\text{vap}=1$\,bar). From the experimental data they correlated the 
front speed using the nominal superheat $\Delta T_\text{nom}$ as the dependent
variable. The nominal superheat is defined as
\begin{equation}
 \Delta T_\text{nom} = T_\text{ini} - T_\text{sat}(p=1\,\text{bar})\,,
\end{equation}
relative to the initial temperature $T_\text{ini}$ of the liquid and the saturation temperature of the fluid at $p=1$\,bar.
For butane this temperature is equal to $T_\text{sat}(p=1\,\text{bar})=272.3$\,K and for propane $T_\text{sat}(p=1\,\text{bar})=230.74$\,K. 
% The numerical velocity of the boiling front $s=U_\text{fr}$ is again calculated as local wave velocity at the interface.

We compare the evaporation front velocities for butane (critical temperature $T_c= 425.125$\,K, molar mass $M=0.0581\,\nicefrac{\text{kg}}{\text{mol}}$) and 
propane ($T_c= 369.89$\,K, $M=0.0441\,\nicefrac{\text{kg}}{\text{mol}}$) during the expansion process 
for various initial superheats. The temperature dependent values for the entropy production 
coefficient $k^\ast$ are taken from \cite{waibel2013mscthesis}, evaluated at the corresponding
initial temperature of the liquid phase. 

\begin{figure}
 \centering
  \pgfplotstableread{Data/datareinke_butane.dat}{\datareinke}
  \pgfplotstableread{Data/experiments_butane.dat}{\dataexperiments}
  \begin{tikzpicture}[scale=1,]
  \begin{axis}[xlabel={Nominal superheat $\Delta T_\text{nom}$ [K]},xmin=0,xmax=80,
               ylabel={Evaporation front velocity $s$ [$\nicefrac{\text{m}}{\text{s}}$]},ymin=0,ymax=1.5,
               height=6cm,width=12cm,grid=major,
               legend pos= north west,legend cell align=left,
               title=Butane]
   \addplot [black,only marks,mark=*,mark size=3pt] table [x={x}, y={u_frexp}] {\dataexperiments};
   \addlegendentry{Experiments \cite{reinke2001explosive}};
   \addplot [mark=none,black,samples=250, domain=30:73] {0.0307*x-0.977};
   \addlegendentry{Exp. correlation \cite{reinke2001explosive}};
   \addplot [black,only marks,mark=triangle*,mark size=3pt] table [x={DeltaT}, y={u_frnum}] {\datareinke};
   \addlegendentry{Present model};
   \end{axis}
   \end{tikzpicture}
  \vspace{1cm}
  \pgfplotstableread{Data/datareinke_propane.dat}{\datareinke}
  \pgfplotstableread{Data/experiments_propane.dat}{\dataexperiments}
  \begin{tikzpicture}[scale=1,]
  \begin{axis}[xlabel={Nominal superheat $\Delta T_\text{nom}$ [K]},xmin=0,xmax=80,
               ylabel={Evaporation front velocity $s$ [$\nicefrac{\text{m}}{\text{s}}$]},ymin=0,ymax=1.5,
               height=6cm,width=12cm,grid=major,
               legend pos= north west,legend cell align=left,
               title=Propane]
   \addplot [black,only marks,mark=*,mark size=3pt] table [x={x}, y={u_frexp}] {\dataexperiments};
   \addlegendentry{Experiments \cite{reinke2001explosive}};
   \addplot [mark=none,black,samples=250, domain=30:72] {0.0292*x-0.782};
   \addlegendentry{Exp. correlation \cite{reinke2001explosive}};
   \addplot [black,only marks,mark=triangle*,mark size=3pt] table [x={DeltaT}, y={u_frnum}] {\datareinke};
   \addlegendentry{Present model};
   \end{axis}
   \end{tikzpicture}
   \caption{Velocity of the boiling front $s$
   for butane (top) and propane (bottom) comparing the correlation and experiments of Reinke\&Yadigaroglu \cite{reinke2001explosive} to the results of the numerical method.}
   \label{fig:reinkeresult}
\end{figure}
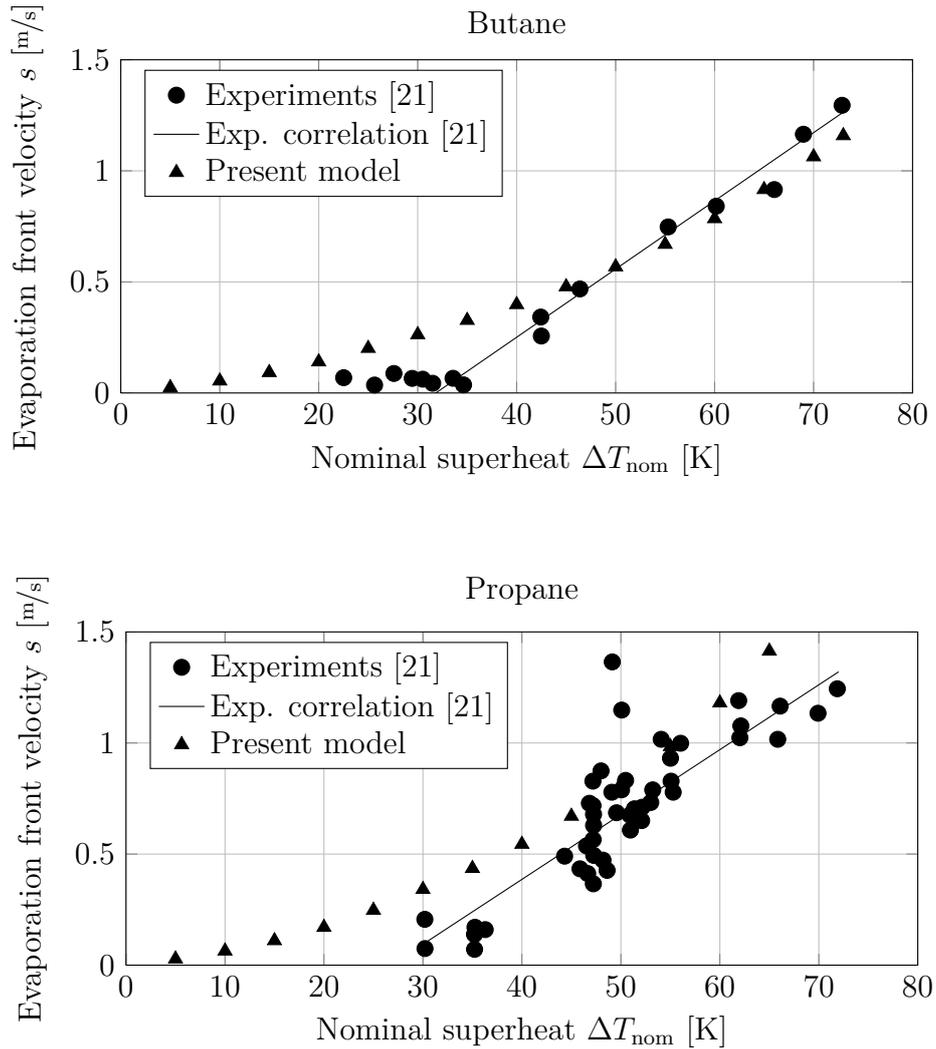

% comparison of the approaches
The numerical results are plotted in Figure~\ref{fig:reinkeresult} together with the experimental measurements of 
Reinke\&Yadigaroglu \cite{reinke2001explosive} 
and their linear correlation in terms of the nominal superheat $\Delta T_\text{nom}$. 
All numerical results for $\Delta T_\text{nom}>40$\,K (butane) and for $\Delta T_\text{nom}>30$\,K (propane) 
are within the 80\,\% confidence limits of the experiments. 
The trend of  increasing front speed for increasing superheats is reproduced by  the numerical approach. 
% Note also the large scatter in the experimental results, especially for the propane experiments.

For low nominal superheats, the velocity of the evaporation wave is slightly 
overestimated. This is due to the occurrence of phase boundaries under non-saturated conditions. 
Such conditions would require a different kinetic relation, see \cite{ABEKNO2006}.
% In the experiments at these conditions no stable evaporation front is observed. Such a wave can be investigated in the numerical approach due to the approximation of the phase transition wave. However, the evaporation front velocities tend to zero which is the correct behavior as for $\Delta T_\text{nom}\rightarrow 0$ the phase equilibrium is retained. For higher temperatures no stable solutions could be found that is agreement with the experimental investigation.

% \corr{Änderungen ab hier}

\subsection{Two-dimensional validation of the numerical method}
\subsubsection{Diagonal interface test case \label{sec:2driemann}}
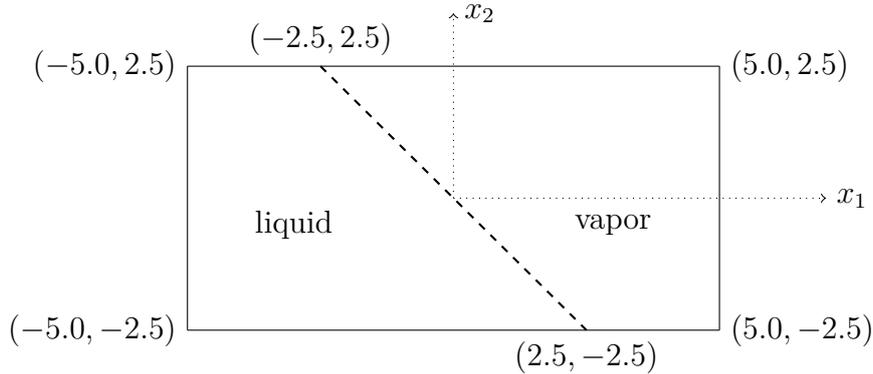
\begin{figure}
 \centering
 \begin{tikzpicture}[scale=0.7]
  \draw[dotted,->] (0,0)--(7,0) node[right] {$x_1$};
  \draw[dotted,->] (0,0)--(0,3.5) node[right] {$x_2$};
  \draw (-5,2.5)--(5,2.5);
  \draw (-5,-2.5)--(5,-2.5);
  \draw (5,-2.5)--(5,2.5);
  \draw (-5,-2.5)--(-5,2.5);
  \draw[thick,dashed] (-2.5,2.5)--(2.5,-2.5);
  \node[right] at (5,2.5) {$(5.0,2.5)$};
  \node[left] at (-5,2.5) {$(-5.0,2.5)$};
  \node[right] at (5,-2.5) {$(5.0,-2.5)$};
  \node[left] at (-5,-2.5) {$(-5.0,-2.5)$};
  \node[above] at (-2.5,2.5) {$(-2.5,2.5)$};
  \node[below] at (2.5,-2.5) {$(2.5,-2.5)$};
  \node at (-3,-0.5) {liquid};
  \node at (3,-0.5) {vapor};
 \end{tikzpicture}
 \caption{Computational domain for the diagonal interface test case. The interface is marked as dashed line. All dimensions are given in millimeters.}
 \label{fig:compdomain}
\end{figure}

% Validation of the multi-dimensional implementation and application of the Riemann solver
The next test problem for validation  uses a one-dimensional  Riemann problem, introduced in 
Section~\ref{sec:1driemann}, as planar solution in two space dimensions. 
We  consider an initial setting such that the initial states are separated by a  diagonal line  at an angle of $-45^\circ$
(see Figure \ref{fig:compdomain}).  The grid is a Cartesian grid, such that the exact solution, in particular the phase boundary,   is not aligned to the grid.\\
%with respect to the grid. A sketch of the computational domain is shown in Figure~\ref{fig:compdomain}
%including the oblique phase interface and the  dimensions. 
%As the waves do not interact with each other, the result on lines normal to the interface is compared to the exact solution obtained by 
%the interface Riemann solution into normal direction. 
%At the borders of the computational domain, the boundary conditions are kept fixed that introduce some numerical disturbances next to the boundaries. 
%The numerical solution along the center line of the computational domain is compared to the exact solution.

For the calculation of the discretization error, we consider the following (generic) Riemann problem with phase transfer
We consider the setting as  in Section \ref{sec:1driemann} for the fluid n-dodecane. The initial states  are given by
\begin{align}
  (\rho,v,p,T)(x_1,x_2,0) &= \begin{cases}
                   (584.08\,\nicefrac{\text{kg}}{\text{m}^3},0\,\nicefrac{\text{m}}{\text{s}}, 1.5\,\text{bar}, 500\,\text{K}) &\text{ if }x_1+x_2\le 0,\\
                   (4.38\,\nicefrac{\text{kg}}{\text{m}^3},0\,\nicefrac{\text{m}}{\text{s}}, 1.0\,\text{bar}, 500\,\text{K}) &\text{ else, }
                 \end{cases}
                 \label{eq:micro_dodecane1} \nonumber\\
 p_\text{sat}(T=500\,\text{K}) &= 1.29\,\text{bar.}               
\end{align}

\begin{figure}
 \centering
 \subfloat[Density]{\includegraphics[width=0.45\textwidth]{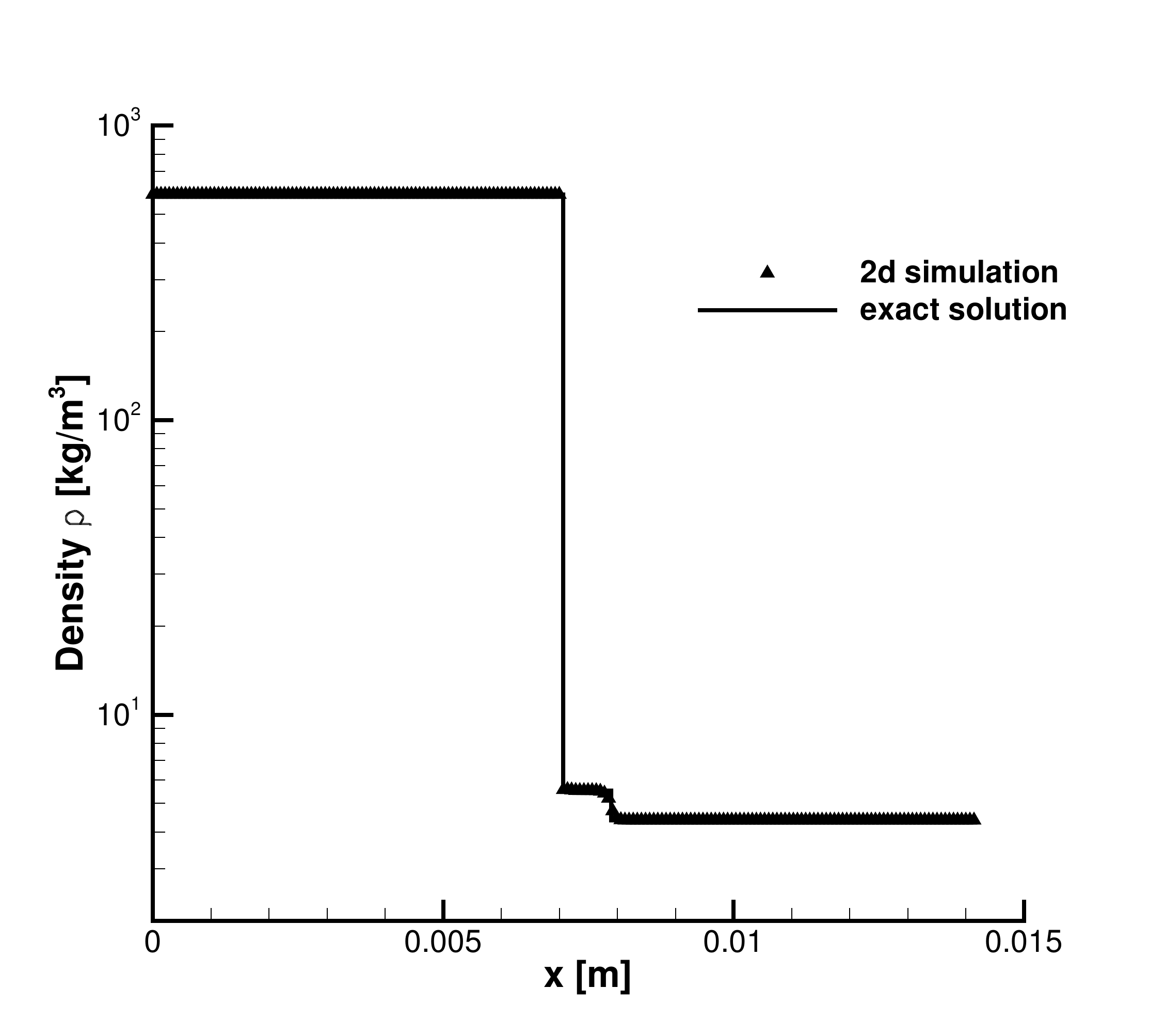}} \hfill
 \subfloat[Pressure]{\includegraphics[width=0.45\textwidth]{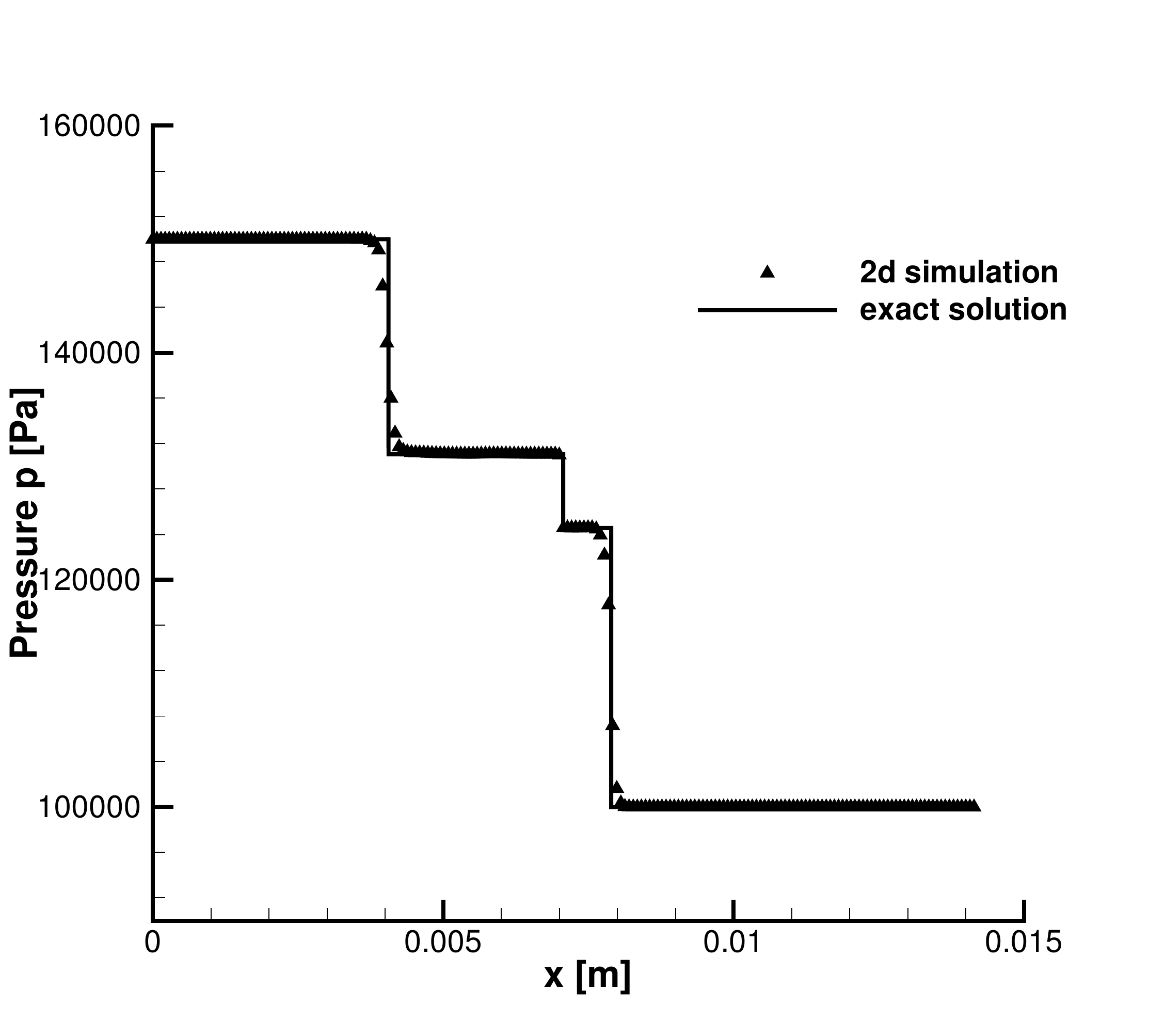}} \\
 \subfloat[Velocity]{\includegraphics[width=0.45\textwidth]{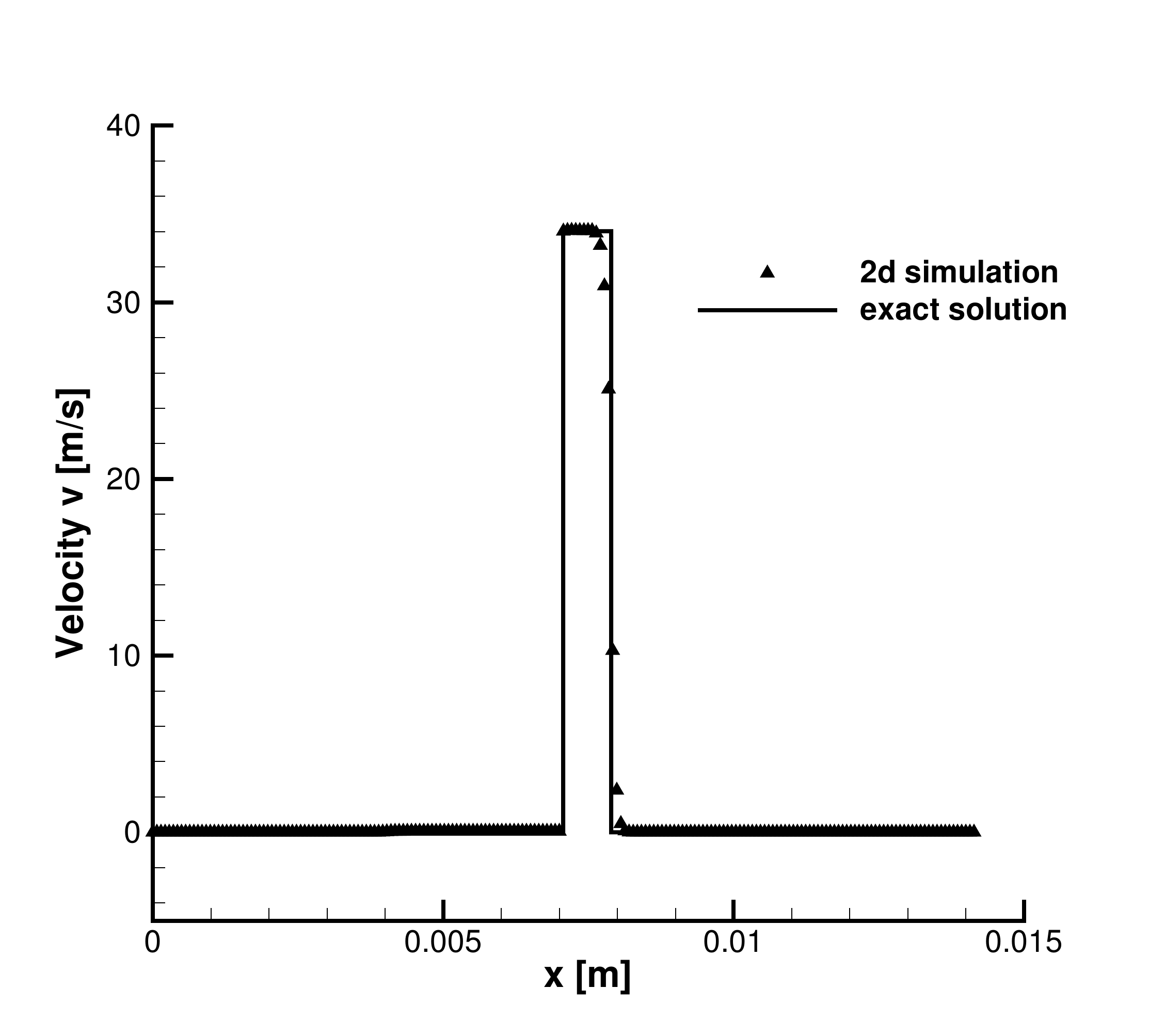}} \hfill
 \subfloat[Temperature]{\includegraphics[width=0.45\textwidth]{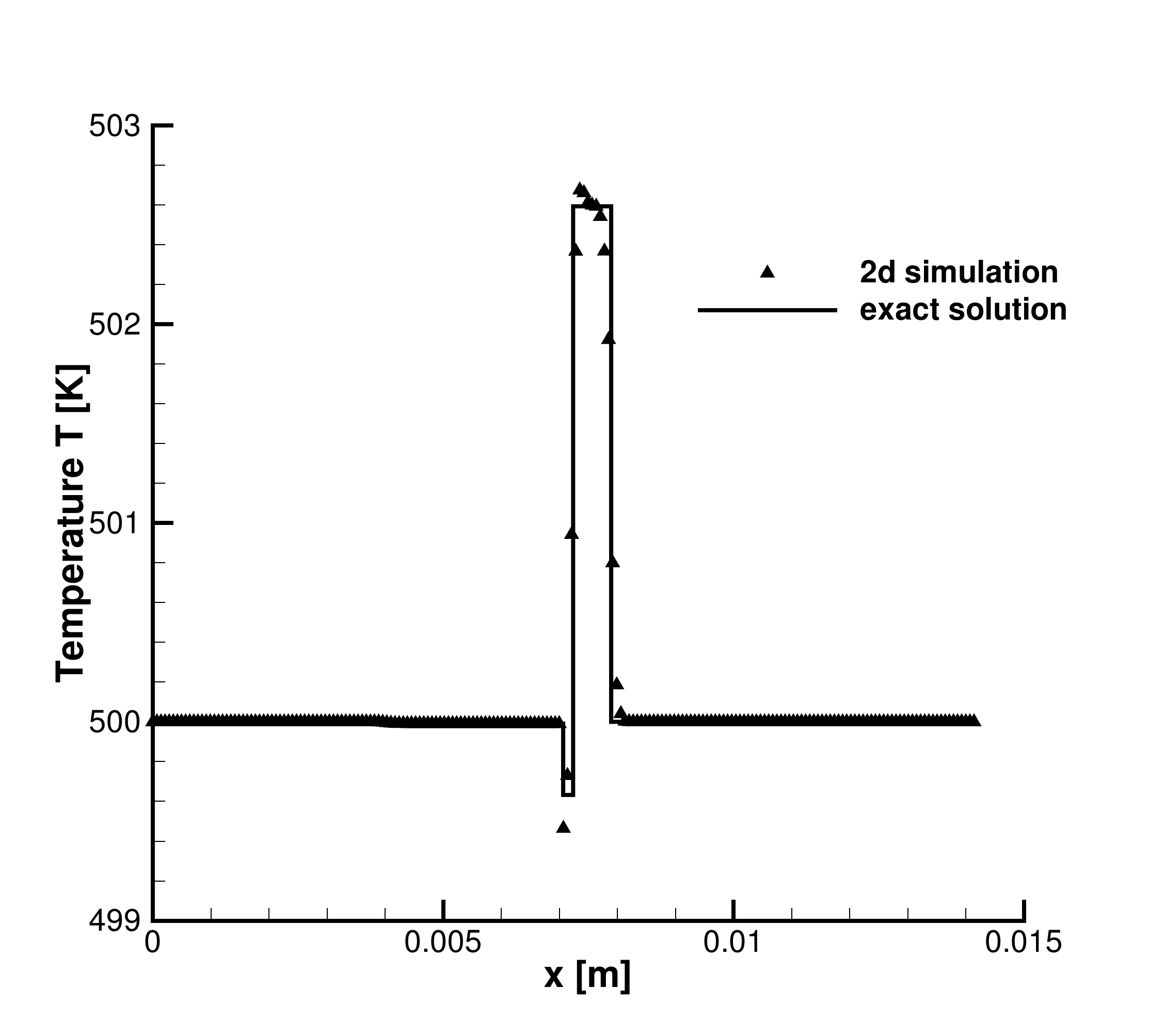}}
 \caption{Two-dimensional Riemann problem at time $t=5\,\mu s$ evaluated on the slice $y=x$ compared to the exact solution normal to the interface. Shown are the results for $160\times80$ degrees of freedom.}
 \label{fig:2driemann_centerline}
\end{figure}

Despite the coarse numerical staircase approximation of the interface, the interface resolution is
coincides well with the exact solution. The numerical solution (cut along the normal)
is shown in Figure~\ref{fig:2driemann_centerline} in comparison to the exact solution. 
Slight disturbances due to the interface approximation are visible directly next to the interface location.
They diminish with 
increasing distance from the interface or by refining the grid.% and do not disturb the solution. 
% The convergence order is retained also for the case of an diagonal phase interface that is not aligned with the phase interface. 

\subsubsection{Circular droplet}%\corr{Es fehlt die Angabe von $\zeta$ }
% comparison to one-dimensional approach with source terms
The previous test case validates the numerical approach for a simple planar interface but does not take  into account surface tension effects. 
%Such an interface does not represent the typical interface configuration in nature and often curved interfaces are present, e.\,g. at a spherical droplet. 
We compare in the following the numerical method for a  two-dimensional circular droplet to the numerical solution  solution for the equations in 
spherical coordinates. % Due to the geometric source terms no analytical solution is available in this case. 
The numerical strategy for the radially symmetric approach is e.\,g. described in \cite{Toro}.\\
The initial conditions for the test case are chosen identical to the test case in Section \ref{sec:2driemann}.
Instead of a diagonal interface geometry, a circular interface with a droplet radius of $r_0=1$\,mm is chosen together with a constant surface tension coefficient of $\zeta=0.01\,\nicefrac{\text{N}}{\text{m}}$, a realistic estimate for n-dodecane. 
In the two-dimensional computation 160 DOF are used to discretize the whole droplet while for the one-dimensional reference simulation a total of 400 DOF is used. Hence, the one-dimensional simulation has a much finer resolution and is regarded as reference solution. In comparison to the previous test case, the states between the waves are no longer constant.

\begin{figure}
 \subfloat[Density]{\includegraphics[width=0.45\textwidth]{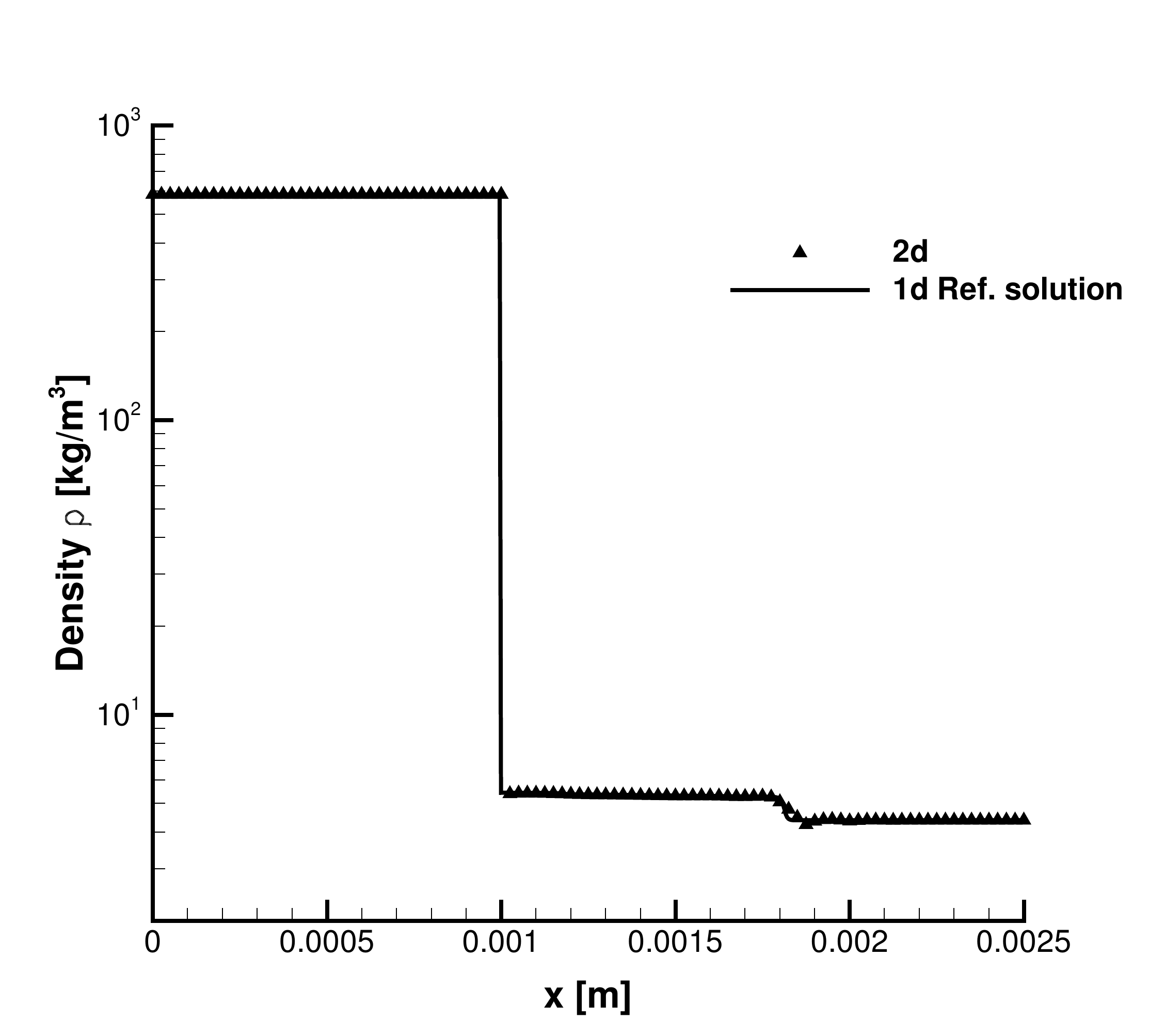}} \hfill
 \subfloat[Pressure]{\includegraphics[width=0.45\textwidth]{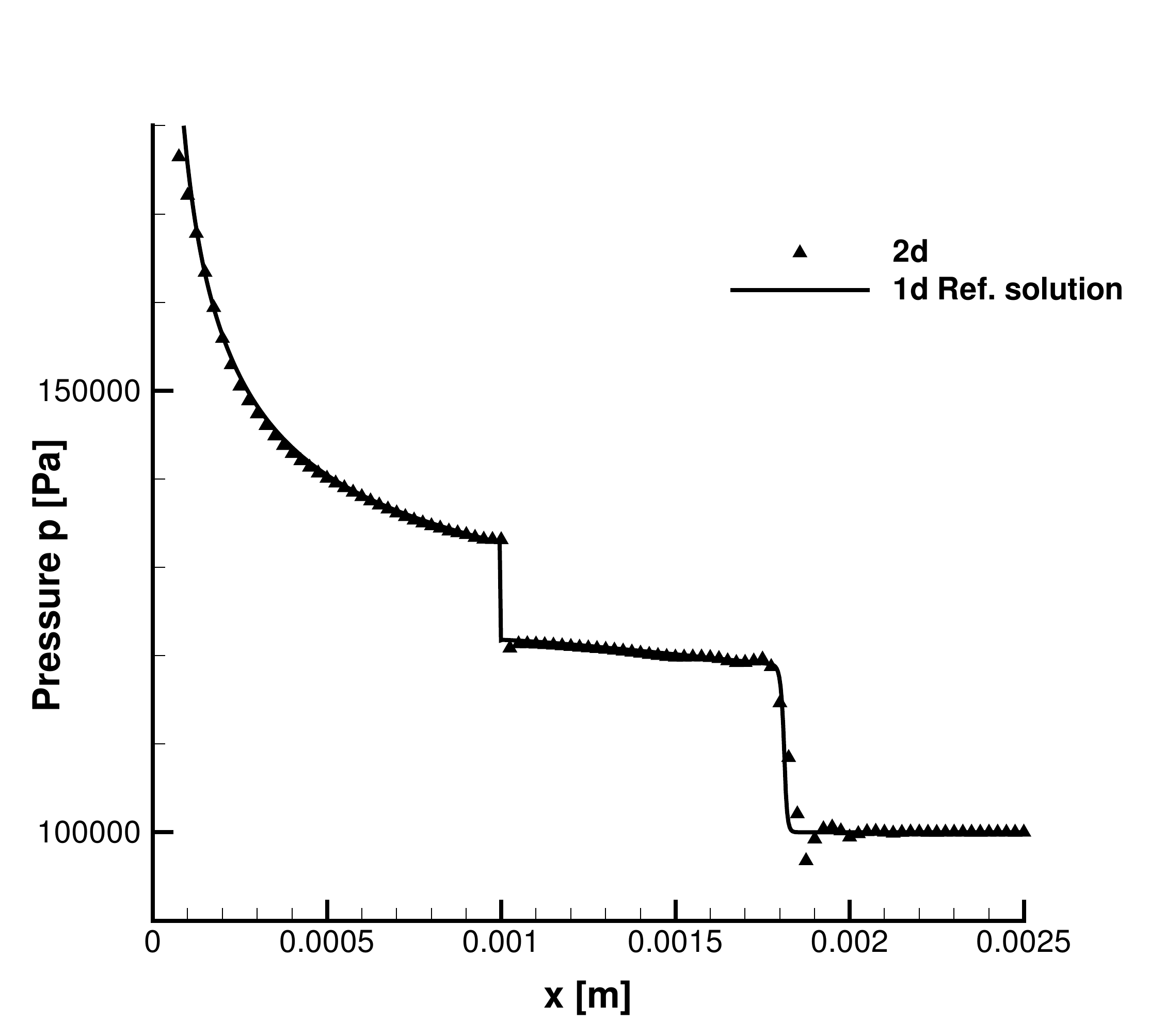}} \\
 \subfloat[Velocity]{\includegraphics[width=0.45\textwidth]{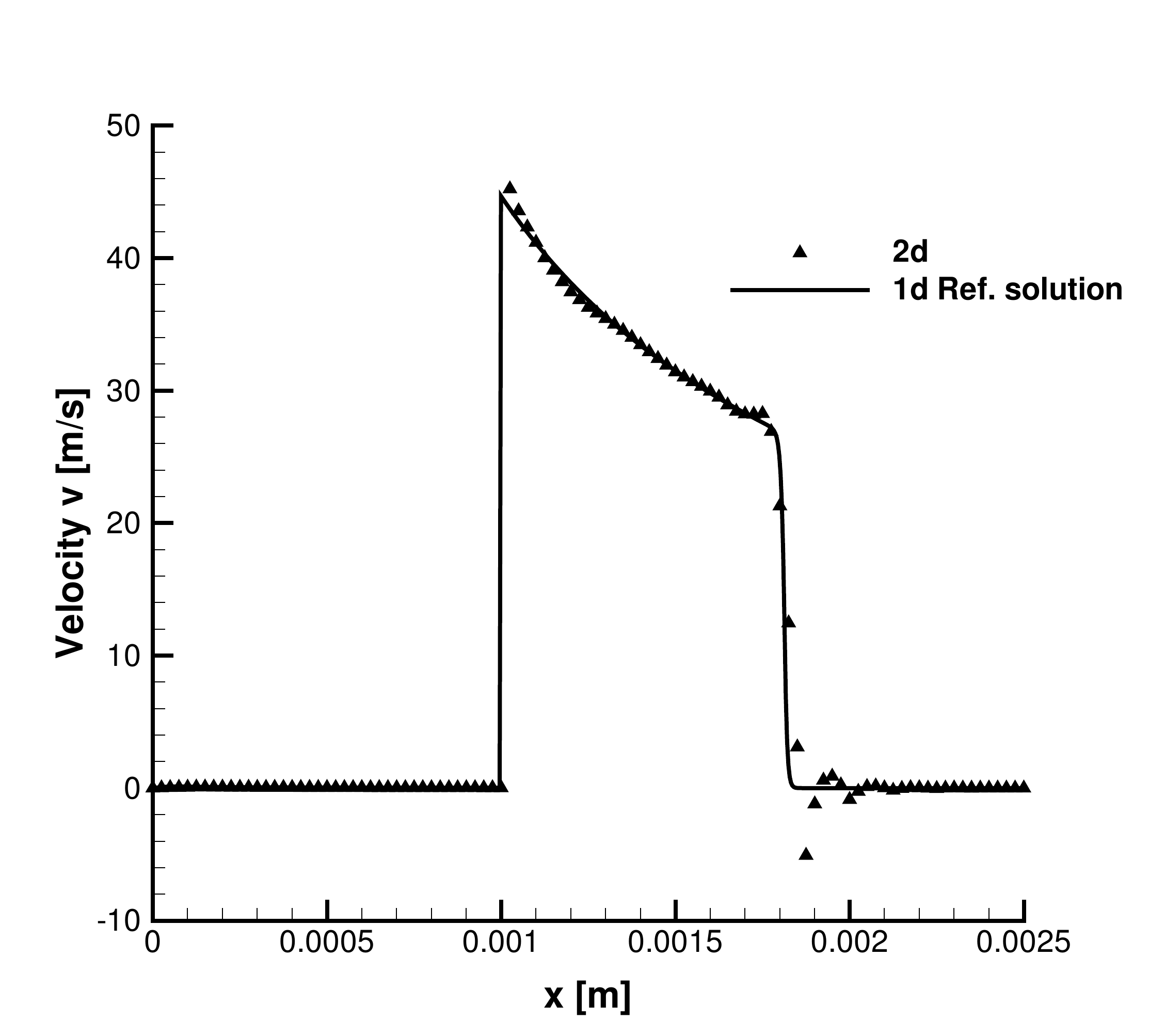}} \hfill
 \subfloat[Temperature]{\includegraphics[width=0.45\textwidth]{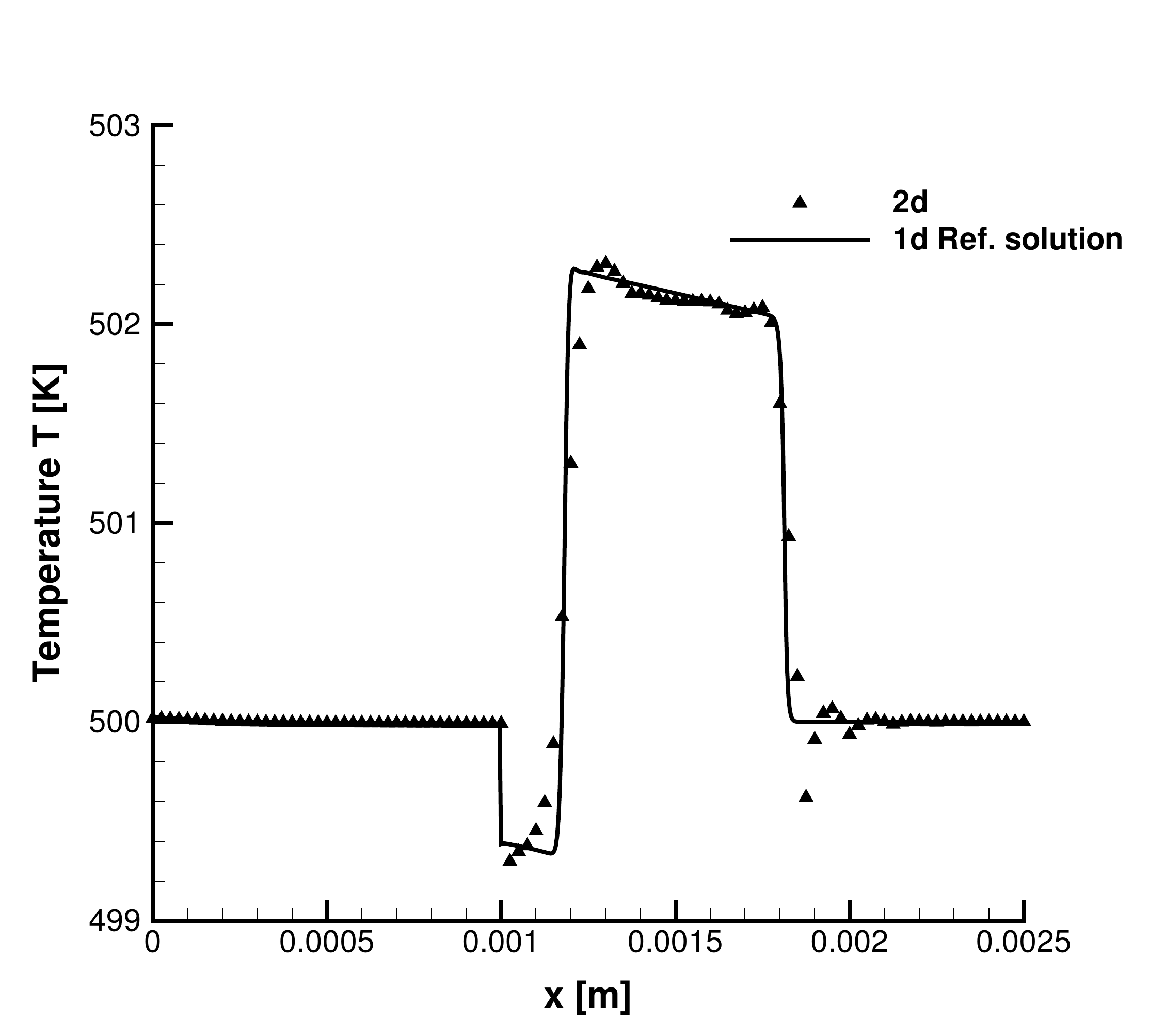}}
 \caption{Circular droplet test case at $t=5.0\cdot10^{-6}$. Shown is a cut along the $x$-axis: Comparison of the radially symmetric
 one-dimensional  approach (solid lines) as reference case to the two-dimensional  simulation (symbols). Note that for the 
 numerical simulation a significantly coarser mesh is used.}
 \label{fig:circdrop}
\end{figure}

\begin{figure}
 \centering
  \begin{tikzpicture}{x=1.0cm,y=1.0cm}
  \node at (0,0) {\includegraphics[width=8cm]{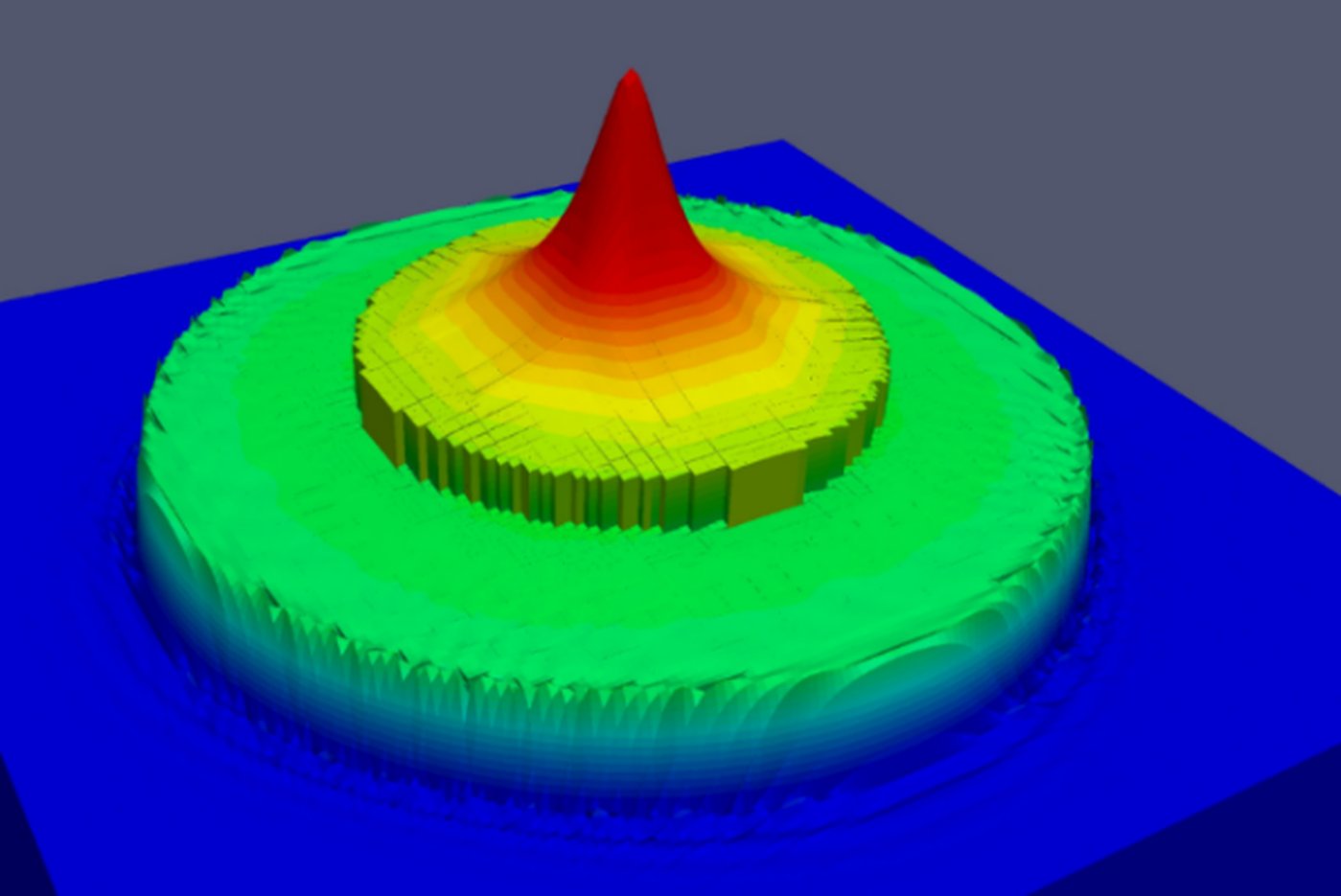}};
  % pressure scale
  \node at (6.0,0) {\includegraphics[height=4cm]{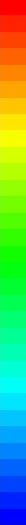}};
  \foreach \y in {0,1,2,3,4,5}{
    \draw[-] (6.15,\y*4./5.-2.) -- (6.3,\y*4./5.-2.);
  }
  \node[right] at (6.3,-2.0) {$1$};
%   \node at (7.,-1.0) {20};
%   \node at (7.5,0.0) {$1.25\cdot10^5$};
%   \node at (7.,1.0) {60};
  \node[right] at (6.3,2.0) {$1.5$};
  \node at (7.,2.5)  {\parbox{3cm}{Pressure $p$ [bar]}};
 \end{tikzpicture}
 \caption{Three-dimensional view of the solution structure of the pressure for the circular droplet test case with evaporation.}
 \label{fig:circdrop3d}
\end{figure}

The numerical results of the one-dimensional radial-symmetric and two-dimensional approach are compared in Figure \ref{fig:circdrop}. 
% This is a first result of the circular geometry approximation.
The results at a one-dimensional slice along the $x_1$-axis is presented in Figure~\ref{fig:circdrop}. The two-dimensional results are in good agreement with the one-dimensional reference simulation assuming spherical symmetry. Some over- and undershoots in the two-dimensional numerical solution are due to the approximation on the coarse grid.  But the multi-dimensional implementation reproduces well the results of the one-dimensional approach assuming spherical symmetry.
A three-dimensional view of the test case is presented in Figure~\ref{fig:circdrop3d} visualizing the 
wave structure of the test case and the sharp resolution of the phase interface. 
%All four distinct waves can be seen in the solution structure, as introduced in Figure~\ref{fig:fan}. % Contact wave is not visible
% In this case the contact discontinuity 
% is hardly visible due to the 
% low temperature difference  \corr{Im Druckfeld sollte doch nichts zu sehen sein}. 

\subsection{Oscillating droplet}
\label{sec:wobblingdrop}
% validation of the inclusion of surface tension forces in the interface model
The inclusion of surface tension forces in the Riemann solver at the phase interface 
is validated for an oscillating droplet test case. In this investigation, the
case with resolved phase transition effects is compared to the case without.
In contrast to the mathematical model and the other numerical examples, we account for viscosity and heat conduction in the bulk phases.
With phase transition, the Riemann solver described in Section \ref{sec:riemannsolver} is used. Without phase transition, a simpler linearized Lax-Curve Riemann solver as described in \cite{FJS2013} is 
applied. The influence of evaporation onto the droplet oscillation frequency has been investigated, 
e.\,g., by Schlottke\&Weigand \cite{schlottke2008direct} in an  incompressible simulation. 
% In their study the evaporation effects only had a minor effect.

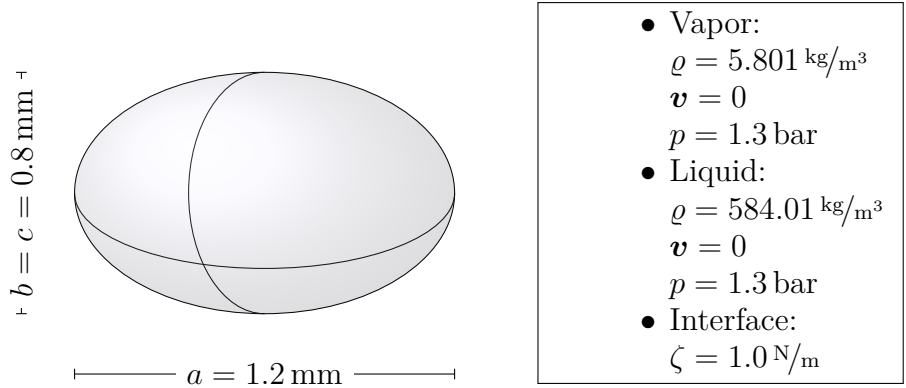
\begin{figure}
 \centering
 \begin{tikzpicture}[x=1cm,y=1cm,scale=2.0]
    \draw (0,0) ellipse (1.25 and 0.8);
    \draw (-1.25,0) arc (180:360:1.25cm and 0.5cm);
%     \draw[dashed] (-1,0) arc (180:0:1cm and 0.5cm);
    \draw (0,0.8) arc (90:270:0.5cm and 0.8cm);
%     \draw[dashed] (0,1) arc (90:-90:0.5cm and 1cm);
    \shade[ball color=blue!10!white,opacity=0.20] (0,0) ellipse (1.25 and 0.8);
    \draw (-1.25,-1.2) -- (1.25,-1.2) node [midway,fill=white] {$a=1.2$\,mm};
    \draw (-1.25,-1.2cm+1pt) -- (-1.25,-1.2cm-1pt);
    \draw (+1.25,-1.2cm+1pt) -- (+1.25,-1.2cm-1pt);
    \draw (-1.6,-0.8) -- (-1.6,0.8) node [midway,fill=white,rotate=90] {$b=c=0.8$\,mm};
    \draw (-1.6cm+1pt,-0.8) -- (-1.6cm-1pt,-0.8cm);
    \draw (-1.6cm+1pt,+0.8) -- (-1.6cm-1pt,+0.8cm);
    \node at (3.0,0) {\framebox{
      \begin{varwidth}{\linewidth}\begin{itemize}
        \item Vapor: \\ $\rho=5.801\,\nicefrac{\text{kg}}{\text{m}^3}$ \\ $\vv v=0$ \\ $p=1.3$\,bar 
        \item Liquid: \\ $\rho=584.01\,\nicefrac{\text{kg}}{\text{m}^3}$ \\ $\vv v =0$ \\ $p=1.3$\,bar 
        \item Interface: \\ $\zeta=1.0\,\nicefrac{\text{N}}{\text{m}}$
      \end{itemize}\end{varwidth}
    }};
 \end{tikzpicture}
 \caption{Initial conditions for the oscillating droplet test case using the fluid n-dodecane
 at an initially constant temperature $T=500$\,K.}
 \label{fig:wobbdrop_ini}
\end{figure}

In this validation problem we investigate an initially deformed droplet with the semi-axes $a=1.2r_0$ and $b=c=0.8r_0$ as 
plotted in Figure \ref{fig:wobbdrop_ini}. The initial radius is chosen as $r_0=1$\,mm and the radius of a corresponding 
spherical droplet is $a_\text{equi}=(a\cdot b \cdot c)^{1/3} =0.9157 r_0$. Due to surface tension 
forces acting at the phase interface, the droplet starts a periodic oscillation that diminishes with time due to
the influence of  viscosity to a sphere. Hence, in this test case, we also account for  viscosity and heat conduction using a standard numerical diffusion flux called BR1 \cite{BassiRebay}. We consider realistic viscosity estimates based on 
the local flow conditions in the bulk phases. For the initial conditions a viscosity ratio of $\mu_\text{liq}/\mu_\text{vap} = 23.73$
and a ratio of the heat transfer coefficient $k_\text{liq}/k_\text{vap} = 3.58$ is present. The values for viscosity and heat transfer
coefficient are estimated based on the correlation of 
Mulero et al. \cite{mulero2012recommended}. In the final stage, a spherical droplet is obtained whose interface pressure 
jump fulfills the Young-Laplace equation.

%Here, this stage is not reached. 
In the numerical simulation the droplet is located within a computational domain of $[-3.5,3.5]^3$\,mm and at the domain boundary wall boundary conditions
% \footnote{was sind outlet boundary conditions und wieso verhalten sich die wie wall boundary conditions?} 
are applied such that impinging waves are reflected at the boundaries. A numerical resolution of 96 DOF in each axis direction (about 900.000 DOF in total) is used to ensure the resolution of all effects within the computational domain.
The oscillation frequency of the droplet obtained is compared to the analytical investigations of Lamb \cite{lamb1932} for droplets without gravitation effects. He found the resonance mode frequency $f_l$ of the $l^\text{th}$ oscillation mode the following analytical relation
\begin{equation}
  f_l^\text{ana} = \sqrt{\frac{\sigma l (l-1)(l+1)}{3\pi \rho_0 V}},
\end{equation}
assuming a droplet with small oscillation amplitudes in vacuum or air neglecting the influence of gravity.
For the oscillation of the ellipsoid, the second mode is of importance with the analytical frequency
\begin{equation}
  f_2^\text{ana} \approx 582\,\text{Hz}\,.
\end{equation}
This analytical model is valid for small oscillations and was derived for an incompressible fluid.

% For a first comparison of the results, the initial temperature in the computational domain is chosen equal as well as an equal pressure. Thus, the influence of the phase transition effects resolved in the interface Riemann solver should be small compared to the surface tension effects. This covers the main aspect of this test case, the validation of the surface tension force in the numerical approach for compressible two-phase flows.
% To be able to validate the surface tension forces the initial conditions are chosen such that evaporation effects should be negligible. Thus, we apply initial states at the phase equilibrium for $T=500$\,K.
The focus of this test case is on the validation of the surface tension forces. Thus, we apply initial conditions at the thermodynamic equilibrium for $T=500$\,K for which evaporation effects should be negligible.

\begin{figure}
  \centering
  \begin{minipage}{0.3\textwidth}
   \includegraphics[width=\textwidth]{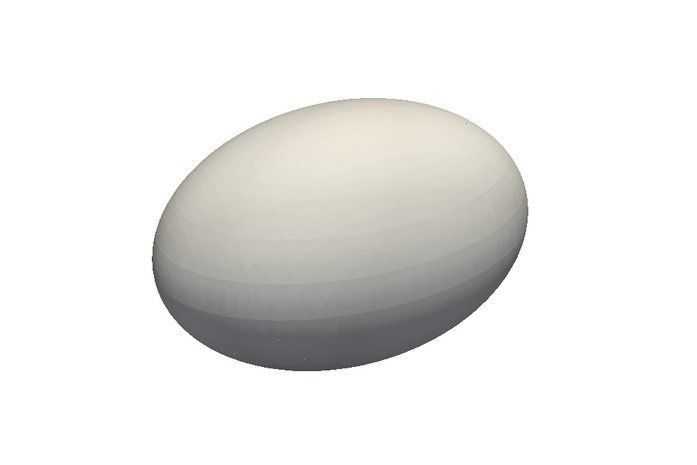}\\
   $t=0$\,ms \\
   \includegraphics[width=\textwidth]{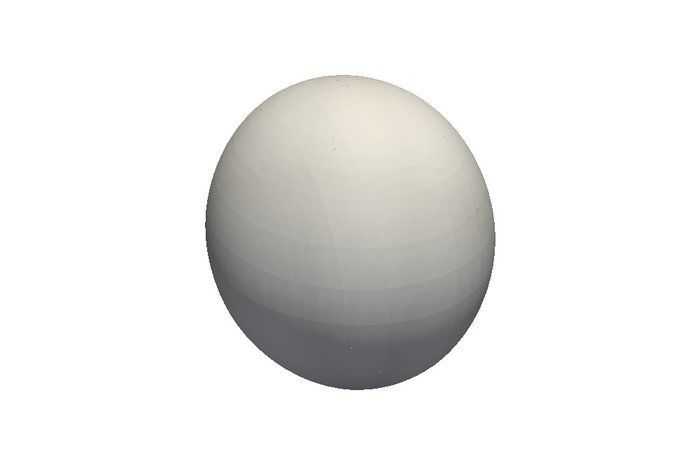}\\
   $t=1.2$\,ms \\
   \includegraphics[width=\textwidth]{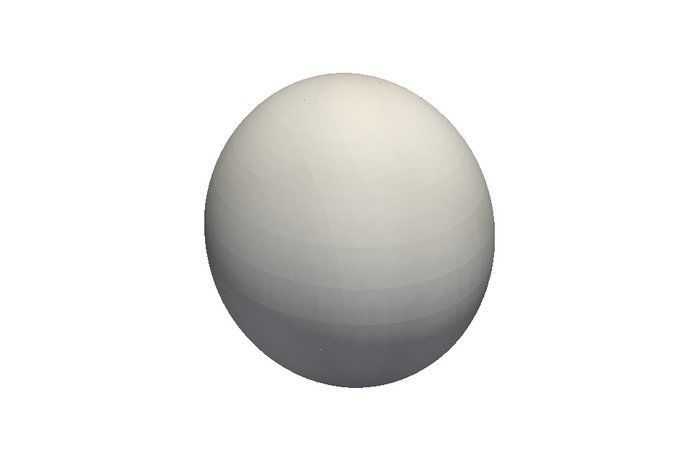} \\
   $t=2.4$\,ms \\
  \end{minipage}\hfil
  \begin{minipage}{0.3\textwidth}
   \includegraphics[width=\textwidth]{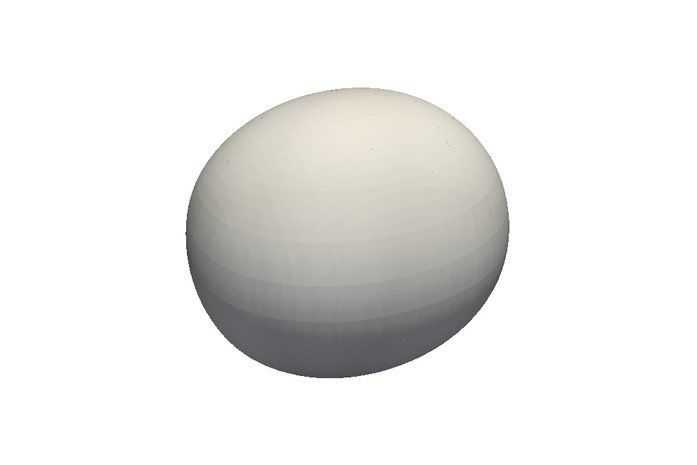}\\
   $t=0.4$\,ms \\
   \includegraphics[width=\textwidth]{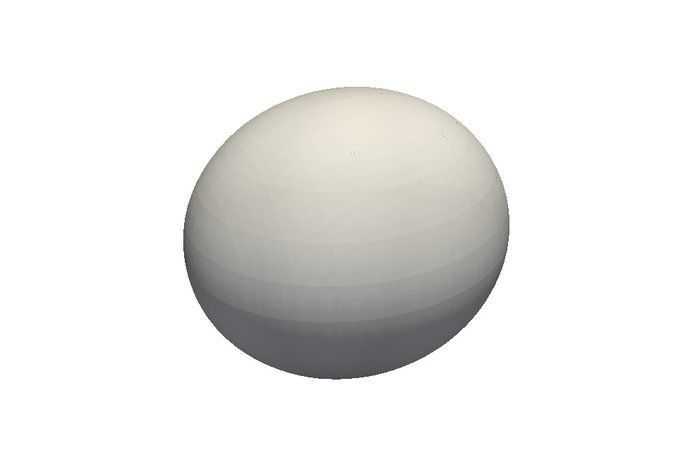}\\
   $t=1.6$\,ms \\
   \includegraphics[width=\textwidth]{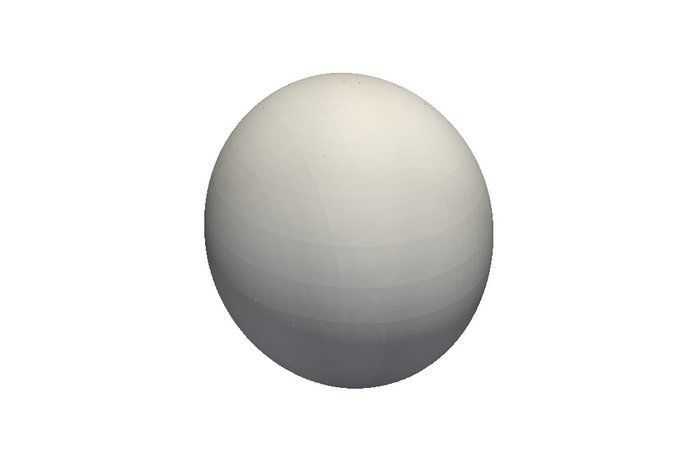}\\
   $t=2.8$\,ms \\
  \end{minipage}\hfil
  \begin{minipage}{0.3\textwidth}
   \includegraphics[width=\textwidth]{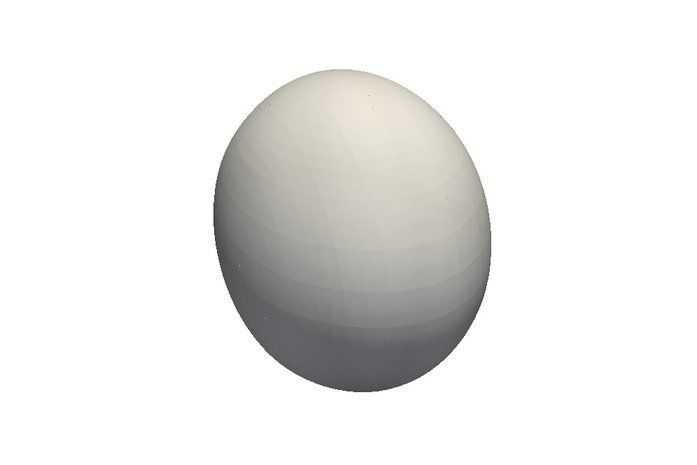}\\
   $t=0.8$\,ms \\
   \includegraphics[width=\textwidth]{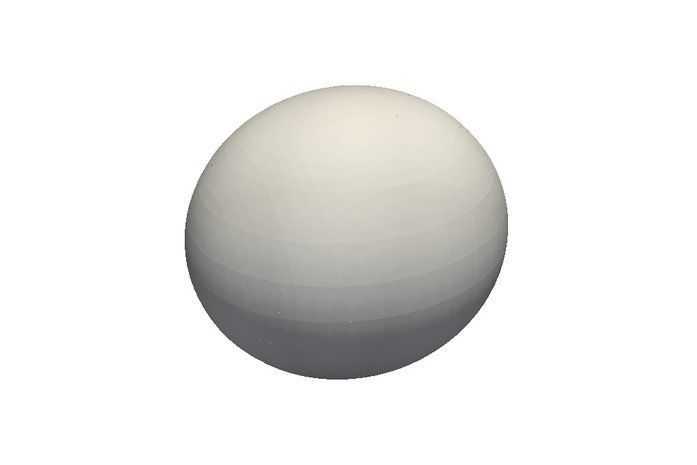}\\
   $t=2.0$\,ms \\
   \includegraphics[width=\textwidth]{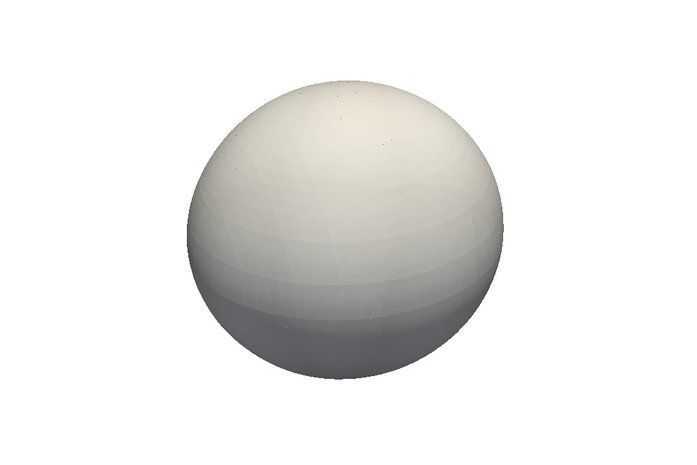}\\
   $t=3.2$\,ms \\
  \end{minipage}
 \caption{3D surface contour for the wobbling n-dodecane droplet. The numerically obtained oscillation frequency is $T_2^\text{num}\approx 1.75$\,ms is in good agreement with the analytical one of $T_2^\text{ana}=1.717$\,ms.}
 \label{fig:wobbdrop}
\end{figure}

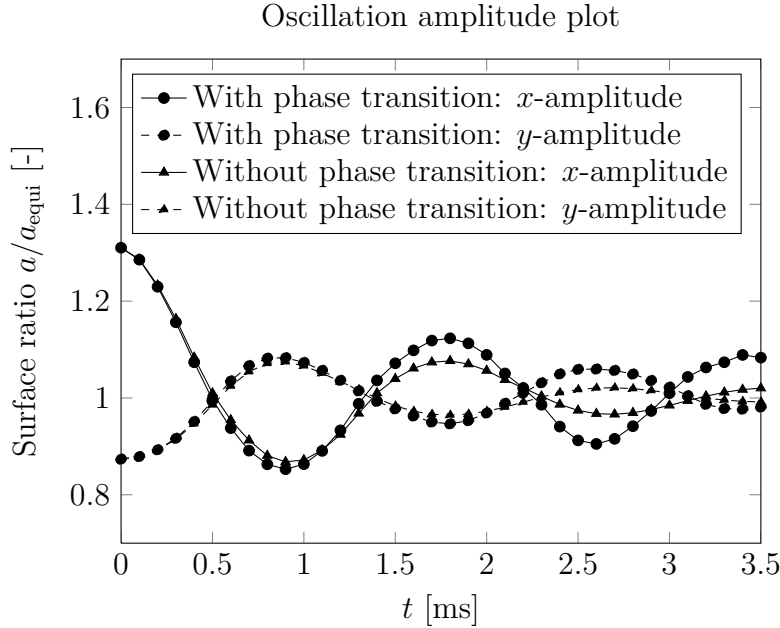
\begin{figure}
  \centering
  \pgfplotstableread{Data/oscillation_hllc.dat}{\dataoszillationHLLC}
  \pgfplotstableread{Data/oscillation_exact.dat}{\dataoszillationExact}
  \begin{tikzpicture}[scale=1,]
   \begin{axis}[xlabel={$t$ [ms]},xmin=0,xmax=3.5,
               ylabel={Surface ratio $a/a_\text{equi}$ [-]},ymin=0.7,ymax=1.7,
               height=8cm,width=10cm,
               legend pos= north east,legend cell align=left,
               title=Oscillation amplitude plot]
   \addplot [black,mark=*,mark size=2pt] table [x expr={\thisrow{time}*1000}, y expr={\thisrow{maxx}/0.0009157713940426659}] {\dataoszillationExact};
   \addlegendentry{With phase transition: $x$-amplitude};
   \addplot [black,dashed,mark=*,mark size=2pt] table [x expr={\thisrow{time}*1000}, y expr={\thisrow{maxy}/0.0009157713940426659}] {\dataoszillationExact};
   \addlegendentry{With phase transition: $y$-amplitude};
   \addplot [black,mark=triangle*,mark size=2pt] table [x expr={\thisrow{time}*1000}, y expr={\thisrow{maxx}/0.0009157713940426659}] {\dataoszillationHLLC};
   \addlegendentry{Without phase transition: $x$-amplitude};
   \addplot [black,dashed,mark=triangle*,mark size=2pt] table [x expr={\thisrow{time}*1000}, y expr={\thisrow{maxy}/0.0009157713940426659}] {\dataoszillationHLLC};
   \addlegendentry{Without phase transition: $y$-amplitude};
    \end{axis}
   \end{tikzpicture}
  \caption{Plot of the oscillation amplitude over time $t$ for an initially ellipsoidal droplet including surface tension effects. Compared is the simulation with and without phase transition.}
  %Plotted is the comparison between the approach without phase transition (LinLax-solver \cite{FJS2013}) and the with phase transition (Exact RS). Note that the $y$- and $z$-amplitudes are identical.}
  \label{fig:wobbdrop_amplitude}
\end{figure}

In Figure~\ref{fig:wobbdrop} the surface ratio during two oscillation periods is visualized and the resulting oscillation amplitude is plotted over time in Figure~\ref{fig:wobbdrop_amplitude}. Due to the influence of viscosity and heat transfer in the bulk phases, the oscillation amplitude diminishes with time. In the final state a spherical droplet with the equilibrium radius $a_\text{equi}=(a\cdot b \cdot c)^{1/3} = 0.9157 r_0$ is reached. Included is the comparison without phase transition that is calculated using the methodology described in \cite{fechter2015multiphasedg} and the linear Riemann solver in \cite{FJS2013}. The oscillation frequency coincides with the analytical frequency of Lamb in both cases. However, due to evaporation and condensation, the droplet mass is changing and as a result the oscillation frequency changes. This has a small effect on the frequency as seen in Figure~\ref{fig:wobbdrop_amplitude}. Another effect is that due to phase transition effects the oscillation amplitude 
does not decay to that extend. This might be an effect of the ejected mass at the interface that acts like an additional interface force that increases the oscillation amplitude.
% Another effect is the development of small jet like structures on the droplet surface due to phase transition effects that change the oscillation amplitude of the droplet. Because of that the amplitude of the oscillation decreases not to that extent due to the viscosity influence in the bulk phases.

\subsection{Shock-Droplet interaction}
% TODO: check why the exact RS does not really converge in this case (probably problems with too large phase transition rates or too extreme conditions)
\label{sec:shockdrop}
This test case is used to validate of correct wave propagation at 
the interface in a compressible flow field. A shock wave of Mach number $1.2$ is impinging onto a
n-dodecane droplet at rest. The impinging shock wave is 
partially reflected on the droplet surface and is partially transmitted into the droplet. 
Inside the droplet the shock wave travels at a higher speed due to the higher sound velocity in the liquid phase. 
Due to the post-shock momentum of the flow, the droplet gets deformed. 

%Note that it is not possible to simulate shocks using an incompressible flow solver.

The initial conditions and the computational domain used for the simulation are defined as in
Figure~\ref{fig:shockdrop_ini}. Initially the shock is placed at $x_s=-1.5$\,mm within the
computational domain of $[-2.5,-5]\times[7.5,5]$\,mm. The shock-droplet interaction is calculated as
two-dimensional test case with 240 DOF in each direction. The initial conditions are chosen so that the 
droplet evaporates at the pre-shock conditions.

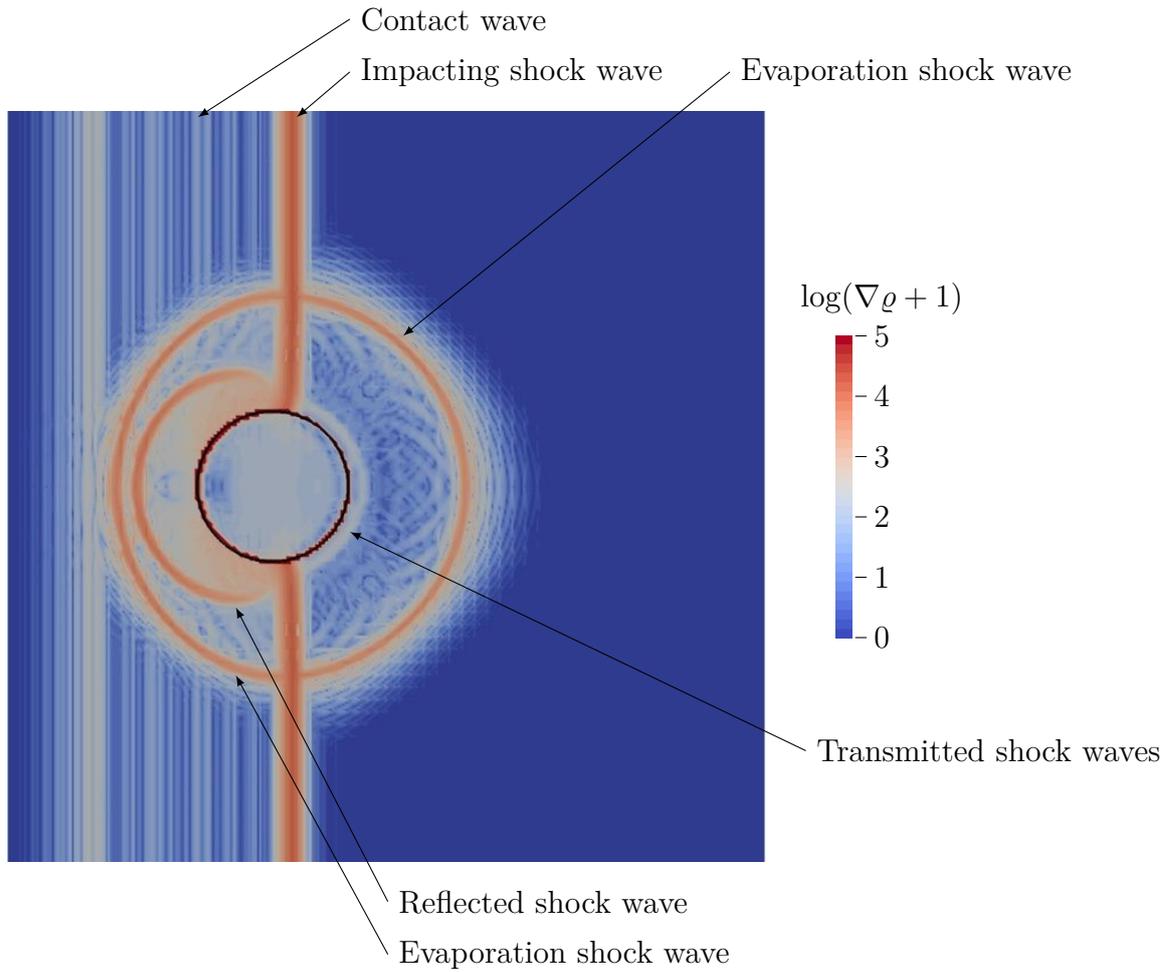
\begin{figure}
 \centering
 \begin{tikzpicture}[x=1cm,y=1cm,scale=1,>=latex]%options/.expand once=latex-latexnew, arrowhead=1cm, line width=1pt]
  \node at (0,0) {\includegraphics[width=10cm]{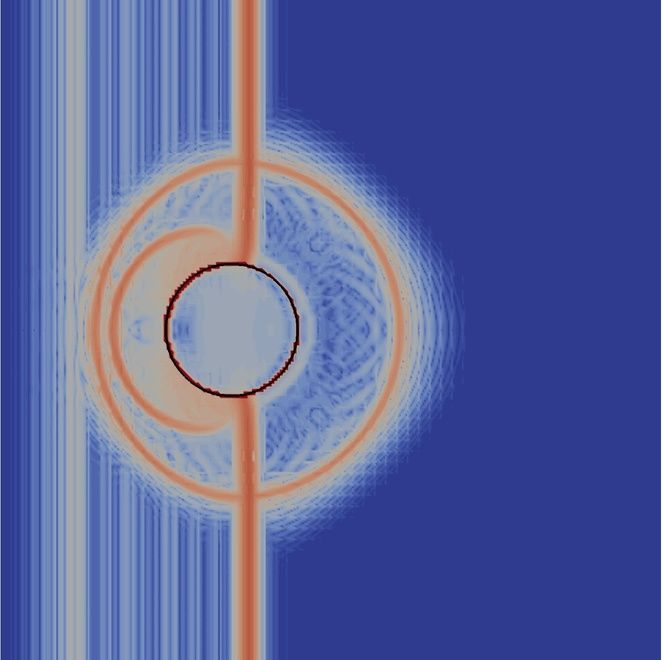}};
  \node at (6.0,0) {\includegraphics[height=4cm]{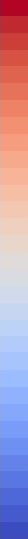}};
  \foreach \y in {0,1,2,3,4,5}{
    \draw[-] (6.15,\y*0.8-2.) -- (6.3,\y*0.8-2.);
  }
  % Legend
  \node at (6.5,-2.) {0};
  \node at (6.5,-1.2) {1};
  \node at (6.5,-0.4) {2};
  \node at (6.5, 0.4) {3};
  \node at (6.5, 1.2) {4};
  \node at (6.5,2.) {5};
  \node at (6.5,2.5)  {$\log(\nabla\rho+1)$};
  % shock description
  \draw[<-] (-1.2,4.9)--(-.5,5.5) node[right] {Impacting shock wave};
  \draw[<-] (-2.5,4.9)--(-.5,6.2) node[right] {Contact wave};
  \draw[<-] (0.2,2.0)--(4.5,5.5) node[right] {Evaporation shock wave};
  \draw[<-] (-0.5,-0.6)--(5.5,-3.5) node[right] {Transmitted shock waves};
  \draw[<-] (-2.0,-1.6)--(0,-5.5) node[right] {Reflected shock wave};
  \draw[<-] (-2.0,-2.5)--(0,-6.2) node[right] {Evaporation shock wave};
 \end{tikzpicture}
 \caption{Illustration of the wave structure for the shock-droplet interaction test case with evaporation at $t=0.1$\,ms.
 The black line shows the position of the phase interface.}
 \label{fig:shockdrop_illustration}
\end{figure}

\begin{figure}
 \centering
   \begin{minipage}{.35\textwidth}
    $t=10\,\mu$s: \\
    \includegraphics[width=\textwidth]{Figures/ShockDroplet/2d_exact_schlieren_0001.jpg} \\
    $t=20\,\mu$s: \\
    \includegraphics[width=\textwidth]{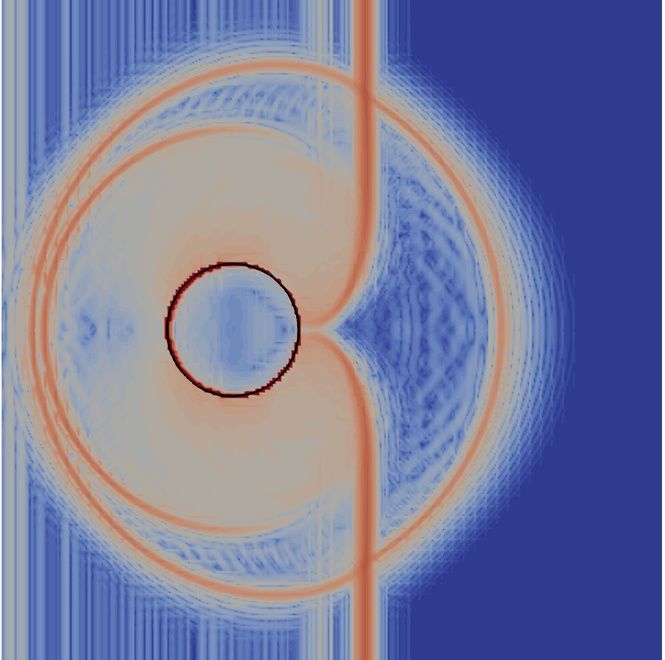} \\
    $t=30\,\mu$s: \\
    \includegraphics[width=\textwidth]{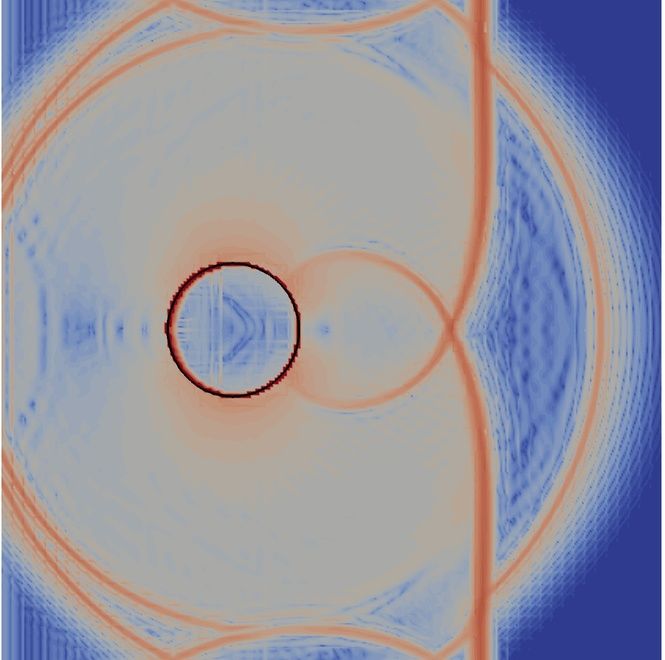} \\[20pt]
    \begin{tikzpicture}[x=1cm,y=1cm,scale=1]
     \node at (0.0,0) {\rotatebox{-90}{\includegraphics[height=4cm]{Figures/ShockDroplet/pressure_3d_scale.jpg}}};
     \foreach \x in {0,1,2,3,4,5}{
       \draw[-] (\x*0.8-2,-0.05) -- (\x*0.8-2,-0.20);
     }
     % Legend
     \node at (-2.,-0.5) {0};
     \node at (-1.2,-0.5) {1};
     \node at (-0.4,-0.5) {2};
     \node at (0.4, -0.5) {3};
     \node at (1.2, -0.5) {4};
     \node at (2.,-0.5) {5};
     \node at (0,-1)  {$\log(\nabla\rho+1)$};    
   \end{tikzpicture}
  \end{minipage}\hfil
  \begin{minipage}{.35\textwidth}
     $t=10\,\mu$s: \\
    \includegraphics[width=\textwidth]{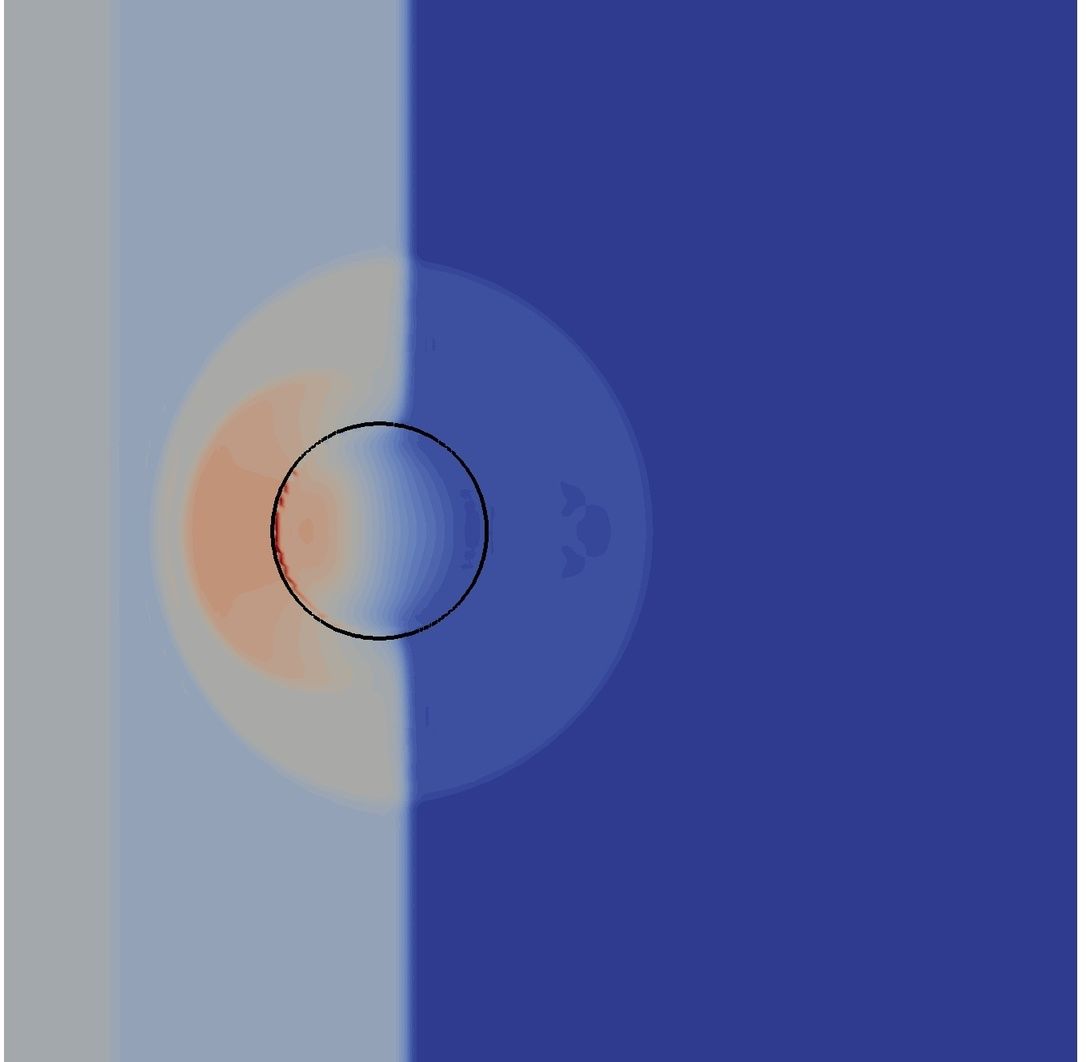} \\
    $t=20\,\mu$s: \\
    \includegraphics[width=\textwidth]{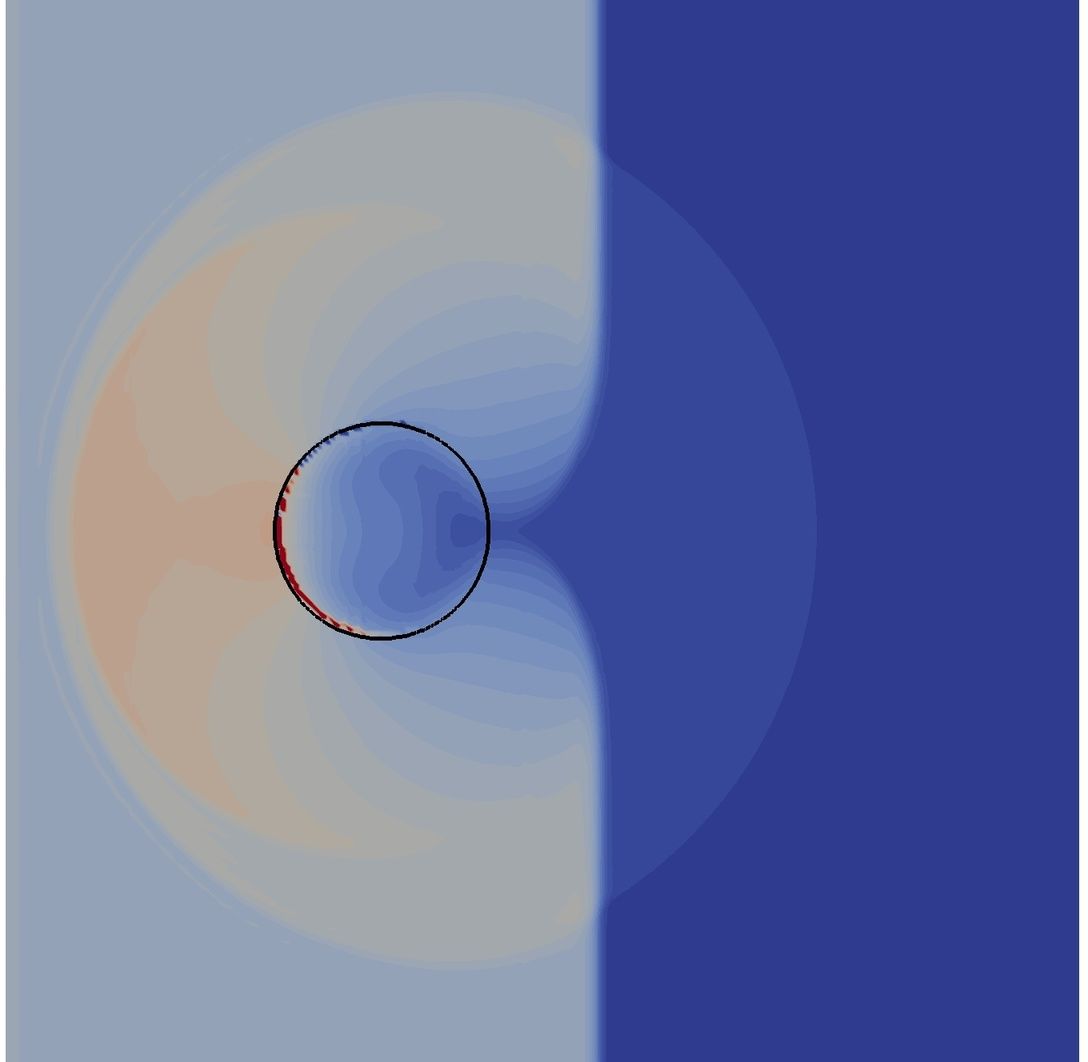} \\
    $t=30\,\mu$s: \\
    \includegraphics[width=\textwidth]{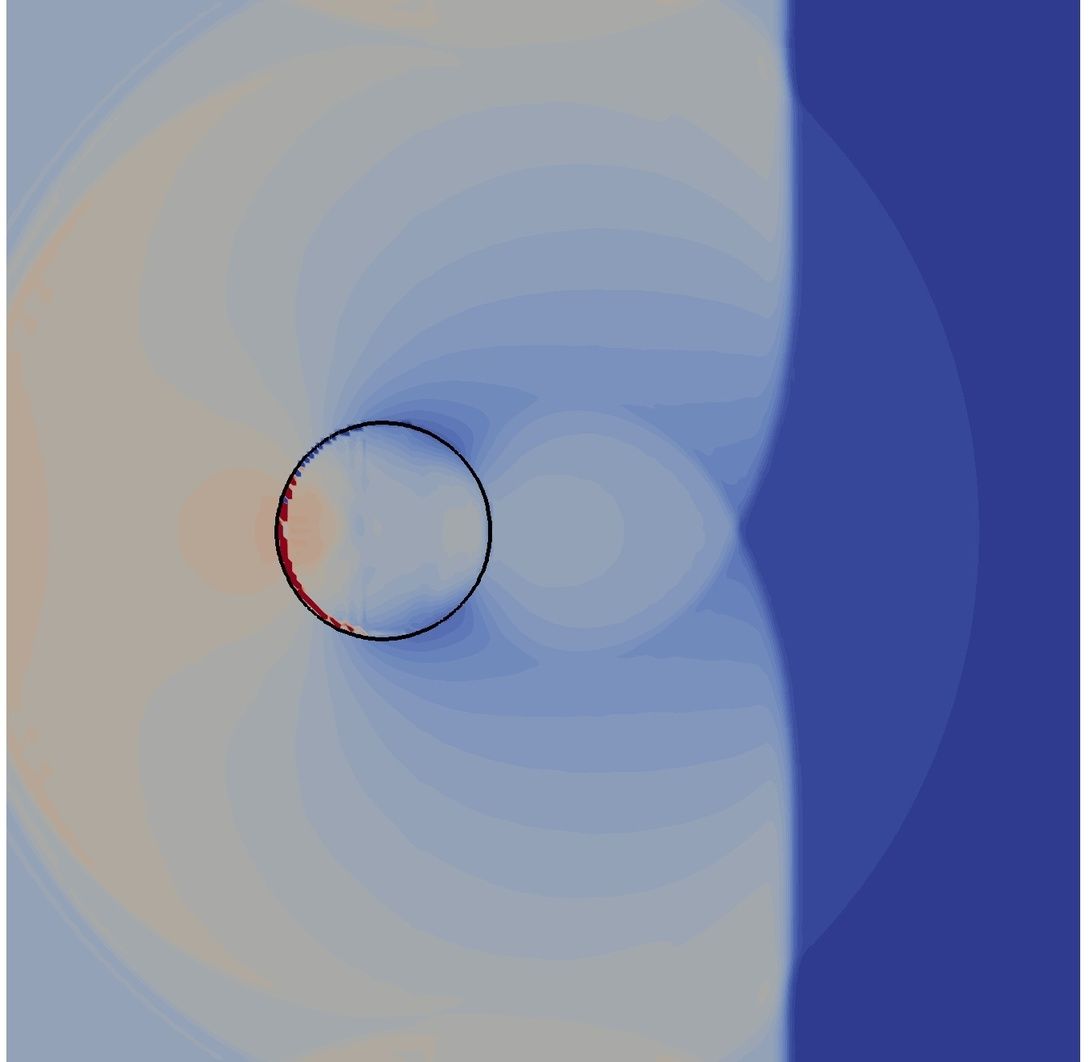} \\[20pt]
    \begin{tikzpicture}[x=1cm,y=1cm,scale=1]
     \node at (0.0,0) {\rotatebox{-90}{\includegraphics[height=4cm]{Figures/ShockDroplet/pressure_3d_scale.jpg}}};
     \foreach \x in {0,1,2,3,4}{
       \draw[-] (\x-2,-0.05) -- (\x-2,-0.20);
     }
     % Legend
     \node at (-2.,-0.5) {1};
     \node at (-1,-0.5) {1.25};
     \node at (-0,-0.5) {1.5};
     \node at (1, -0.5) {1.75};
     \node at (2, -0.5) {2};
     \node at (0,-1)  {$p$ [bar]};   
   \end{tikzpicture}
  \end{minipage}\hfil
 \caption{The shock-droplet interaction for different times. The black solid line indicates the interface position as determined by the level-set method. Left: Logarithmic density gradient visualization. Right: Pressure visualization.}
 \label{fig:shockdrop}
\end{figure}

\begin{figure}
 \centering
   \begin{minipage}{.35\textwidth}
    $t=40\,\mu$s: \\
    \includegraphics[width=\textwidth]{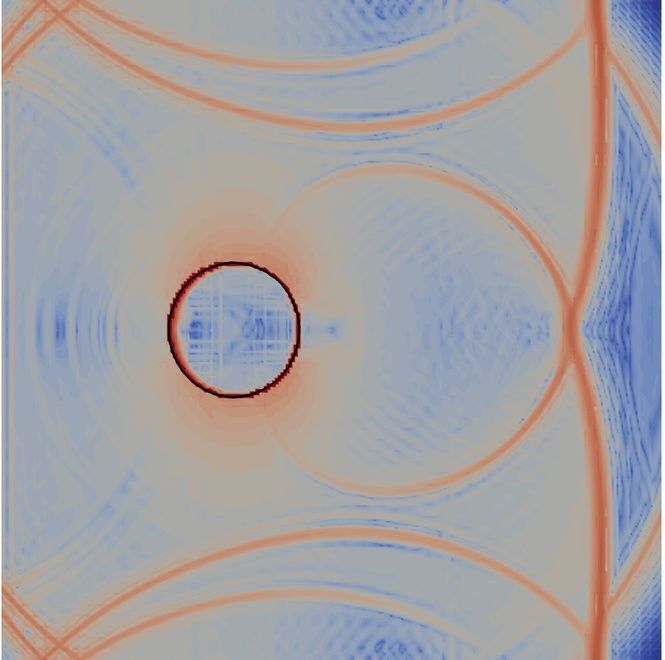} \\
    $t=50\,\mu$s: \\
    \includegraphics[width=\textwidth]{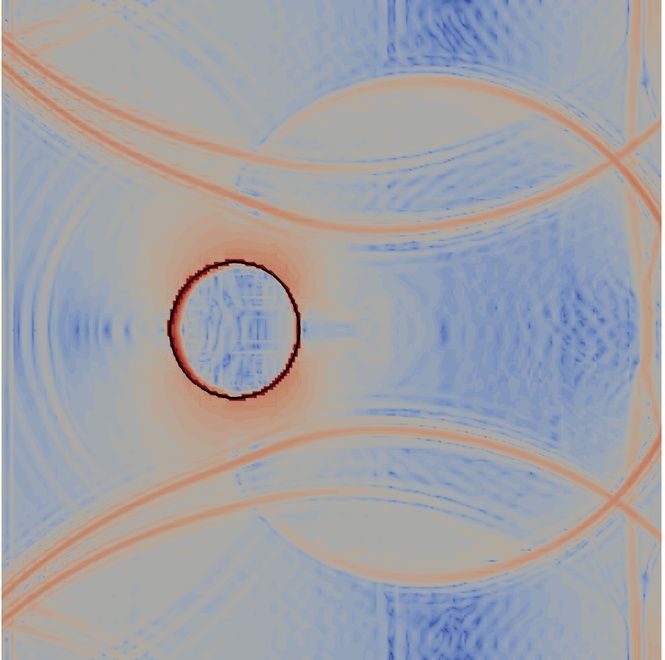} \\
    $t=60\,\mu$s: \\
    \includegraphics[width=\textwidth]{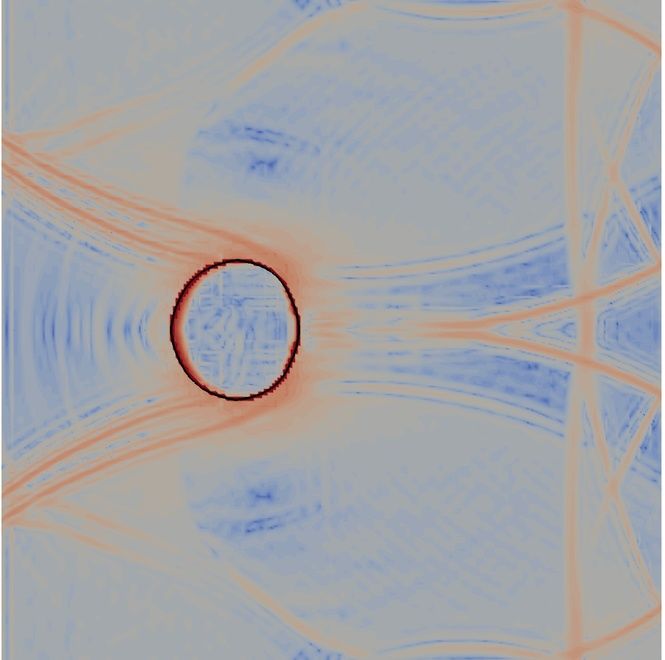} \\[20pt]
    \begin{tikzpicture}[x=1cm,y=1cm,scale=1]
     \node at (0.0,0) {\rotatebox{-90}{\includegraphics[height=4cm]{Figures/ShockDroplet/pressure_3d_scale.jpg}}};
     \foreach \x in {0,1,2,3,4,5}{
       \draw[-] (\x*0.8-2,-0.05) -- (\x*0.8-2,-0.20);
     }
     % Legend
     \node at (-2.,-0.5) {0};
     \node at (-1.2,-0.5) {1};
     \node at (-0.4,-0.5) {2};
     \node at (0.4, -0.5) {3};
     \node at (1.2, -0.5) {4};
     \node at (2.,-0.5) {5};
     \node at (0,-1)  {$\log(\nabla\rho+1)$};    
   \end{tikzpicture}
  \end{minipage}\hfil
  \begin{minipage}{.35\textwidth}
     $t=40\,\mu$s: \\
    \includegraphics[width=\textwidth]{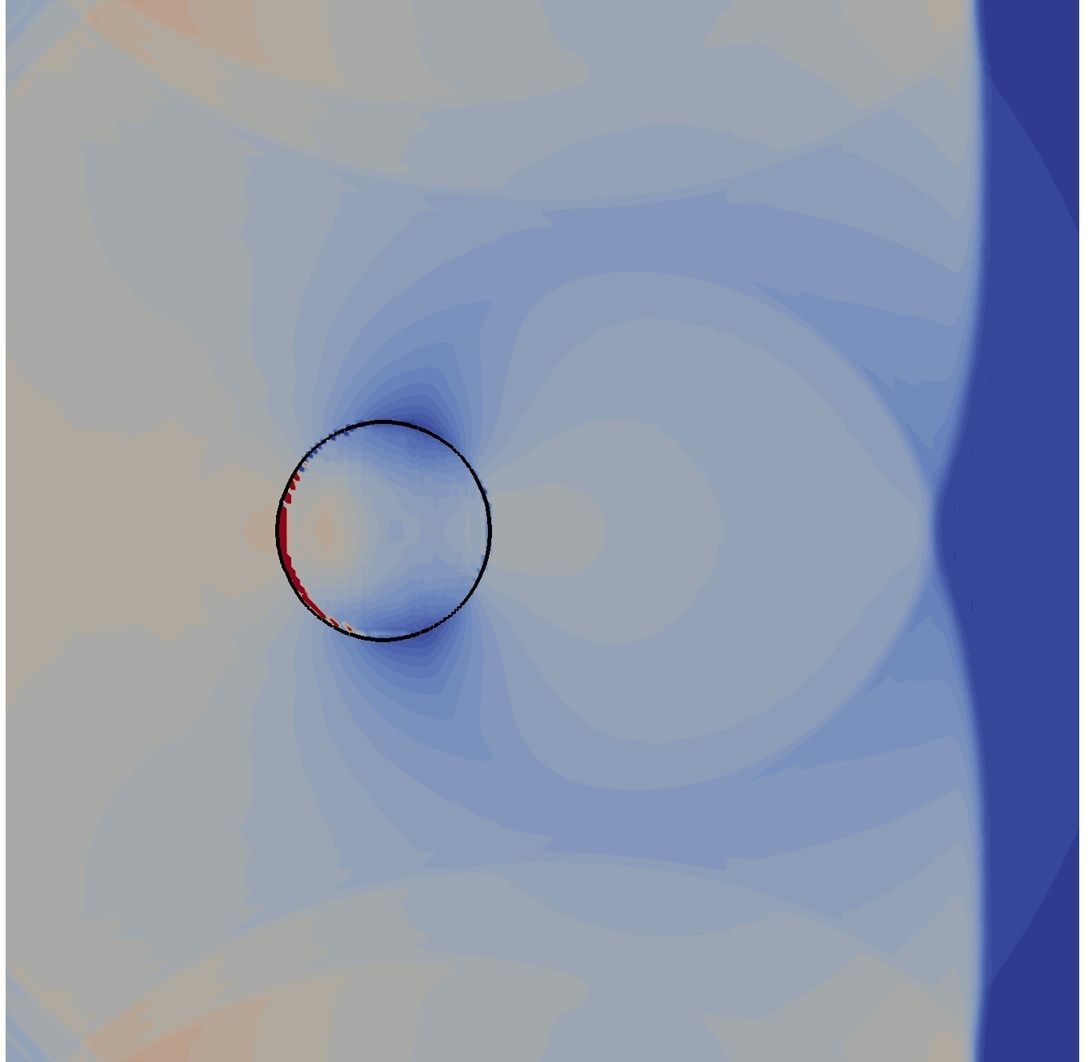} \\
    $t=50\,\mu$s: \\
    \includegraphics[width=\textwidth]{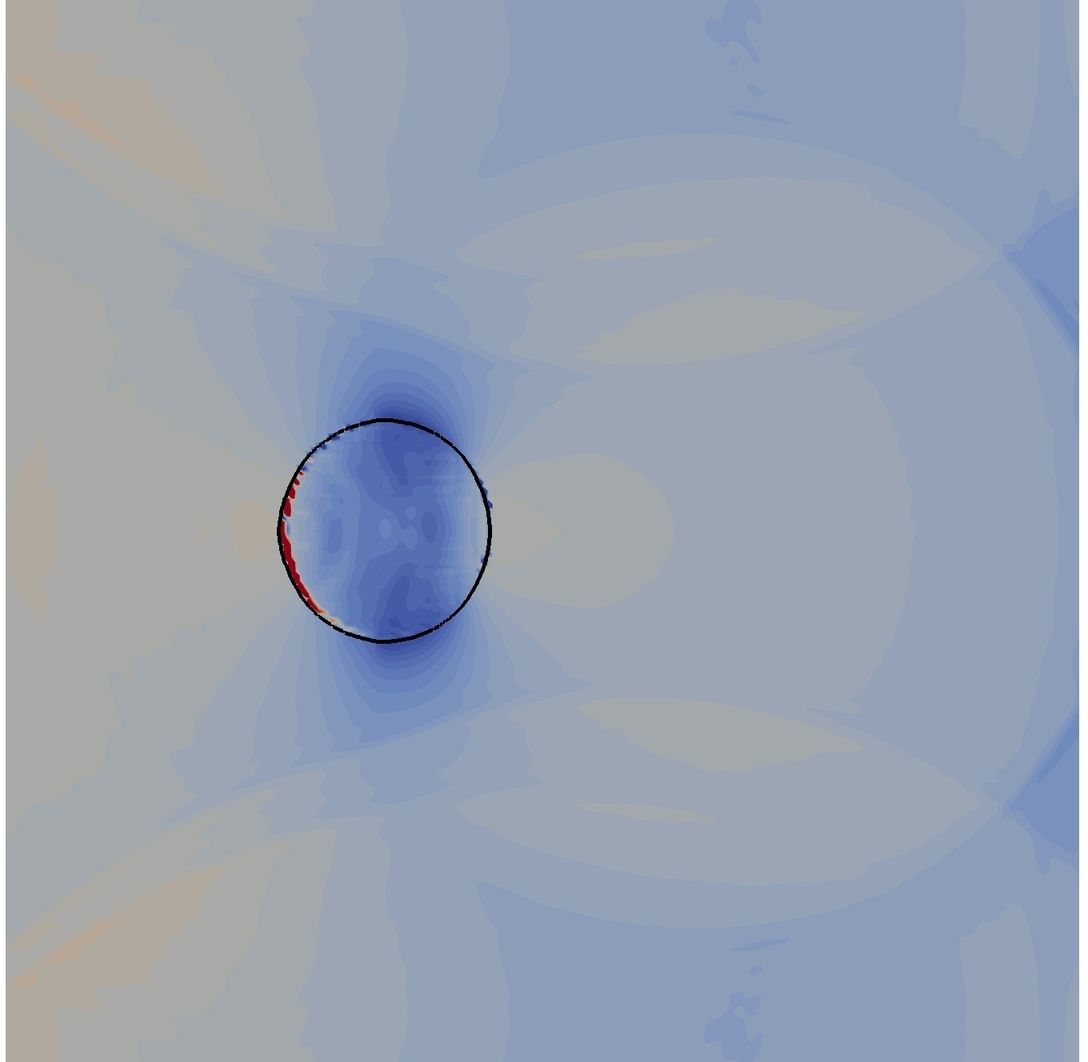} \\
    $t=60\,\mu$s: \\
    \includegraphics[width=\textwidth]{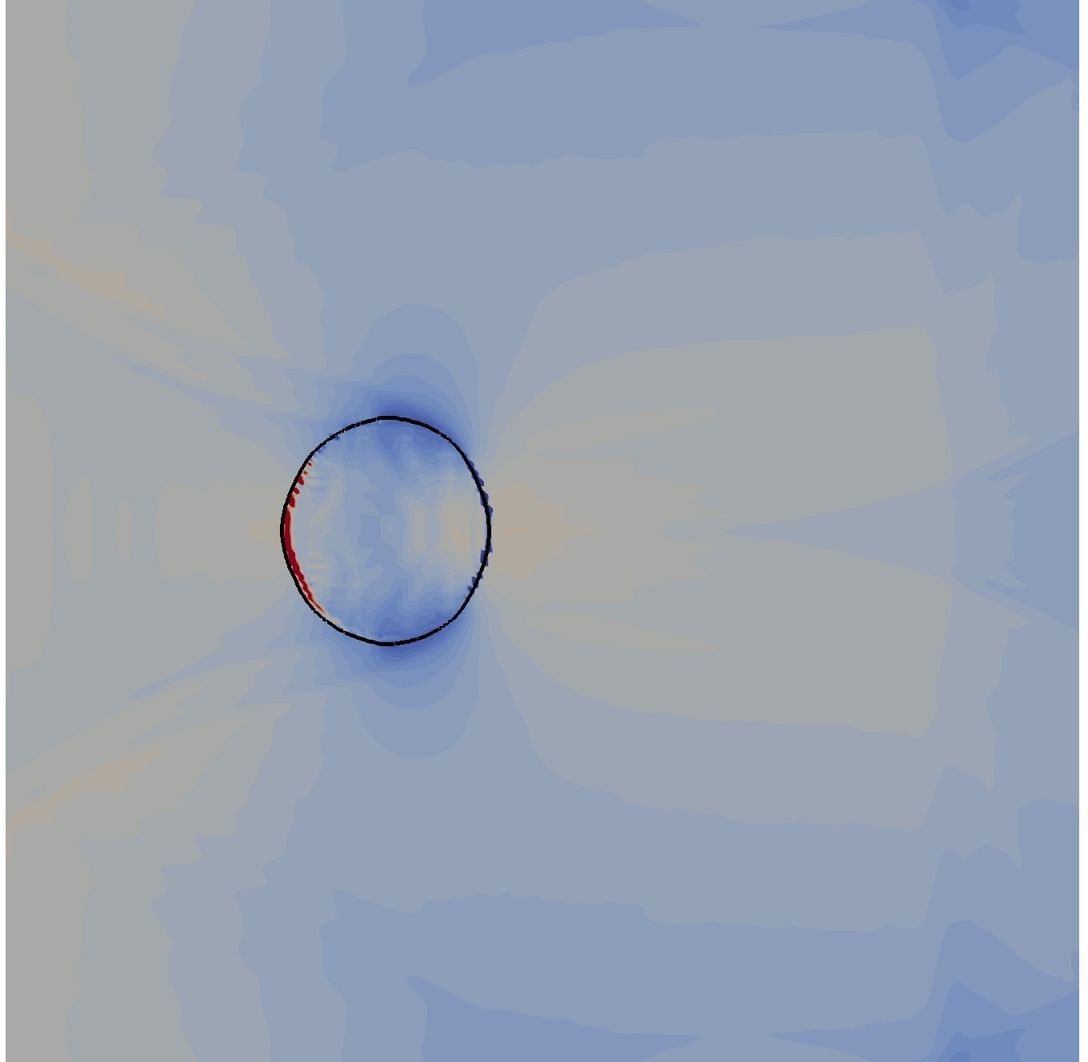} \\[20pt]
    \begin{tikzpicture}[x=1cm,y=1cm,scale=1]
     \node at (0.0,0) {\rotatebox{-90}{\includegraphics[height=4cm]{Figures/ShockDroplet/pressure_3d_scale.jpg}}};
     \foreach \x in {0,1,2,3,4}{
       \draw[-] (\x-2,-0.05) -- (\x-2,-0.20);
     }
     % Legend
     \node at (-2.,-0.5) {1};
     \node at (-1,-0.5) {1.25};
     \node at (-0,-0.5) {1.5};
     \node at (1, -0.5) {1.75};
     \node at (2, -0.5) {2};
     \node at (0,-1)  {$p$ [bar]};    
   \end{tikzpicture}
  \end{minipage}\hfil
 \caption{The shock-droplet interaction for different times. The black solid line indicates the interface position as determined by the level-set method. Left: Logarithmic density gradient visualization. Right: Pressure visualization.}
 \label{fig:shockdrop2}
\end{figure}

 Figure~\ref{fig:shockdrop_illustration} provides an explanation of the shock structures that are visible  in the shock-droplet interaction.
The results at different time instances plotted in Figures~\ref{fig:shockdrop} and \ref{fig:shockdrop2} reproduce the characteristics of shock-droplet interactions without resolved phase transition effects. Due to the resolved phase transition effects an additional shock wave can be seen in the solution. This wave is due to the initial evaporation of the droplet in the pre-shock state. The numerical approach detects the evaporation and condensation regimes accordingly on the droplet surface. Due to the impinging shock wave, the surrounding gas conditions are changed so that condensation instead of evaporation takes place. 
Note that evaporation and condensation may occur at the same time
at different locations on the droplet surface.

The shock Mach number was chosen to $1.2$.
%due to the additional constraint imposed by the kinetic relation that describes the phase transition rate. 
Stronger shock waves have been investigated without resolved phase transition effects, see e.\,g. \cite{FJS2013,fechter2015multiphasedg,hu2009hllc}. 
In this case, locally negative pressures occur that are not permitted by the two-phase Riemann solver.
%a result of a non-thermodynamic equilibrium state. Theses states are not permitted by the chosen kinetic relation. In the case of phase transitions this causes strong problems in the evaluation of the kinetic relation and in the estimation of the mass transfer rate. 
%If the shock impinges onto an evaporating droplet pressure and temperature increase and the surrounding vapor starts to condensate on the surface. 
Note that, this is a more advanced test case compared to incompressible investigations of evaporating droplets. 
% in a crossflow environment, i.\,e., evaporation within a moving vapor flow as compressible effects have to be considered, as e.\,g.  in \cite{schlottke2008direct} \footnote{den Satz verstehe ich nicht}. 

\begin{figure}
 \centering
 \begin{tikzpicture}
   \draw (0,0) rectangle (6,4);
   \shade[ball color=blue!10!white,opacity=0.20] (2,2) circle (0.5);
   \draw (2,2) circle (0.5);
%   \foreach \y in {0.25,0.75,...,3.75}
%     \draw[->] (-0.5,\y)--(0,\y);
   \draw[thick] (1,4)--(1,0) node[below] {$x_s$};
   \node[above] at (0,4.0) {Inflow boundary};
   \draw (2,2)--(3,3) node[right] {Droplet};
   \draw (1.5,0)--(2.0,-0.5) node[below] {Symmetry boundary};
   \node[above] at (6,4.0) {Outflow boundary};
   \draw[->] (6,2)--(7,2) node[right] {$x_1$};
   \draw[->] (2,4)--(2,5) node[left] {$x_2$};
  \end{tikzpicture} \newline
  \begin{raggedright}
  Initial conditions:\\
  \indent Vapor: \[
    (\rho,{\vv{v}},p) = \begin{cases} 
			(6.272\,\nicefrac{\text{kg}}{\text{m}^3}, (47.85,0,0)^\transp\,\nicefrac{\text{m}}{\text{s}}, 1.45\,\text{bar})&\text{ if }x_1\le x_s \\
			(4.383\,\nicefrac{\text{kg}}{\text{m}^3}, (0,0,0)^\transp\,\nicefrac{\text{m}}{\text{s}}, 1.0\,\text{bar})&\text{ else.}
			\end{cases} \]
  \indent Liquid:\[
    (\rho,{\vv{v}},p) = (584.01\,\nicefrac{\text{kg}}{\text{m}^3}, (0,0,0)^\transp, 1.3\,\text{bar})
    \]
  \end{raggedright}
 \caption{Initial setting for the shock-droplet interaction test case
 for a n-dodecane droplet interacting with a Mach $1.2$ shock wave. The initial shock position is $x_s=-0.15$\,mm.}
 \label{fig:shockdrop_ini}
\end{figure}

\clearpage
\section{Conclusion and Outlook}
\label{sec:conclusion}
In the paper we described the construction and validation of a numerical method for the sharp-interface tracking of compressible two-phase flows with phase transition and surface tension. The sharp-interface resolution uses a ghost-fluid method in which the coupling of the fluids uses the exact solution of a two-phase Riemann problem that takes into account phase transition and surface tension effects. To obtain a unique solution in the case of a phase transition, this solution satisfies a kinetic relation, which provides information from the micro-scale and controls the entropy change across the phase interface. The proposed method allows a flexible investigation of interface geometries and is not restricted to spherical droplets.

Due to the computational effort, the here considered sharp interface approach can handle a small number of interfaces only, but these with a consistent thermodynamic modeling. The proper physical transition or jump conditions can be imposed within the two-phase Riemann problem. To capture  the mass transfer properly, thermodynamic information from the micro-scale in the form of the kinetic relation is needed. Inherently, we assume for all the applications under consideration that the width of the physical interface is smaller than the grid cell size. This motivates the approximation of the interface as a discontinuity. If this is not valid, then a diffuse interface treatment with a proper thermodynamic modelling should be the better approach.   
Our numerical approach showed good results in the validation against experimental investigations of rapid evaporation processes for alkanes. The multi-dimensional approach has been applied to several examples including a compressible shock-droplet interaction. The numerical approximation showed good results for all these cases.

Future work goes in several directions. Currently, the approach handles one fluid in the gas and liquid case. We will extend this to multiple gas components. In this paper we restricted ourselves mainly to the Euler equations. For the application \ref{sec:wobblingdrop} only, diffusion processes have been taken into account by adding a standard numerical diffusion flux. Besides this, the influences of viscosity and heat conduction at the two-phase interface were neglected. It will be considered in the future, because these effects may be very important for the behavior of the interface in other physical situations. A suitable model for the diffusion fluxes in a sharp interface approach may be also based on local Riemann problems as it was done in \cite{gassner2008dgrp2} for single phase fluid flow.

\section*{Acknowledgements}
The authors gratefully acknowledge the support of Joachim Gross and Christoph Klink concerning the estimation of the entropy production term
by means of the DFT. The work was partially supported by the German Research Foundation (DFG) through  
SFB~TRR~75 ``Droplet dynamics under extreme ambient conditions'' and the grant RO 2222/4-1.

\end{document}